\documentclass[journal,onecolumn,12pt]{IEEEtran}

\usepackage{cite}

\ifCLASSINFOpdf
\else
\fi
\usepackage[cmex10]{amsmath}
\usepackage{amssymb}

\usepackage{epsfig}
\usepackage{psfrag}

\usepackage[caption=false,font=footnotesize]{subfig}

\begin{document}
\title{On the Achievable Rate of Stationary Rayleigh Flat-Fading Channels with Gaussian Inputs}

\author{Meik~D\"orpinghaus,~\IEEEmembership{Member,~IEEE,}~and~Heinrich~Meyr,~\IEEEmembership{Life Fellow,~IEEE}
\thanks{This work has been supported by the UMIC (Ultra High Speed Mobile Information and Communication) research centre. The material in this paper was presented in part at the International Symposium on Information Theory and its Applications (ISITA) 2008, Auckland, New Zealand, December 2008, and at the International Symposium on Information Theory and its Applications (ISITA) 2010, Taichung, Taiwan, October 2010.}
\thanks{M. D\"orpinghaus was with the Institute for Integrated Signal Processing Systems, RWTH Aachen University, 52056 Aachen, Germany and is now with the Institute for Theoretical Information Technology, RWTH Aachen University, 52056 Aachen, Germany (e-mail: doerpinghaus@ti.rwth-aachen.de).}
\thanks{H. Meyr is with the Institute for Integrated Signal Processing Systems, RWTH Aachen University, 52056 Aachen, Germany (e-mail: meyr@iss.rwth-aachen.de).}}

\maketitle

\begin{abstract}
In this work, we consider a discrete-time stationary Rayleigh flat-fading channel with unknown channel state information at transmitter and receiver side. The law of the channel is presumed to be known to the receiver. In addition, we assume the power spectral density of the fading process to be compactly supported. For independent identically distributed (i.i.d.) zero-mean proper Gaussian input distributions, we investigate the achievable rate. One of the main contributions of the present paper is the derivation of two new upper bounds on the achievable rate with zero-mean proper Gaussian input symbols. The first one holds only for the special case of a rectangular power spectral density and depends on the SNR and the spread of the power spectral density. Together with a lower bound on the achievable rate, which is achievable with i.i.d.\ zero-mean proper Gaussian input symbols, we have found a set of bounds which is tight in the sense that their difference is bounded. Furthermore, we show that the high SNR slope is characterized by a \emph{pre-log} of $1-2f_{d}$, where $f_{d}$ is the normalized maximum Doppler frequency. This pre-log is equal to the high SNR pre-log of the peak power constrained capacity. Furthermore, we derive an alternative upper bound on the achievable rate with i.i.d.\ input symbols which is based on the one-step channel prediction error variance. The novelty lies in the fact that this bound is not restricted to peak power constrained input symbols like known bounds, e.g., in \cite{SetWanHajLap07Arxiv}. Therefore, the derived upper bound can also be used to evaluate the achievable rate with i.i.d.\ proper Gaussian input symbols. In addition, we compare the derived bounds on the achievable rate with i.i.d.\ zero-mean proper Gaussian input symbols with bounds on the peak power constrained capacity given in \cite{SetWanHajLap07Arxiv}, \cite{SetHaj05}, and \cite{La05}. Finally, we compare the achievable rate with i.i.d.\ zero-mean proper Gaussian input symbols with the achievable rate using synchronized detection in combination with a solely pilot based channel estimation.
\end{abstract}


\IEEEpeerreviewmaketitle

\section{Introduction}
\IEEEPARstart{I}{n} this paper, we consider a stationary Rayleigh flat-fading channel with temporal correlation. We assume that the channel state information is unknown to the transmitter and the receiver, while the receiver is aware of the channel law. The capacity of this scenario is particularly important, as it applies to many realistic mobile communication systems. In order to acquire channel state information, the temporal correlation of the channel can be exploited by the system, e.g., by inserting training sequences at transmit side. While these training sequences can be understood as a specific type of code \cite{La05}, we are interested in the achievable rate on this channel irrespective of the use of training sequences.

The capacity of fading channels where the channel state information is unknown, i.e., sometimes referred to as noncoherent capacity, has received a lot of attention in the recent literature. E.g., \cite{MaHo99} considers a block fading channel model, where the channel is assumed to be constant over a block of $N$ symbols and changes independently from block to block. This model is non-stationary and, therefore, different from the one we consider in the present work. On the other hand, in \cite{BaFoMe01a} and \cite{Med00} the achievable rate of time-continuous fading channels has been examined under the assumption of the use of training sequences for channel tracking and a coherent detection based on the acquired channel estimate. Furthermore, in \cite{La05} the asymptotic high SNR capacity of a stationary Gaussian fading channel has been investigated, whereas in \cite{EtTs06} an approximate behavior of the capacity for different SNR regimes has been considered. In addition, in \cite{SetWanHajLap07Arxiv} bounds on the capacity for temporally correlated Rayleigh fading channels with a peak power constraint have been derived with specific emphasis on the low SNR regime. Furthermore, the case of frequency selective stationary fading channels has been discussed, e.g., in \cite{DurSchuBoelSha08IT_pub} and \cite{KochLapArxiv09}. 

The main goal of the present work is the investigation of the achievable rate with independent identically distributed (i.i.d.) zero-mean proper Gaussian input symbols on noncoherent stationary discrete-time Rayleigh flat-fading channels. On the one hand, i.i.d.\ zero-mean proper Gaussian input symbols are capacity achieving in case the channel state is perfectly known at the receiver. Even though they are not capacity achieving for the given noncoherent scenario \cite{LapSha02}, the achievable rate with i.i.d.\ zero-mean proper Gaussian input symbols is highly interesting, as in many cases the capacity-achieving input distribution becomes peaky and, thus, impractical for real system design. In contrast, i.i.d.\ zero-mean proper Gaussian input distributions serve well to upper-bound the achievable rate with practical modulation and coding schemes, see also \cite{RusekLozanoJindalArxivISIT2010} and \cite{Ungerboeck82}. In \cite{RusekLozanoJindalArxivISIT2010} the achievable rate with i.i.d.\ Gaussian inputs has been studied for the block fading channel. Furthermore, in \cite{ChenHajekKoetterMadhow04} bounds on the mutual information with Gaussian input distributions have been derived for a Gauss-Markov fading channel, whose PSD has an unbounded support. The results in \cite{ChenHajekKoetterMadhow04} indicate that at moderate SNR and/or slow fading, Gaussian inputs still work well. In contrast to these publications, in the present work we study the achievable rate with i.i.d.\ zero-mean proper Gaussian input symbols for the case of a discrete-time stationary Rayleigh flat-fading channel, where the power spectral density (PSD) of the channel fading process is characterized by a compact support with a normalized maximum Doppler frequency $f_{d}<0.5$, i.e., \emph{nonregular} fading \cite{Doob90}, as it, e.g., corresponds to the widely used Jakes' model \cite{Jakes75}. 

\subsection{Contributions}
Within the present work, we consider a discrete-time stationary Rayleigh flat-fading channel with a compactly supported PSD. The channel fading process is assumed to be nonregular. Furthermore, it is assumed that the channel state information is unknown to the transmitter and the receiver, while the receiver is aware of the channel law. In this context we obtain the following.

In Section~\ref{SectionBoundsMain}, we give an upper bound on the achievable rate with i.i.d.\ zero-mean proper Gaussian input symbols for the special case of a rectangular PSD, depending on the SNR and the spread of the PSD. Especially, the therefor used lower bound on the conditional output entropy rate $h'(\mathbf{y}|\mathbf{x})$ for a rectangular PSD is, to the best of our knowledge, new. The particularity of the given lower bound on $h'(\mathbf{y}|\mathbf{x})$ lies in the fact that its derivation is not based on a peak power constraint, enabling its evaluation for Gaussian input symbols. The assumption of a rectangular PSD is usually made in typical communication system design. For comparison, we give a lower bound on the achievable rate, holding for an arbitrary PSD with compact support, which is already known from \cite{DengHaim04}. With this lower and this upper bound on the achievable rate with i.i.d.\ zero-mean proper Gaussian inputs, we have found a set of bounds which is tight in the sense that its difference is bounded by $(1+2f_{d})\gamma$ [nat/channel use] for all SNRs, where $\gamma\approx 0.57721$ is the Euler constant and $f_{d}$ is the normalized maximum Doppler frequency. Furthermore, in Section~\ref{SectCompPeakBounds}, we discuss the relation of the bounds on the achievable rate with i.i.d.\ zero-mean proper Gaussian input symbols to bounds on the peak power constrained capacity given in \cite{SetHaj05} and \cite{SetWanHajLap07Arxiv}.

We show, in Section~\ref{Sect_HighSNR_SISO}, that the asymptotic high SNR slope (\emph{pre-log}) of the achievable rate with i.i.d.\ zero-mean proper Gaussian input symbols is given by $1-2f_{d}$. This exactly corresponds to the high SNR behavior of the peak power constrained capacity as discussed in \cite{La05}. Furthermore, in Section~\ref{Section_Comp_Lapidoth}, we compare the bounds on the achievable rate with i.i.d.\ zero-mean proper Gaussian input symbols to the high SNR asymptotes on the peak power constrained capacity given in \cite{La05}.

Additionally, in Section~\ref{Section_AlternativeIIDBound}, we derive an alternative upper bound on the achievable rate with i.i.d.\ input symbols. This upper bound, whose derivation relies on the assumption on i.i.d.\ input symbols, is based on the one-step channel prediction error variance, and, thus, is linked to a physical interpretation. Compared to this, the bounds given in Section~\ref{SectionBoundsMain} are based on a purely mathematical derivation. There exist already bounds on the capacity with peak power constrained input distributions, which are based on the one-step channel prediction error variance, see, e.g., \cite{La05}. However, for the derivation of the channel prediction based capacity bounds in \cite{La05}, the peak power constraint has been required for technical reasons. Differently, our new upper bound based on the channel prediction error variance is not restricted to peak power constrained input symbols, enabling its evaluation for Gaussian inputs. However, due to the restriction to i.i.d.\ input symbols, we only get an upper bound on the achievable rate and not on the capacity. We evaluate this upper bound on the achievable rate with i.i.d.\ input symbols, on the one hand, for zero-mean proper Gaussian input symbols. In contrast to the upper bound on the achievable rate with i.i.d.\ zero-mean proper Gaussian input symbols derived in Section~\ref{SectionBoundsMain}, which holds only for a rectangular PSD of the fading process, the upper bound based on the channel prediction error variance holds for an arbitrary PSD with compact support. 

On the other hand, and this is out of the main focus of the present paper, we also evaluate the upper bound on the achievable rate with i.i.d.\ input symbols based on the channel prediction error variance for peak power constrained input symbols, and compare this new upper bound to capacity bounds given in \cite{SetWanHajLap07Arxiv}.

Finally, we compare the achievable rate with i.i.d.\ zero-mean proper Gaussian input symbols to the achievable rate with synchronized detection and a solely pilot based channel estimation. Using synchronized detection with a solely pilot based channel estimation, the channel is estimated solely based on pilot symbols and then, in a second separate step, the channel estimate is used for coherent detection. Thus, this comparison shows, how far such systems stay below the achievable rate with Gaussian code books, using no pilot symbols. This comparison might give an indication of the possible gain of advanced receivers using a joint processing of pilot and data symbols, in comparison to the separate processing. Such an instance of a joint processing of pilots and data symbols is, e.g., the approach of iterative code-aided channel estimation and decoding \cite{Valenti2001}, where the channel estimation is iteratively enhanced based on reliability information on the data symbols delivered by the decoder. The enhanced channel estimation, then in a further iteration allows for an enhanced decoding, and so on.

The rest of this paper is organized as follows. In Section~\ref{Sect_SysModelSISO}, we introduce the channel model including a discussion of its limitations. Section~\ref{SectionBoundsMain} presents the derivation of the bounds on the achievable rate with i.i.d.\ zero-mean proper Gaussian input symbols, which are based on a purely mathematical derivation. This includes the discussion of their tightness, the evaluation of the high SNR behavior, the comparison to the high SNR asymptotes for the capacity given in \cite{La05}, and the discussion of their relation to the bounds on the peak power constrained capacity given in \cite{SetHaj05} and \cite{SetWanHajLap07Arxiv}. Afterwards, in Section~\ref{Section_AlternativeIIDBound} an upper bound on the achievable rate with i.i.d.\ input symbols based on the channel prediction error variance is derived and evaluated for proper Gaussian inputs and for peak power constrained inputs. Subsequently, in Section~\ref{Sect_SyncDet} the achievable rate with i.i.d.\ zero-mean proper Gaussian input symbols is compared to the achievable rate with synchronized detection and a solely pilot based channel estimation, before we give a conclusion in Section~\ref{Sect_Conclusion}.

\section{System Model}\label{Sect_SysModelSISO}
We consider an ergodic discrete-time jointly proper Gaussian \cite{NeeserMassey1993} flat-fading channel, whose output at time $k$ is given by
\begin{align}
y_{k}&=h_{k} \cdot x_{k}+ n_{k}\label{EqBasic_IORealtion_SISOflat}
\end{align}
where $x_{k} \in \mathbb{C}$ is the complex-valued channel input, $h_{k} \in \mathbb{C}$ represents the channel fading coefficient, and $n_{k} \in \mathbb{C}$ is additive white Gaussian noise. The processes $\{h_{k}\}$, $\{x_{k}\}$, and $\{n_{k}\}$ are assumed to be mutually independent.

We assume that the noise $\{n_{k}\}$ is a sequence of i.i.d.\ proper Gaussian random variables of zero-mean and variance $\sigma_{n}^{2}$. The stationary channel fading process $\{h_{k}\}$ is zero-mean jointly proper Gaussian. In addition, the fading process is time-selective and characterized by its autocorrelation function
\begin{align}
r_{h}(l)&=\mathrm{E}[ h_{k+l}\cdot h^{*}_{k}]. \label{Corr_h}
\end{align}
Its variance is given by $r_{h}(0)=\sigma_{h}^{2}$.

The normalized PSD of the channel fading process is defined by
\begin{align}
S_{h}(f)&=\sum_{l=-\infty}^{\infty} r_{h}(l)e^{-j 2\pi l f}, \quad |f|< 0.5 \label{PowerSpec}
\end{align}
where we assume that the PSD exists and use the definition $j=\sqrt{-1}$. Here, the frequency $f$ is normalized with respect to the symbol duration $T_{\textrm{Sym}}$. In the following, we use this normalized PSD and, thus, refer to it as PSD for simplification. For a jointly proper Gaussian process, the existence of the PSD implies ergodicity \cite{Sethuraman2005}. As the channel fading process $\{h_{k}\}$ is assumed to be stationary, $S_{h}(f)$ is real-valued. Because of the limitation of the velocity of the transmitter, the receiver, and of objects in the environment, the spread of the PSD is limited, and we assume it to be compactly supported within the interval $[-f_{d},f_{d}]$, with $0<f_{d}< 0.5$, i.e., $S_{h}(f)=0$ for $f\notin [-f_{d},f_{d}]$.
The parameter $f_{d}$ corresponds to the normalized maximum Doppler shift and, thus, indicates the dynamics of the channel. To ensure ergodicity, we exclude the case $f_{d}=0$. Following the definition given in \cite{Doob90}, this fading channel is sometimes referred to as \emph{nonregular}.

For technical reasons, in some of the proofs, i.e., for the calculation of the upper bound on the achievable data rate in Section~\ref{SectionBoundsMain}, we restrict to autocorrelation functions $r_{h}(l)$ which are absolutely summable, i.e.,
\begin{align}
\sum_{l=-\infty}^{\infty}|r_{h}(l)|&< \infty \label{CorrAbsoluteSummableNew}
\end{align}
instead of the more general class of square summable autocorrelation functions, i.e.,
\begin{align}
\sum_{l=-\infty}^{\infty}|r_{h}(l)|^{2}&< \infty. \label{CorrSquareSummableNew}
\end{align}
The assumption of absolutely summable autocorrelation functions is not a severe restriction. E.g., the important rectangular PSD, see below in (\ref{Rect_PSD}), can be arbitrarily closely approximated by a PSD with the shape corresponding to the transfer function of a raised cosine filter, whose autocorrelation function is absolutely summable, see Appendix~\ref{Appendix_Approx_RectRC}. Therefore, in the rest of this work, we often evaluate the derived bounds on the achievable rate for a rectangular PSD of the channel fading process, although some of the derivations are based on the assumption of an absolutely summable autocorrelation function.

A common model concerning the temporal correlation of the channel fading process $S_{h}(f)$ is the Jakes' model \cite{Jakes75}, for which the corresponding PSD of the discrete-time fading process $S_{h}(f)$ in (\ref{PowerSpec}) is given by
\begin{align}
S_{h}(f)\big|_{\textrm{Jakes}}&=\left\{\begin{array}{ll}
\frac{\sigma_{h}^{2}}{\pi\sqrt{f_{d}^{2}-f^{2}}} & \textrm{for } |f|<f_{d}\\
0 & \textrm{for } f_{d}\le |f| \le 0.5
\end{array}\right. .\label{Jakes_PSD}
\end{align}
This PSD can be derived analytically for a dense scatterer environment with a vertical receive antenna and a constant azimuthal gain, a uniform distribution of signals arriving at all angles with phases being independently distributed over all angles, i.e., in the interval $[0, 2\pi)$, based on a sum of sinusoids \cite{Jakes75}.

Often the Jakes' PSD in (\ref{Jakes_PSD}) is approximated by the following rectangular PSD
\begin{align}
S_{h}(f)\big|_{\textrm{Rect}}&=\left\{\begin{array}{ll}
\frac{\sigma_{h}^{2}}{2f_{d}} & \textrm{for } |f|\le f_{d}\\
0 & \textrm{for } f_{d}< |f| \le 0.5
\end{array}\right. .\label{Rect_PSD}
\end{align}
For the derivation of the upper bound on the achievable rate in Section~\ref{SectionBoundsMain} we restrict to rectangular PSDs for mathematical tractability. 

Typical fading channels, as they are observed in mobile communication environments, are characterized by relatively small normalized Doppler frequencies $f_{d}$ in the regime of $f_{d} \ll 0.1$. Therefore, the restriction to channels with $f_{d}<0.5$, i.e., \emph{nonregular} fading, in the present work is reasonable. Although in practical scenarios the observed channel dynamics are very small, within this work, we consider the range of $0<f_{d}<0.5$ to get a thorough understanding of the behavior of the bounds on the achievable rate.

\subsection{Matrix-Vector Notation}\label{Sect_MatrixVector}
We base the derivation of bounds on the achievable rate on the following matrix-vector notation of the system model:
\begin{align}
\mathbf{y}&=\mathbf{H}\cdot \mathbf{x}+\mathbf{n}=\mathbf{X}\cdot \mathbf{h}+\mathbf{n} \label{Mod1}
\end{align}
where the vectors $\mathbf{x}$ is defined as
\begin{align}
\mathbf{x}&=[x_{1}, \hdots, x_{N} ]^T.
\end{align}
The vectors $\mathbf{y}$ and $\mathbf{n}$ are defined analogously. The matrix $\mathbf{H}$ is diagonal and defined as $\mathbf{H}=\textrm{diag}(\mathbf{h})$ with $\mathbf{h}=[h_{1}, \hdots, h_{N}]^{T}$. Here the $\textrm{diag}(\cdot)$ operator generates a diagonal matrix whose diagonal elements are given by the argument vector. The diagonal matrix $\mathbf{X}$ is given by
$\mathbf{X}=\textrm{diag}(\mathbf{x})$. The quantity $N$ is the number of considered symbols. Later on, we investigate the case of $N\rightarrow \infty$ to evaluate the achievable rate.

Using this vector notation, we express the temporal correlation of the fading process by the correlation matrix
\begin{align}
\mathbf{R}_{h}&=\mathrm{E}[\mathbf{h} \mathbf{h}^{H}]\label{Cha_Autocov_matrix}
\end{align}
which has a Hermitian Toeplitz structure.

Concerning the input distribution, unless otherwise stated, we make the assumption that the symbols $x_{k}$ are i.i.d.\ zero-mean proper Gaussian distributed with an average power $\sigma_{x}^{2}$. Thus, the average SNR is given by\footnote{Remark: Only in case there is no peak power constraint, as, e.g., in the case of Gaussian input symbols, the average SNR $\rho$ is in general equal to the actual average SNR. In contrast, in case of an additional peak power constraint the achievable rate is in general not maximized by using the maximum average transmit power $\sigma_{x}^{2}$. Thus, in this case $\rho$ does not necessarily correspond to the actual average SNR and, therefore, we name it \emph{nominal} average SNR when considering a peak power constraint. For further discussion see Section~\ref{Section_Comp_Lapidoth}  and Section~\ref{Section_Peak_UpperBound}.} 
\begin{align}
\rho&=\frac{\sigma_{x}^{2}\sigma_{h}^{2}}{\sigma_{n}^{2}}.\label{Def_SNR}
\end{align}

\subsection{Limitations of the Symbol Rate Discrete-Time Model}\label{SISOFlatModelLimit}
To discuss the limitations of the symbol rate discrete-time model given in (\ref{EqBasic_IORealtion_SISOflat}), we start from the underlying appropriately bandlimited continuous-time model, where the channel output is given by
\begin{align}
y(t)&=h(t)\cdot s(t)+n(t)\label{SISOtimecontinuousModel}
\end{align}
with $h(t)$ being the continuous-time channel fading process, i.e., the corresponding discrete-time process $h_{k}$ is given by
\begin{align}
h_{k}&=h(k T_{\textrm{Sym}})
\end{align}
where $T_{\textrm{Sym}}$ is the symbol duration. Analogously, the continuous-time and the discrete-time additive noise and channel output processes are related by
\begin{align}
n_{k}&=n(k T_{\textrm{Sym}})\\
y_{k}&=y(k T_{\textrm{Sym}}).
\end{align}
The continuous-time transmit process $s(t)$ is given by
\begin{align}
s(t)&=\sum_{k=-\infty}^{\infty} x_{k} \cdot g(t-k T_{\textrm{Sym}})
\end{align}
where $g(t)$ is the transmit pulse. We assume the use of bandlimited transmit pulses, which, therefore, have an infinite impulse response. In typical systems, often root-raised cosine pulses are used such that in combination with the matched filter at the receiver intersymbol interference is minimized. Their normalized frequency response $G(f)$ is given by
\begin{align}
G(f)&=\sqrt{G_{\textrm{RC}}(f)}
\end{align}
with $G_{\textrm{RC}}(f)$ being the transfer function of the raised cosine filter
\begin{align}
G_{\textrm{RC}}(f)&=\left\{\begin{array}{ll}
T_{\textrm{Sym}} & \textrm{for } |f|\le \frac{1-\beta_{\textrm{ro}}}{2}\\
&\\
\frac{T_{\textrm{Sym}}}{2}\left[1+\cos\left(\frac{\pi}{\beta_{\textrm{ro}}}\left[|f|-\frac{1-\beta_{\textrm{ro}}}{2}\right]\right)\right] & \textrm{for } \frac{1-\beta_{\textrm{ro}}}{2}<|f|\le \frac{1+\beta_{\textrm{ro}}}{2}\\
&\\
0 & \textrm{otherwise}
\end{array}\right.
\end{align}
where $0\le \beta_{\textrm{ro}}\le 1$ is the roll-off factor.

The continuous-time input/output relation in (\ref{SISOtimecontinuousModel}) has the following stochastic representation in frequency domain
\begin{align}
S_{y}(f)&=S_{h}(f) \star S_{s}(f) +S_{n}(f)
\end{align}
where $\star$ denotes convolution and $S_{y}(f)$, $S_{h}(f)$, $S_{s}(f)$, and $S_{n}(f)$ are the normalized power spectral densities of the continuous-time processes $y(t)$, $h(t)$, $s(t)$, and $n(t)$, e.g.,
\begin{align}
S_{s}(f)&=\int_{-\infty}^{\infty}\mathrm{E}\left[s(t+\tau)s^{*}(t)\right]e^{-j2\pi f \tau}d\tau
\end{align}
and correspondingly for the other PSDs. Here, we always assume normalization with $1/T_{\textrm{Sym}}$.

We are interested in the normalized bandwidth of the component $S_{h}(f) \star S_{s}(f)$, i.e., the component containing information on the transmitted sequence $\left\{x_{k}\right\}$. The normalized bandwidth of the transmit signal $s(t)$ directly corresponds to the normalized bandwidth of the transmit pulse $g(t)$, which is at least $1$ (corresponding to $\beta_{\textrm{ro}}=0$). The normalized bandwidth of the channel fading process is given by $2f_{d}$. Thus, the normalized bandwidth of the component $S_{h}(f) \star S_{s}(f)$ is given by at least $1+2f_{d}$. To get a sufficient statistic, we would have to sample the channel output $y(t)$ at least with a frequency of $\frac{1+2f_{d}}{T_{Sym}}$ (for $\beta_{\textrm{ro}}=0$). As the discrete-time channel output process $\{y_{k}\}$ is a sampled version of $y(t)$ with the rate $1/T_{\textrm{Sym}}$, the discrete-time observation process $\{y_{k}\}$ is not a sufficient statistic of $y(t)$. This shows the limitations of the symbol rate discrete-time system model in (\ref{EqBasic_IORealtion_SISOflat}). As in typical systems the normalized maximum Doppler frequency $f_{d}$ is small in comparison to the symbol rate $1/T_{\textrm{Sym}}$, the amount of discarded information is negligible. Besides this, in typical systems channel estimation is also performed at symbol rate and, therefore, also exhibit the loss due to the lack of a sufficient statistic. In addition, the majority of the current literature on the study of the capacity of stationary Rayleigh fading channels, e.g., \cite{La05} or \cite{SetWanHajLap07Arxiv}, is based on symbol rate discrete-time input-output relations and therefore do not ask the question about a sufficient statistic. Nevertheless, these considerations should be kept in mind in the following, especially, as we examine the derived bounds not only for very small values of $f_{d}$.

\section{Bounds on the Achievable Rate}\label{SectionBoundsMain}
The information theoretic capacity is defined by
\begin{align}
C&=\lim_{N\rightarrow \infty}\sup_{\mathcal{P}}\frac{1}{N}\mathcal{I}(\mathbf{y};\mathbf{x}) \quad \textrm{[nat/cu]}\label{DefInforTheoreticCapacity}
\end{align}
where $\mathcal{I}(\mathbf{y};\mathbf{x})$ is the mutual information between $\mathbf{x}$ and $\mathbf{y}$, where \emph{cu} is an abbreviation for \emph{channel use}, and where the supremum is taken over the set $\mathcal{P}$ containing all input distributions with an average power constrained to $\sigma_{x}^{2}$, i.e.,
\begin{align}
\mathcal{P}&=\left\{p(\mathbf{x})\left|\mathbf{x}\in \mathbb{C}^{N}, \frac{1}{N}\mathrm{E}\left[\mathbf{x}^{H}\mathbf{x}\right]\le \sigma_{x}^{2}\right.\right\}.\label{SetDef_General}
\end{align}
As the PSD of the fading process in (\ref{PowerSpec}) is assumed to exist, and as the channel fading process is jointly proper Gaussian, the channel fading process is ergodic. Therefore, the information theoretic capacity given in (\ref{DefInforTheoreticCapacity}) and the operational capacity coincide \cite{Sethuraman2005}, i.e., for each rate $R<C$ there exist a code for which the probability of an erroneously decoded codeword approaches zero in the limit of an infinite codeword length. 

As it has already been discussed, the main focus of the present paper is not the discussion of the capacity but of the achievable rate with i.i.d.\ zero-mean proper Gaussian input symbols. As this kind of input distribution is in general not capacity achieving, we use the term \emph{achievable rate} $R$, which then directly corresponds to the mutual information rate $\mathcal{I}'(\mathbf{y};\mathbf{x})$, i.e.,
\begin{align}
R&=\mathcal{I}'(\mathbf{y};\mathbf{x})=\lim_{N\rightarrow
\infty}\frac{1}{N}\mathcal{I}(\mathbf{y};\mathbf{x})\label{Basic2}
\end{align}
where, as described in Section~\ref{Sect_SysModelSISO}, the elements of $\mathbf{x}$ are i.i.d.\ zero-mean proper Gaussian distributed. 

\subsection{The Mutual Information Rate $\mathcal{I}'(\mathbf{y};\mathbf{x})$}
In general, by means of the chain rule, the mutual information rate in (\ref{Basic2}) can be expanded as \cite{Biglieri1998}
\begin{align}
\mathcal{I}'(\mathbf{y};\mathbf{x})&=\mathcal{I}'(\mathbf{y};\mathbf{x}|\mathbf{h})-\mathcal{I}'(\mathbf{x};\mathbf{h}|\mathbf{y})
\label{MutInf}
\end{align}
where $\mathcal{I}'(\mathbf{y};\mathbf{x}|\mathbf{h})$ is the mutual information rate in case the channel is known at the receiver, i.e., the mutual information rate of the coherent channel, and $\mathcal{I}'(\mathbf{x};\mathbf{h}|\mathbf{y})$ is the penalty due to the channel uncertainty. It is interesting to note that the penalty term can be further separated as follows:
\begin{align}
\mathcal{I}'(\mathbf{x};\mathbf{h}|\mathbf{y})&\stackrel{(a)}{=}\mathcal{I}'(\mathbf{y},\mathbf{x};\mathbf{h})-\mathcal{I}'(\mathbf{y};\mathbf{h})\nonumber\\
&\stackrel{(b)}{=}\mathcal{I}'(\mathbf{y};\mathbf{h}|\mathbf{x})+\mathcal{I}'(\mathbf{h};\mathbf{x})-\mathcal{I}'(\mathbf{y};\mathbf{h})\nonumber\\
&\stackrel{(c)}{=}\mathcal{I}'(\mathbf{y};\mathbf{h}|\mathbf{x})-\mathcal{I}'(\mathbf{y};\mathbf{h})\label{Basic4}
\end{align}
where for (a) and (b) we use the chain rule for mutual information and for (c) we exploit the fact that the mutual information between the channel fading process described by $\mathbf{h}$ and the input sequence $\mathbf{x}$ is zero due to the independency of $\mathbf{h}$ and $\mathbf{x}$ and, thus,
\begin{align}
\mathcal{I}'(\mathbf{h};\mathbf{x})&=0.\label{Basic5}
\end{align}
Obviously, with (\ref{Basic4}) the penalty term corresponds to the difference between the knowledge on the channel $\mathbf{h}$ that can be obtained from the observation $\mathbf{y}$ while knowing the transmit sequence $\mathbf{x}$ in comparison to not knowing it.

However, the derivation of the bounds on the mutual information rate $\mathcal{I}'(\mathbf{y};\mathbf{x})$ in the present work is based on the following straightforward separation of $\mathcal{I}'(\mathbf{y};\mathbf{x})$ into the differential entropy rates
\begin{align}
\mathcal{I}'(\mathbf{y};\mathbf{x})&=h'(\mathbf{y})-h'(\mathbf{y}|\mathbf{x}).\label{Mut_first_limes}
\end{align}
where $h'(\cdot)$ denotes the differential entropy rate
\begin{align}
h'(\cdot )&=\lim_{N\rightarrow \infty}\frac{1}{N}h(\cdot ).
\end{align}

In Section~\ref{Sect_hy}, we give a lower and an upper bound on the channel output entropy rate $h'(\mathbf{y})$, which are independent of the PSD of the channel fading process $S_{h}(f)$. In Section~\ref{Sect_hyx}, we derive an upper bound and the lower bound on $h'(\mathbf{y}|\mathbf{x})$. The upper bound, which is already known from \cite{DengHaim04}, holds for an arbitrary PSD of the channel fading process with compact support. For the lower bound on $h'(\mathbf{y}|\mathbf{x})$ we find a closed form expression only for the special case of a rectangular PSD.

\subsection{The Received Signal Entropy Rate $h'(\mathbf{y})$}\label{Sect_hy}
\subsubsection{Lower Bound on $h'(\mathbf{y})$}\label{SubSect_LowBound_hy} 
The mutual information with perfect channel state information at the receiver can be upper-bounded by
\begin{align}
\mathcal{I}(\mathbf{y};\mathbf{x}|\mathbf{h})&=h(\mathbf{y}|\mathbf{h})-h(\mathbf{y}|\mathbf{h},\mathbf{x})\nonumber\\
&\le h(\mathbf{y})-h(\mathbf{y}|\mathbf{h},\mathbf{x}).
\end{align}
Here, we make use of the fact that conditioning reduces entropy. Thus, we can lower-bound the entropy rate $h'(\mathbf{y})$ by
\begin{align}
h'(\mathbf{y})&\ge
\mathcal{I}'(\mathbf{y};\mathbf{x}|\mathbf{h})+h'(\mathbf{y}|\mathbf{h},\mathbf{x}).\label{BoundCondEntrop}
\end{align}

The mutual information rate in case the channel is known at the receiver, i.e., the first term on the RHS of (\ref{BoundCondEntrop}), is given by
\begin{align}
\mathcal{I}'(\mathbf{y};\mathbf{x}|\mathbf{h})&=\frac{1}{N}\mathrm{E}_{\mathbf{h}} \left[
\mathrm{E}_{\mathbf{y,x}} \left[\left. \log\left(
\frac{p(\mathbf{y}|\mathbf{h},\mathbf{x})}{p(\mathbf{y}|\mathbf{h})}\right)
\right|\mathbf{h}\right]\right]\nonumber\\
&\stackrel{(a)}{=}\mathrm{E}_{h_{k}} \left[
\mathrm{E}_{y_{k},x_{k}} \left[\left. \log\left(
\frac{p(y_{k}|h_{k},x_{k})}{p(y_{k}|h_{k})}\right)
\right|h_{k}\right]\right]\nonumber\\
&=\mathcal{I}(y;x|h)\nonumber\\
&\stackrel{(b)}{=}\mathrm{E}_{h_{k}}\left[\log\left(
1+\rho \frac{|h_{k}|^{2}}{\sigma_{h}^{2}}\right)
\right]\nonumber\\
&=\int_{z=0}^{\infty}\log\left(1+\rho z\right)e^{-z}dz\label{PerfKnow_Gen}
\end{align}
where (a) is based on the fact that due to conditioning on the channel fading vector $\mathbf{h}$ the channel uses become independent and we furthermore assume i.i.d.\ input symbols.\footnote{All logarithms in this paper are to the base $e$ and, unless stated otherwise, all rates are in \emph{nat}.} Therefore, we can drop the time index for ease of notation. Finally, (b) holds for i.i.d.\ zero-mean proper Gaussian inputs which are capacity achieving in the coherent case. Thus, the RHS of (\ref{PerfKnow_Gen}) is the capacity of the coherent channel. Obviously, the coherent capacity is independent of the temporal correlation of the channel, see, e.g., \cite{Med00}.

The second term on the RHS of (\ref{BoundCondEntrop}) originates from AWGN and, thus, can be calculated as
\begin{align}
h'(\mathbf{y}|\mathbf{h},\mathbf{x})&= \log \left( \pi e \sigma_{n}^{2} \right). \label{NoiseEntropy}
\end{align}

Hence, a lower bound on the entropy rate $h'(\mathbf{y})$ is given by
\begin{align}
h'(\mathbf{y})&\ge h_{L}'(\mathbf{y})=\int_{z=0}^{\infty}\log\left[\pi e \left(\sigma_{n}^{2}+\sigma_{x}^{2}\sigma_{h}^{2} z \right)\right]e^{-z} dz.\label{low_hy_basic}
\end{align}

\subsubsection{Upper Bound on $h'(\mathbf{y})$}\label{SubSect_UppBound_hy} 
In this section, we give an upper bound on the entropy rate $h'(\mathbf{y})$. First, we make use of the fact that the entropy $h(\mathbf{y})$ of a zero-mean complex random vector $\mathbf{y}$ of dimension $N$ with nonsingular correlation matrix $\mathbf{R}_{y}=\mathrm{E}[\mathbf{y}\mathbf{y}^{H}]$ is upper-bounded by \cite{NeeserMassey1993}
\begin{align}
h(\mathbf{y})&\le \log\left[(\pi e)^{N}\det(\mathbf{R}_{y})\right]\nonumber\\
&\stackrel{(a)}{=}N\log\left[\pi e\left(\sigma_{x}^{2}\sigma_{h}^{2}+\sigma_{n}^{2}\right)\right].\label{UpBoundEntropProp}
\end{align}
Here (a) follows from the fact that $\mathbf{R}_{y}$ is diagonal and given by
\begin{align}
\mathbf{R}_{y}&=(\sigma_{x}^{2}\sigma_{h}^{2}+\sigma_{n}^{2})\mathbf{I}_{N}.
\end{align}
due to the assumption on i.i.d.\ input symbols. Nevertheless, the upper bound on $h(\mathbf{y})$ in (\ref{UpBoundEntropProp}) also holds without the assumption on independent input symbols, which can be easily verified by using Hadamard's inequality instead of the equality in (a).

Hence, with (\ref{UpBoundEntropProp}) the upper bound $h'_{U}(\mathbf{y})$ on the entropy rate
$h'(\mathbf{y})$ is given by
\begin{align}
h'(\mathbf{y})&\le h'_{U}(\mathbf{y})=\log\left(\pi e
\left(\sigma_{x}^{2}\sigma_{h}^{2}+\sigma_{n}^{2}\right)\right).\label{h_U2}
\end{align}

In Appendix~\ref{hy_upp_num_uncorr}, we give another upper bound on $h'(\mathbf{y})$ for the case of zero-mean proper Gaussian inputs based on numerical integration to calculate $h(y_{k})$, i.e., the output entropy at an individual time instant, see also \cite{Perera}. As this bound can only be evaluated numerically using Hermite polynomials and Simpson's rule or by Monte Carlo integration, we do not further consider it here.

\subsection{The Entropy Rate $h'(\mathbf{y}|\mathbf{x})$}\label{Sect_hyx} 
In this section, we give an upper bound and a lower bound on the conditional channel output entropy rate $h'(\mathbf{y}|\mathbf{x})$. The probability density of $\mathbf{y}$ conditioned on $\mathbf{x}$ is zero-mean proper Gaussian. Therefore, its entropy is
\begin{align}
h(\mathbf{y}|\mathbf{x})&=\mathrm{E}_{\mathbf{x}}\left[\log\left((\pi e)^{N}
\det(\mathbf{R}_{y|x})\right)\right]\label{h_y_x_zero_new}
\end{align}
where the covariance matrix $\mathbf{R}_{y|x}$ is given by
\begin{align}
\mathbf{R}_{y|x}&=\mathrm{E}_{\mathbf{h},\mathbf{n}}\left[\mathbf{y} \mathbf{y}^{H}\big|\mathbf{x}
\right]\nonumber\\
&=\mathrm{E}_{\mathbf{h}}\left[\mathbf{X}\mathbf{h}\mathbf{h}^{H} \mathbf{X}^{H}\big|\mathbf{x}\right]+\sigma_{n}^{2}
\mathbf{I}_{N}\nonumber\\&=\mathbf{X} \mathbf{R}_{h}\mathbf{X}^{H}+ \sigma_{n}^{2}
\mathbf{I}_{N}.\label{ForDiss_PDF_ycondX}
\end{align}
As the channel correlation matrix $\mathbf{R}_{h}$ is Hermitian and, thus, normal, the spectral decomposition theorem applies, i.e.,
\begin{align}
\mathbf{R}_{h}&=\mathbf{U}\mathbf{\Lambda}_{h}\mathbf{U}^{H}\label{Corr_Rh__}
\end{align}
where the diagonal matrix $\mathbf{\Lambda}_{h}=\textrm{diag}\left(\lambda_{1},\hdots,\lambda_{N}\right)$ contains the eigenvalues $\lambda_i$ of $\mathbf{R}_{h}$ and the matrix $\mathbf{U}$ is unitary.

\subsubsection{Upper Bound on $h'(\mathbf{y}|\mathbf{x})$}\label{Sect_hyx_Upper} 
The following upper-bounding of $h(\mathbf{y}|\mathbf{x})$ is already known from \cite{DengHaim04}. Making use of (\ref{Corr_Rh__}), Jensen's inequality and the concavity of the log function, we can upper-bound $h(\mathbf{y}|\mathbf{x})$ in (\ref{h_y_x_zero_new}) as follows:
\begin{align}
h(\mathbf{y}|\mathbf{x})&=\mathrm{E}_{\mathbf{x}}\left[\log \det
\left(\frac{1}{\sigma_{n}^{2}}\mathbf{X}\mathbf{U}\mathbf{\Lambda}_{h}\mathbf{U}^{H}\mathbf{X}^{H}+\mathbf{I}_{N}\right)\right]+N
\log(\pi e \sigma_{n}^{2})\label{h_y_x_first}\\
&\stackrel{(a)}{=} \mathrm{E}_{\mathbf{x}}\left[\log \det
\left(\frac{1}{\sigma_{n}^{2}}\mathbf{X}^{H}\mathbf{X}\mathbf{U}\mathbf{\Lambda}_{h}\mathbf{U}^{H}+\mathbf{I}_{N}\right)\right]+N
\log(\pi e \sigma_{n}^{2})\label{h_y_x_second}\\
&\stackrel{(b)}{\le}\log \det
\left(\frac{\sigma_{x}^{2}}{\sigma_{n}^{2}}\mathbf{U}\mathbf{\Lambda}_{h}\mathbf{U}^{H}+\mathbf{I}_{N}\right)+N
\log(\pi e \sigma_{n}^{2})
\nonumber\\
&=\log \det \left(\frac{\sigma_{x}^{2}}{\sigma_{n}^{2}}\mathbf{\Lambda}_{h}+\mathbf{I}_{N}\right)+N \log(\pi
e \sigma_{n}^{2})
\nonumber\\
&=\sum_{i=1}^{N}\log \left(\frac{\sigma_{x}^{2}}{\sigma_{n}^{2}}\lambda_{i}+1\right)+N \log(\pi e
\sigma_{n}^{2}).\label{hyx_finite}
\end{align}
For (a) the following relation is used
\begin{align}
&\det(\mathbf{AB}+\mathbf{I})=\det(\mathbf{BA}+\mathbf{I})\label{DetEqual}
\end{align}
which holds as $\mathbf{A}\mathbf{B}$ has the same eigenvalues as $\mathbf{B}\mathbf{A}$ for $\mathbf{A}$ and $\mathbf{B}$ being square matrices \cite[Theorem 1.3.20]{Horn}. For (b) we have used the fact that $\log\det(\cdot)$ is concave on the set of positive definite matrices\footnote{For the special case of independent transmit symbols, (b) can also be shown in two steps by using Jensen's inequality and in a second step expressing the determinant by a Laplacian expansion by minors to calculate the expectation, i.e.,
\begin{align}
\mathrm{E}_{\mathbf{x}}\left[\log \det
\left(\frac{1}{\sigma_{n}^{2}}\mathbf{X}^{H}\mathbf{X}\mathbf{U}\mathbf{\Lambda}_{h}\mathbf{U}^{H}+\mathbf{I}_{N}\right)\right]&\le \log \mathrm{E}_{\mathbf{x}}\left[\det
\left(\frac{1}{\sigma_{n}^{2}}\mathbf{X}^{H}\mathbf{X}\mathbf{U}\mathbf{\Lambda}_{h}\mathbf{U}^{H}+\mathbf{I}_{N}\right)\right]\nonumber\\
&=\log \det
\left(\frac{\sigma_{x}^{2}}{\sigma_{n}^{2}}\mathbf{U}\mathbf{\Lambda}_{h}\mathbf{U}^{H}+\mathbf{I}_{N}\right).\nonumber
\end{align}}.

To calculate the bound on the entropy rate $h'(\mathbf{y}|\mathbf{x})$, we consider the case $N\rightarrow\infty$, i.e., the dimension of the matrix $\mathbf{\Lambda}_{h}$ grows without bound. As $\mathbf{R}_{h}$ is Hermitian Toeplitz, we can evaluate (\ref{hyx_finite}) using Szeg\"o's theorem on the asymptotic eigenvalue distribution of Hermitian Toeplitz matrices \cite{Gray_ToeplitzReview}, \cite{Szgo58}. Consequently,
\begin{align}
\lim_{N\rightarrow \infty}\frac{1}{N}\sum_{i=1}^{N}\log
\left(\frac{\sigma_{x}^{2}}{\sigma_{n}^{2}}\lambda_{i}+1\right)&=\int_{-\frac{1}{2}}^{\frac{1}{2}}
\log\left(S_{h}(f)\frac{\sigma_{x}^{2}}{\sigma_{n}^{2}}+1\right)df.
\end{align}
Notice that due to the assumption that the PSD exists, the autocorrelation function of the channel fading process is square summable, see (\ref{CorrSquareSummableNew}), and, thus, Szeg\"o's theorem can be applied.

Hence, we get the following upper bound:
\begin{align}
h'(\mathbf{y}|\mathbf{x})&\le
h'_{U}(\mathbf{y}|\mathbf{x})\nonumber\\
&=\int_{-\frac{1}{2}}^{\frac{1}{2}}
\log\left(S_{h}(f)\frac{\sigma_{x}^{2}}{\sigma_{n}^{2}}+1\right)df+\log(\pi e
\sigma_{n}^{2}).\label{hyxUpp_final}
\end{align}

At this point, it is interesting to note that for constant modulus (CM) input symbols the differential entropy rate $h'(\mathbf{y}|\mathbf{x})$ is equal to the upper bound $h'_{U}(\mathbf{y}|\mathbf{x})$, i.e.,
\begin{align}
h'(\mathbf{y}|\mathbf{x})\big|_{\textrm{CM}}&=h'_{U}(\mathbf{y}|\mathbf{x})\label{hyx_CM_U_equal}
\end{align}
as in this case (\ref{h_y_x_second}) simplifies as
\begin{align}
\mathbf{X}^{H}\mathbf{X}\big|_{\textrm{CM}}&=\sigma_{x}^{2}\mathbf{I}
\end{align}
and, thus, (b) succeeding (\ref{h_y_x_second}) holds with equality.

\subsubsection{Lower Bound on $h'(\mathbf{y}|\mathbf{x})$ for a Rectangular PSD}\label{Sect_Lowhyx} 
In this section, we give a new lower bound on the entropy rate $h'(\mathbf{y}|\mathbf{x})$ for the special case of a rectangular PSD, which is a common approximation of the actual PSD in typical system design.

For the derivation of this lower bound, we derive a circulant matrix which is asymptotically equivalent to the Toeplitz matrix $\mathbf{R}_{h}$. Hereby, we follow a specific approach as shown in \cite{Gray_ToeplitzReview}, where the circulant matrix is constructed by sampling the PSD of the channel fading process. For the discussion of asymptotical equivalency, we write $\mathbf{R}_{h}^{(N)}$ instead of $\mathbf{R}_{h}$, where the superscript $(N)$ denotes the size of the square matrix $\mathbf{R}_{h}$.\looseness-1

Let the first column of the circulant matrix $\mathbf{C}_{h}^{(N)}$ be given by
\begin{align}
&\left(\begin{array}{cccc}
c_{0}^{(N)} & c_{1}^{(N)} & \hdots & c_{N-1}^{(N)}
\end{array}\right)^{T}
\end{align}
where again the superscript $(N)$ denotes the size of the square matrix $\mathbf{C}_{h}^{(N)}$. The elements $c_{k}^{(N)}$ are given by
\begin{align}
c_{k}^{(N)}&=\frac{1}{N}\sum_{l=0}^{N-1}\tilde{S}_{h}\left(\frac{l}{N}\right)e^{j2\pi k\frac{l}{N}}\label{DefCircEntries}
\end{align}
where $\tilde{S}_{h}(f)$ is the periodic continuation of $S_{h}(f)$ given in (\ref{PowerSpec}), i.e., 
\begin{align}
\tilde{S}_{h}(f)&=\sum_{k=-\infty}^{\infty}\delta(f-k)\star S_{h}(f)\label{defineContPSD}
\end{align}
and $S_{h}(f)$ being zero outside the interval $|f|\le 0.5$ for which it is defined in (\ref{PowerSpec}). 

As we assume that the autocorrelation function of the channel fading process is absolutely summable, the PSD of the channel fading process $\tilde{S}_{h}(f)$ is Riemann integrable, and it holds that
\begin{align}
\lim_{N\rightarrow \infty} c_{k}^{(N)}&=\lim_{N\rightarrow \infty} \frac{1}{N}\sum_{l=0}^{N-1}\tilde{S}_{h}\left(\frac{l}{N}\right)e^{j2\pi k\frac{l}{N}}\nonumber\\
&=\int_{-\frac{1}{2}}^{\frac{1}{2}}S_{h}(f)e^{j2\pi k f}df=r_{h}(k)
\end{align}
with $r_{h}(k)$ given by (\ref{Corr_h}).

As the eigenvectors of a circulant matrix are given by a discrete Fourier transform (DFT), the eigenvalues $\tilde{\lambda}_{k}^{(N)}$ with $k=1,\hdots,N$ of the circulant matrix $\mathbf{C}_{h}^{(N)}$ are given by
\begin{align}
\tilde{\lambda}_{k}^{(N)}&=\sum_{l=0}^{N-1}c_{l}^{(N)}e^{-j2\pi \frac{(k-1) l}{N}}\nonumber\\
&=\sum_{l=0}^{N-1}\left( \frac{1}{N}\sum_{m=0}^{N-1}\tilde{S}_{h}\left(\frac{m}{N}\right)e^{j2\pi \frac{m l}{N}}\right)e^{-j2\pi \frac{l (k-1)}{N}}\nonumber\\
&=\sum_{m=0}^{N-1}\tilde{S}_{h}\left(\frac{m}{N}\right)\left\{\frac{1}{N}\sum_{l=0}^{N-1}e^{j2\pi \frac{l(m-(k-1))}{N}}\right\}\nonumber\\
&=\tilde{S}_{h}\left(\frac{k-1}{N}\right).\label{DefEigenvalCirculant_EquivAsym}
\end{align}

Consequently, the spectral decomposition of the circulant matrix $\mathbf{C}_{h}^{(N)}$ is given by 
\begin{align}
\mathbf{C}_{h}^{(N)}&=\mathbf{F}^{(N)}\tilde{\mathbf{\Lambda}}_{h}^{(N)}\left(\mathbf{F}^{(N)}\right)^{H}\label{SpectralDecompCirc}
\end{align}
where the matrix $\mathbf{F}^{(N)}$ is a unitary DFT matrix, i.e., its elements are given by
\begin{align}
\left[\mathbf{F}^{(N)}\right]_{k,l}&=\frac{1}{\sqrt{N}}e^{j2\pi\frac{(k-1)(l-1)}{N}}.\label{DefDFT}
\end{align}
Furthermore, the matrix $\tilde{\mathbf{\Lambda}}_{h}^{(N)}$ is diagonal with the elements $\tilde{\lambda}_{k}^{(N)}$ given in (\ref{DefEigenvalCirculant_EquivAsym}). 

In \cite[Lemma 4.6]{Gray_ToeplitzReview} it is shown that the circulant matrix $\mathbf{C}_{h}^{(N)}$ and the Toeplitz matrix $\mathbf{R}_{h}^{(N)}$ are asymptotically equivalent if the autocorrelation function $r_{h}(l)$ is absolutely summable. In the context of proving this lemma it is shown that the weak norm of the difference of $\mathbf{R}_{h}^{(N)}$ and $\mathbf{C}_{h}^{(N)}$ converges to zero as $N\rightarrow \infty$, i.e.,
\begin{align}
\lim_{N\rightarrow \infty}\left|\mathbf{R}_{h}^{(N)}-\mathbf{C}_{h}^{(N)}\right|&=0\label{WeakAsympEquivCircToplitz_add}
\end{align}
where the weak norm of a matrix $\mathbf{B}$ is defined as
\begin{align}
|\mathbf{B}|&=\left(\frac{1}{N}\textrm{Tr}\left[\mathbf{B}^{H}\mathbf{B}\right]\right)^{\frac{1}{2}}.
\end{align}
The convergence of the weak norm of the difference \mbox{$\mathbf{R}_{h}^{(N)}-\mathbf{C}_{h}^{(N)}$} towards zero is required later on.

By the construction of the circulant matrix $\mathbf{C}_{h}^{(N)}$, the eigenvalues $\tilde{\lambda}_{k}$ of $\mathbf{C}_{h}^{(N)}$ are given by (\ref{DefEigenvalCirculant_EquivAsym}), i.e.,
\begin{align}
\tilde{\lambda}_{k}^{(N)}&=\left\{\begin{array}{ll}
S_{h}\left(f=\frac{k-1}{N}\right) & \textrm{for } 1\le k \le\lceil\frac{N}{2}\rceil\\
S_{h}\left(f=\frac{k-1}{N}-1\right) & \textrm{for } \lceil\frac{N}{2}\rceil < k \le N
\end{array}\right. .
\end{align}
Thus, if the PSD of the channel fading process $S_{h}(f)$ is rectangular, the eigenvalues of the circulant matrix $\mathbf{C}_{h}^{(N)}$ are given by
\begin{align}
\tilde{\lambda}_{k}^{(N)}&=\left\{\begin{array}{ll}
\frac{\sigma_{h}^{2}}{2\cdot f_{d}} & \textrm{for } 1\le k\le f_{d}N+1 \\
& \hspace{0.5cm}\vee \quad(1-f_{d})N+1\le k\le N\\
0 & \textrm{otherwise}
\end{array}\right. .\label{EigRectSpec}
\end{align}
This means that the eigenvalues of $\mathbf{R}_{h}^{(N)}$ corresponding to frequencies $|f|>f_{d}$ become zero for $N\rightarrow \infty$. 

Now, we apply the asymptotic equivalence of $\mathbf{R}_{h}^{(N)}$ and $\mathbf{C}_{h}^{(N)}$ to lower-bound the entropy rate $h'(\mathbf{y}|\mathbf{x})$ given by
\begin{align}
h'(\mathbf{y}|\mathbf{x})&=\lim_{N\rightarrow \infty}\frac{1}{N}h(\mathbf{y}|\mathbf{x})
\end{align}
with $h(\mathbf{y}|\mathbf{x})$ given in (\ref{h_y_x_second}). Thus, we have to show that
\begin{align}
\lim_{N\rightarrow \infty}\frac{1}{N}\mathrm{E}_{\mathbf{x}}\left[\log \det
\left(\frac{1}{\sigma_{n}^{2}}\mathbf{X}^{H}\mathbf{X}\mathbf{R}_{h}^{(N)}+\mathbf{I}_{N}\right)\right]
&=\lim_{N\rightarrow \infty}\frac{1}{N}\mathrm{E}_{\mathbf{x}}\left[\log \det
\left(\frac{1}{\sigma_{n}^{2}}\mathbf{X}^{H}\mathbf{X}\mathbf{C}_{h}^{(N)}+\mathbf{I}_{N}\right)\right].\label{AsympEqProof2}
\end{align}
To prove (\ref{AsympEqProof2}), we have to show that the matrices
\begin{align}
\mathbf{K}_{1}^{(N)}&=\frac{1}{\sigma_{n}^{2}}\mathbf{X}^{H}\mathbf{X}\mathbf{R}_{h}^{(N)}+\mathbf{I}_{N}\\
\mathbf{K}_{2}^{(N)}&=\frac{1}{\sigma_{n}^{2}}\mathbf{X}^{H}\mathbf{X}\mathbf{C}_{h}^{(N)}+\mathbf{I}_{N}
\end{align}
are asymptotically equivalent \cite[Theorem 2.4]{Gray_ToeplitzReview}. This means that we have to show that both matrices are bounded in the strong norm, and that the weak norm of their difference converges to zero for $N\rightarrow \infty$ \cite[Section 2.3]{Gray_ToeplitzReview}.

Concerning the condition with respect to the strong norm we have to show that
\begin{align}
\left\Vert\mathbf{K}_{1}^{(N)}\right\Vert&< \infty\\
\left\Vert\mathbf{K}_{2}^{(N)}\right\Vert&< \infty
\end{align}
with the strong norm of the matrix $\mathbf{B}$ defined by
\begin{align}
\left\Vert\mathbf{B}\right\Vert^{2}&=\max_{k}\gamma_{k}
\end{align}
where $\gamma_{k}$ are the eigenvalues of the Hermitian nonnegative definite matrix $\mathbf{B}\mathbf{B}^{H}$. The diagonal matrix $\mathbf{X}^{H}\mathbf{X}$ contains the transmit powers of the individual transmit symbols. In the case of Gaussian input distributions, for a given $\epsilon>0$, there exists a finite value $M(\epsilon)$ such that the transmit power is smaller than $M(\epsilon)$ with probability $1-\epsilon$. In addition, the strong norms of $\mathbf{R}_{h}^{(N)}$ and $\mathbf{C}_{h}^{(N)}$ are bounded, too. Concerning the boundedness of the eigenvalues of the Hermitian Toeplitz matrix $\mathbf{R}_{h}^{(N)}$ see \cite[Lemma 4.1]{Gray_ToeplitzReview}. Thus, the strong norms of $\mathbf{K}_{1}^{(N)}$ and $\mathbf{K}_{2}^{(N)}$ are asymptotically almost surely bounded.

Furthermore, for the weak norm of the difference \mbox{$\mathbf{K}_{1}^{(N)}-\mathbf{K}_{2}^{(N)}$} we get for $N\rightarrow \infty$
\begin{align}
\left|\mathbf{K}_{1}^{(N)}-\mathbf{K}_{2}^{(N)}\right|&=\left|\frac{1}{\sigma_{n}^{2}}\mathbf{X}^{H}\mathbf{X}\left(\mathbf{R}_{h}^{(N)}-\mathbf{C}_{h}^{(N)}\right)\right|\nonumber\\
&\stackrel{(a)}{\le} \frac{1}{\sigma_{n}^{2}}\left\Vert\mathbf{X}^{H}\mathbf{X}\right\Vert\left|\mathbf{R}_{h}^{(N)}-\mathbf{C}_{h}^{(N)}\right| 
\end{align}
where for (a) we have used \cite[Lemma 2.3]{Gray_ToeplitzReview}.

Based on the above argumentation that $\left\Vert\mathbf{X}^{H}\mathbf{X}\right\Vert$ is asymptotically almost surely bounded, we get for $N\rightarrow \infty$
\begin{align}
\lim_{N\rightarrow \infty}\left|\mathbf{K}_{1}^{(N)}-\mathbf{K}_{2}^{(N)}\right|&\le \lim_{N\rightarrow \infty}\frac{1}{\sigma_{n}^{2}}\left\Vert\mathbf{X}^{H}\mathbf{X}\right\Vert\left|\mathbf{R}_{h}^{(N)}-\mathbf{C}_{h}^{(N)}\right| \nonumber\\
&=0
\end{align}
due to (\ref{WeakAsympEquivCircToplitz_add}). Thus, we have proved that (\ref{AsympEqProof2}) holds and we can express the entropy rate $h'(\mathbf{y}|\mathbf{x})$ by
\begin{align}
h'(\mathbf{y}|\mathbf{x})&=\lim_{N\rightarrow\infty}\frac{1}{N}\mathrm{E}_{\mathbf{x}}\left[\log \det
\left(\frac{1}{\sigma_{n}^{2}}\mathbf{X}^{H}\mathbf{X}\mathbf{C}_{h}^{(N)}+\mathbf{I}_{N}\right)\right]+\log(\pi e \sigma_{n}^{2})\nonumber\\
&=\lim_{N\rightarrow\infty}\frac{1}{N}\mathrm{E}_{\mathbf{x}}\left[\log \det\left(\frac{1}{\sigma_{n}^{2}}\mathbf{X}^{H}\mathbf{X}\mathbf{F}\tilde{\mathbf{\Lambda}}_{h}\mathbf{F}^{H}+\mathbf{I}_{N}\right)\right]+\log(\pi e \sigma_{n}^{2}).\label{AsympEquiProoefend}
\end{align}
Here, $\mathbf{F}\tilde{\mathbf{\Lambda}}_{h}\mathbf{F}^{H}$ is the spectral decomposition of the circulant matrix $\mathbf{C}_{h}^{(N)}$, see (\ref{SpectralDecompCirc}) (from here on we again omit the superscript $(N)$ for ease of notation). Thus, $\tilde{\mathbf{\Lambda}}_{h}$ is a diagonal matrix containing the eigenvalues $\tilde{\lambda}_{k}$ as given in (\ref{EigRectSpec}) and the matrix $\mathbf{F}$ is a unitary matrix with the eigenvectors of $\mathbf{C}_{h}^{(N)}$ on its columns.

To calculate a lower bound on $h'(\mathbf{y}|\mathbf{x})$, we transform the term with the expectation operation at the RHS of (\ref{AsympEquiProoefend}) as follows:
\begin{align}
\mathrm{E}_{\mathbf{x}}\left[\log \det
\left(\frac{1}{\sigma_{n}^{2}}\mathbf{X}^{H}\mathbf{X}\mathbf{F}\tilde{\mathbf{\Lambda}}_{h}\mathbf{F}^{H}+\mathbf{I}_{N}\right)\right]
&\stackrel{(a)}{=}\mathrm{E}_{\mathbf{x}}\left[\log \det
\left(\frac{1}{\sigma_{n}^{2}}\tilde{\mathbf{\Lambda}}_{h}\mathbf{F}^{H}\mathbf{X}^{H}\mathbf{X}\mathbf{F}+\mathbf{I}_{N}\right)\right]\nonumber\\
&\stackrel{(b)}{=}\mathrm{E}_{\mathbf{x}}\left[\log
\det\left(\frac{\sigma_{h}^{2}}{2f_{d}\sigma_{n}^{2}}\tilde{\mathbf{F}}^{H}\mathbf{X}^{H}\mathbf{X}\tilde{\mathbf{F}}+\mathbf{I}_{2\lfloor
 f_{d}N\rfloor+1}\right)\right]\label{hyx_Rect}
\end{align}
where for (a) we have used (\ref{DetEqual}). For (b) the eigenvalue distribution in (\ref{EigRectSpec}) is used, and the matrix $\tilde{\mathbf{F}}$ is given by
\begin{align}
\tilde{\mathbf{F}}&=\left[\mathbf{f}_{1},\hdots,\mathbf{f}_{\lfloor f_{d}N+1\rfloor},\mathbf{f}_{\lceil (1-f_{d})N+1\rceil},\hdots,\mathbf{f}_{N}\right]\in
\mathbb{C}^{N\times(2\lfloor  f_{d}N \rfloor+1)}
\end{align}
where the $\mathbf{f}_{i}$ are the orthonormal columns of the unitary matrix $\mathbf{F}$. I.e., $\tilde{\mathbf{F}}$ contains the eigenvectors corresponding to the non-zero eigenvalues of $\mathbf{C}_{h}^{(N)}$. Now, we apply the following inequality given in
\cite[Lemma~1]{GuoShaVer06}.\\
\emph{Lemma 1:} Let $\mathbf{A}\in \mathbb{C}^{m\times n}$ with orthonormal rows and $m\le n$. Then
\begin{align}
\log \det\left( \mathbf{A}\textrm{ diag}\left(p_{1},\hdots,p_{n}\right)\mathbf{A}^{H}\right)
&\ge\textrm{trace}\left[\mathbf{A}\textrm{ diag}(\log p_{1},\hdots,\log
p_{n})\mathbf{A}^{H}\right]\label{Inequal_Lemma1}
\end{align}
if $p_{1},\hdots,p_{n}>0$.\\
With Lemma~1, we can lower-bound (\ref{hyx_Rect}) such that
\begin{align}
&\mathrm{E}_{\mathbf{x}}\left[\log
\det\left(\frac{\sigma_{h}^{2}}{2f_{d}\sigma_{n}^{2}}\tilde{\mathbf{F}}^{H}\mathbf{X}^{H}\mathbf{X}\tilde{\mathbf{F}}+\mathbf{I}_{2\lfloor
 f_{d}N\rfloor+1}\right)\right]\nonumber\\
&\quad\ge \mathrm{E}_{\mathbf{x}}\Bigg[\textrm{trace}\Bigg[
\tilde{\mathbf{F}}^{H}\textrm{diag}\Bigg(\log\left(\frac{\sigma_{h}^{2}|x_{1}|^{2}}{2f_{d}\sigma_{n}^{2}}+1\right),\hdots,\log\left(\frac{\sigma_{h}^{2}|x_{N}|^{2}}{2f_{d}
\sigma_{n}^{2}}+1\right)\Bigg)\tilde{\mathbf{F}}\Bigg]\Bigg] \nonumber\\
&\quad=\textrm{trace}\Bigg[\tilde{\mathbf{F}}^{H}\textrm{diag}\Bigg(\mathrm{E}_{x}\log\left(\frac{\sigma_{h}^{2}|x_{1}|^{2}}{2f_{d}\sigma_{n}^{2}}+1\right),\hdots,\mathrm{E}_{x}\log\left(\frac{\sigma_{h}^{2}|x_{N}|^{2}}{2f_{d}\sigma_{n}^{2}}+1\right)\Bigg)\tilde{\mathbf{F}}\Bigg]\nonumber\\
&\quad\stackrel{(a)}{=}\sum_{k=1}^{2\lfloor f_{d}N\rfloor+1}
\mathrm{E}_{x}\log\left(\frac{\sigma_{h}^{2}}{2f_{d}\sigma_{n}^{2}}|x|^{2}+1\right)\end{align}
where (a) results, because all $x_{k}$ are identically distributed and because the columns of $\tilde{\mathbf{F}}$ are orthonormal. Hence, with (\ref{AsympEquiProoefend}) a lower bound on the entropy rate is given by
\begin{align}
h'(\mathbf{y}|\mathbf{x})&\ge \lim_{N\rightarrow \infty}\frac{1}{N}\sum_{k=1}^{2\lfloor f_{d}N\rfloor+1}
\mathrm{E}_{x}\log\left(\frac{\sigma_{h}^{2}}{2f_{d}\sigma_{n}^{2}}|x|^{2}+1\right)+\log(\pi e\sigma_{n}^{2}) \nonumber\\
&=2f_{d}\mathrm{E}_{x}\log\left(\frac{\sigma_{h}^{2}}{2f_{d}\sigma_{n}^{2}}|x|^{2}+1\right)+ \log(\pi e\sigma_{n}^{2})=h'_{L}(\mathbf{y}|\mathbf{x}).\label{hyxLow_final_}
\end{align}
Thus, we have found a lower bound on the entropy rate $h'(\mathbf{y}|\mathbf{x})$ for identically distributed (i.d.) input distributions. To the best of our knowledge, this is the only known lower bound on the entropy rate $h'(\mathbf{y}|\mathbf{x})$ which is not based on a peak power constraint. 

For independently identically distributed (i.i.d.) zero-mean proper Gaussian input symbols the lower bound $h'_{L}(\mathbf{y}|\mathbf{x})$ becomes
\begin{align}
h'_{L}(\mathbf{y}|\mathbf{x})&=2f_{d}\int_{z=0}^{\infty}\log\left(\frac{\sigma_{h}^{2}\sigma_{x}^{2}}{2f_{d}\sigma_{n}^{2}}z+1\right)e^{-z}dz+ \log(\pi e\sigma_{n}^{2}).\label{h_yx_Lower_Gaussian}
\end{align}

\subsubsection*{Discussion on the Assumption of a Rectangular PSD}
For the case of constant modulus (CM) input distributions, it can be shown that the rectangular PSD maximizes $h'(\mathbf{y}|\mathbf{x})$ among all PSDs with a compact support interval $[-f_{d},f_{d}]$ and a channel power $\sigma_{h}^{2}$. For the proof of this statement, we have to calculate $\sup_{S_{h}(f)\in\mathcal{S}}h'(\mathbf{y}|\mathbf{x})\big|_{\textrm{CM}}$ where the set $\mathcal{S}$ of PSDs is given by
\begin{align}
\mathcal{S}&=\left\{ S_{h}(f) =0 \textrm{ for } f_{d}< |f| \le 0.5, \int_{-\frac{1}{2}}^{\frac{1}{2}}S_{h}(f)df=\sigma_{h}^{2}\right\}.
\end{align}
With (\ref{hyxUpp_final}) and (\ref{hyx_CM_U_equal}), we get
\begin{align}
\sup_{S_{h}(f)\in\mathcal{S}}h'(\mathbf{y}|\mathbf{x})\big|_{\textrm{CM}}&=\sup_{S_{h}(f)\in \mathcal{S}}\int_{-\frac{1}{2}}^{\frac{1}{2}}
\log\left(\pi e\left(S_{h}(f)\sigma_{x}^{2}+\sigma_{n}^{2}\right)\right)df\nonumber\\
&=\sup_{S_{h}(f)\in \mathcal{S}}\int_{-f_{d}}^{f_{d}}
\log\left(\pi e\left(S_{h}(f)\sigma_{x}^{2}+\sigma_{n}^{2}\right)\right)df\label{DiscAssRectPSDWorst-1}\\
&\stackrel{(a)}{=}\int_{-f_{d}}^{f_{d}}
\log\left(\pi e\left(\frac{\sigma_{h}^{2}}{2f_{d}}\sigma_{x}^{2}+\sigma_{n}^{2}\right)\right)df\label{DiscAssRectPSDWorst}
\end{align}
i.e., the PSD $S_{h}(f)$ which maximizes $h'(\mathbf{y}|\mathbf{x})$ is rectangular
\begin{align}
S_{h}(f)&=\left\{\begin{array}{ll} \frac{\sigma_{h}^{2}}{2f_{d}} & \textrm{for } |f|\le f_{d}\\
0& \textrm{otherwise} \end{array} \right. .
\end{align}
The last step in (\ref{DiscAssRectPSDWorst}) can be proven as follows. The argument of the supremum in (\ref{DiscAssRectPSDWorst-1}) is concave on the convex set $\mathcal{S}$. To find the $S_{h}(f)$ that maximizes the supremum in (\ref{DiscAssRectPSDWorst-1}), we define the functional
\begin{align}
J(S_{h})&=\int_{-f_{d}}^{f_{d}}
\log\left(\pi e\left(S_{h}(f)\sigma_{x}^{2}+\sigma_{n}^{2}\right)\right)df+c \left(\int_{-f_{d}}^{f_{d}}S_{h}(f)df-\sigma_{h}^{2}\right)
\end{align}
where $c$ is a constant and the last term accounts for the constraint
\begin{align}
\int_{-\frac{1}{2}}^{\frac{1}{2}}S_{h}(f)df&=\sigma_{h}^{2}.\label{Cond_Power_PSD_Rect_Assum}
\end{align}
For the $S_{h}(f)$ that maximizes, the following equation must be fulfilled for each $f$ within the interval $[-f_{d}, f_{d}]$
\begin{align}
\frac{\partial J}{\partial S_{h}(f)}&=\frac{\sigma_{x}^{2}}{S_{h}(f)\sigma_{x}^{2}+\sigma_{n}^{2}}+c=0.
\end{align}
As this equation has to be fulfilled for each $f$ and constant $c$, $S_{h}(f)$ must be constant within the interval $[-f_{d}, f_{d}]$. As the second derivative of $J$ with respect to $S_{h}(f)$ is negative for all $S_{h}(f)$ included in $\mathcal{S}$, the given extremum is a maximum. Thus, with (\ref{Cond_Power_PSD_Rect_Assum}), (\ref{DiscAssRectPSDWorst}) follows.

We conjecture that a rectangular PSD of the channel fading process maximizes $h'(\mathbf{y}|\mathbf{x})$ for any i.i.d.\ input distribution with an average power $\sigma_{x}^{2}$, including the case of i.i.d.\ zero-mean proper Gaussian input symbols. Concerning this discussion see also \cite[Section~IV-A]{ChenVera2007}. Consequently, the lower bound in (\ref{hyxLow_final_}) then holds only for a rectangular PSD. As this lower bound on $h'(\mathbf{y}|\mathbf{x})$ is finally used for the upper bound on $\mathcal{I}'(\mathbf{y};\mathbf{x})$, following the preceding conjecture, we get an upper bound on the achievable rate for a given maximum Doppler spread $f_{d}$ for the worst case PSD.

\subsection{The Achievable Rate}\label{Sect_AchievRate_SISO} 
Based on the upper and lower bounds on $h'(\mathbf{y})$ and $h'(\mathbf{y}|\mathbf{x})$, we are now able to give upper and lower bounds on the achievable rate with i.i.d.\ zero-mean proper Gaussian input symbols.

\subsubsection{Lower Bound}\label{Sect_LowerAchieRateMath}
With (\ref{Mut_first_limes}), (\ref{low_hy_basic}), and (\ref{hyxUpp_final}), we get the following lower bound on the capacity 
\begin{align}
\mathcal{I}'(\mathbf{y};\mathbf{x})&\ge h'_{L}(\mathbf{y})-h'_{U}(\mathbf{y}|\mathbf{x})\nonumber\\
&=\int_{z=0}^{\infty}\log\left(1+\rho z \right)e^{-z} dz-\int_{-\frac{1}{2}}^{\frac{1}{2}}
\log\left(1+\rho \frac{S_{h}(f)}{\sigma_{h}^{2}}\right)df\nonumber\\
&=\mathcal{I}'_{L}(\mathbf{y};\mathbf{x})\label{MutLow_unmo}
\end{align} 
where $\rho$ is the average SNR as defined in (\ref{Def_SNR}). Notice that lower bounds on the achievable rate are also lower bounds on the capacity. Therefore, in the context of this lower bound we use the term \emph{capacity} in the following. The lower bound in (\ref{MutLow_unmo}) is achievable with i.i.d.\ zero-mean proper Gaussian input symbols. As already stated, the lower bound on the capacity given in (\ref{MutLow_unmo}) is already known from \cite{DengHaim04}. The bound in (\ref{MutLow_unmo}) holds for an arbitrary PSD of the channel fading process with compact support. For the special case of a rectangular PSD as given in (\ref{Rect_PSD}) the lower bound in (\ref{MutLow_unmo}) becomes
\begin{align}
\mathcal{I}_{L}'(\mathbf{y};\mathbf{x})\big|_{\textrm{Rect}}&=\int_{0}^{\infty} \log\left(\rho
z+1\right)e^{-z}dz-2f_{d}\log\left(\frac{\rho}{2f_{d}}+1\right).\label{MutLow_unmo_rect}
\end{align}

As the mutual information rate is nonnegative, we can further modify the lower bound in (\ref{MutLow_unmo}) as follows:
\begin{align}
\mathcal{I}_{L_{mod}}'(\mathbf{y};\mathbf{x})&=\max\{\mathcal{I}_{L}'(\mathbf{y};\mathbf{x}),0\}.\label{MutLow}
\end{align}

\subsubsection{Upper Bound}\label{Sect_UpperAchieRateMath}
Using (\ref{Mut_first_limes}), (\ref{h_U2}), and (\ref{h_yx_Lower_Gaussian}) we can upper-bound the achievable rate with i.i.d.\ zero-mean proper Gaussian input symbols and a rectangular PSD of the channel fading process by
\begin{align}
\mathcal{I}'(\mathbf{y};\mathbf{x})&\ge h'_{U}(\mathbf{y})-h'_{L}(\mathbf{y}|\mathbf{x})\nonumber\\
&=\log\left(\rho+1\right)-2f_{d}\int_{0}^{\infty}\log\left(\frac{\rho}{2f_{d}}z+1\right)e^{-z}dz\nonumber\\
&=\mathcal{I}'_{U}(\mathbf{y};\mathbf{x})
.\label{MutUpp_unmo}
\end{align}
Notice that for the derivation of this upper bound the assumption on independent input symbols has not been used. On the other hand, the restriction to identically distributed input symbols is required for the calculation of $h'_{L}(\mathbf{y}|\mathbf{x})$ and, thus, for $\mathcal{I}'_{U}(\mathbf{y};\mathbf{x})$.

To the best of our knowledge, the upper bound in (\ref {MutUpp_unmo}) is new. Most other available upper bounds on the capacity hold only for input distributions with a peak power constraint and become loose for high peak-to-average power ratios, see , e.g., \cite{SetHaj05} and \cite{SetWanHajLap07Arxiv}. However, it has to be stated that the peak power constrained upper bounds in \cite{SetHaj05} and \cite{SetWanHajLap07Arxiv} are upper bounds on capacity and hold for an arbitrary PSD of the channel fading process. 

As the mutual information rate in case of perfect channel state information at the receiver $\mathcal{I}'(\mathbf{x};\mathbf{y}|\mathbf{h})$ always upper-bounds the mutual information rate in the absence of channel state information, i.e.,
\begin{align}
\mathcal{I}'(\mathbf{y};\mathbf{x})&\le
\mathcal{I}'(\mathbf{y};\mathbf{x}|\mathbf{h})\label{MutforPerfUpperBounds}
\end{align}
we can modify the upper bound in (\ref{MutUpp_unmo}) as follows:
\begin{align}
&\mathcal{I}_{U_{mod}}'(\mathbf{y};\mathbf{x})=\min\{\mathcal{I}_{U}'(\mathbf{y};\mathbf{x}),\mathcal{I}'(\mathbf{x};\mathbf{y}|\mathbf{h})\}\label{MutUpp}
\end{align}
with $\mathcal{I}'(\mathbf{x};\mathbf{y}|\mathbf{h})$ given in (\ref{PerfKnow_Gen}).

\subsubsection{Tightness of Bounds}
In the following we study the tightness of the given bounds on the achievable rate for i.i.d.\ zero-mean proper Gaussian input symbols.

\begin{figure}
\centering
 \psfrag{fd}[cc][tc][1.2]{$f_{d}$}
 \psfrag{bits}[cc][cc][1.2]{[bit/cu]}
 \psfrag{upper bound xxx}[cl][cl][1.1]{lower bound (\ref{MutLow_unmo_rect})/(\ref{MutLow})}
 \psfrag{lower bound xxxxxxxxxx}[cl][cl][1.1]{upper bound (\ref{MutUpp_unmo})/(\ref{MutUpp})}
 \psfrag{0dB}[cc][cc][1.1]{$0$\,dB}
 \psfrag{6dB}[cc][cc][1.1]{$6$\,dB}
 \psfrag{12dB}[cc][cc][1.1]{$12$\,dB}
    \includegraphics[width=0.8\columnwidth]{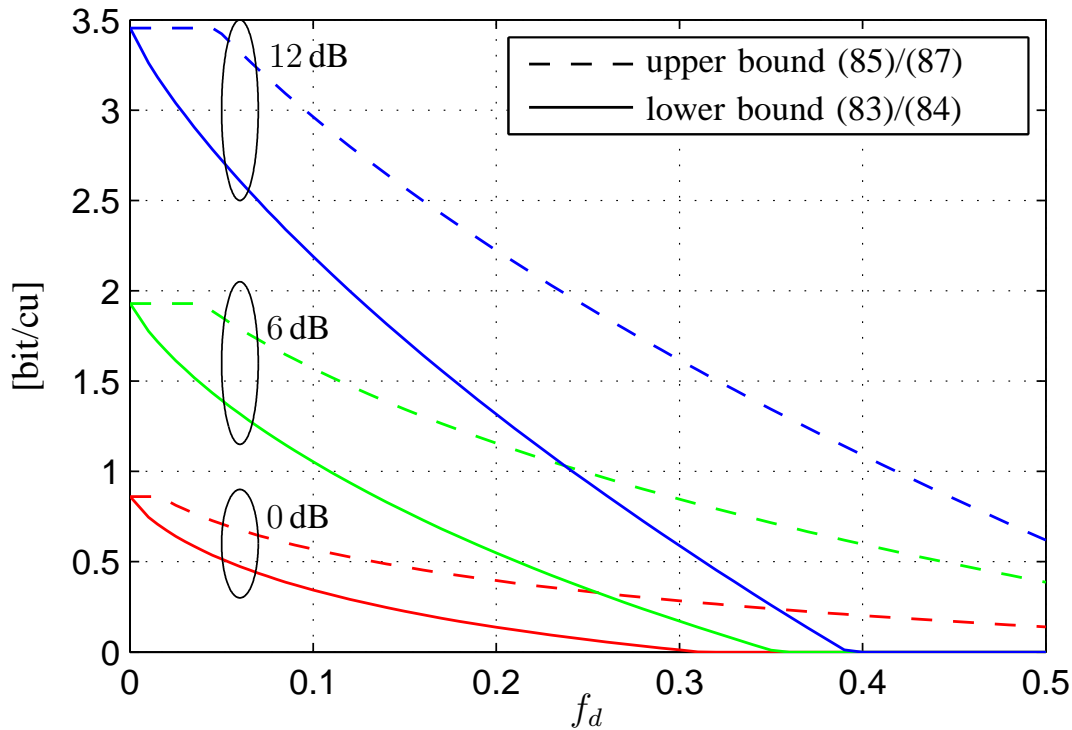}
    \caption{Upper and lower bound on the achievable rate with i.i.d.\ zero-mean proper Gaussian input distribution on a Rayleigh flat-fading channel with a rectangular PSD in bits per channel use (cu) over $f_{d}$}
    \label{FigUppLowMutInf}
\end{figure}

Fig.~\ref{FigUppLowMutInf} shows the upper bound (\ref{MutUpp_unmo})/(\ref{MutUpp}) and the lower bound (\ref{MutLow_unmo_rect})/(\ref{MutLow}) on the achievable rate with i.i.d.\ zero-mean proper Gaussian input symbols as a function of the channel dynamics, which is characterized by $f_{d}$, in case the PSD of the channel fading process is rectangular for different SNRs. Obviously, the achievable rate strongly decreases with increasing channel dynamics, i.e., $f_{d}$. Furthermore, the gap between the upper and the lower bound depends on the SNR and gets larger with an increasing SNR. In the following, we study the tightness of the given bounds analytically. This examination will show that the gap between the upper and the lower bound is bounded.

To evaluate the tightness of the upper and the lower bound on the achievable rate with i.i.d.\ zero-mean proper Gaussian input symbols, we first evaluate the tightness of the upper and the lower bound on the channel output entropy rate $h'(\mathbf{y})$. Afterwards, we evaluate the tightness of the upper and lower bound on $h'(\mathbf{y}|\mathbf{x})$.

The difference between the upper bound $h'_{U}(\mathbf{y})$ and the lower bound $h'_{L}(\mathbf{y})$ in (\ref{h_U2}) and (\ref{low_hy_basic}) is given by
\begin{align}
\Delta_{h'(\mathbf{y})}&=h'_{U}(\mathbf{y})-h'_{L}(\mathbf{y})\nonumber\\
&=\log\left(1+\rho\right)-\int_{0}^{\infty} \log\left(1+\rho
z\right)e^{-z}dz.\label{Diff_UppLow_hy_simple}
\end{align}
Fig.~\ref{Fig_Dif_hyU_hyL_new_ana_mod_bound_py} in the Appendix~\ref{hy_upp_num_uncorr} shows this difference. For $\rho\rightarrow 0$ the difference $\Delta_{h'(\mathbf{y})}$ converges to zero. And for $\rho\rightarrow \infty$ the difference is given by
\begin{align}
&\lim_{\rho\rightarrow \infty}\Delta_{h'(\mathbf{y})}=\gamma\approx 0.57721
\quad[\textrm{nat/cu}]\label{limit_tightness}
\end{align}
where $\gamma$ is the Euler constant. The limit in (\ref{limit_tightness}) can be found in \cite{LapSha02}.

The difference $\Delta_{h'(\mathbf{y})}$ monotonically increases with the SNR, as 
\begin{align}
\frac{\partial \Delta_{h'(\mathbf{y})}}{\partial \rho}
&=\frac{1}{1+\rho}-\int_{0}^{\infty}\frac{z}{1+\rho z}e^{-z}dz\nonumber\\
&\stackrel{(a)}{\ge}\frac{1}{1+\rho}-\frac{1}{1+\rho}= 0\label{AlternativeProofEquationDelta}
\end{align}
where for (a) we have used that $\frac{z}{1+\rho z}$ is concave in $z$ and, thus, we can apply Jensen's inequality. Thus, $\Delta_{h'(\mathbf{y})}$ is bounded by
\begin{align}
0&\le \Delta_{h'(\mathbf{y})} \le \gamma.
\end{align}

In addition, the difference between the upper bound and the lower bound on
$h'(\mathbf{y}|\mathbf{x})$ in case of a rectangular PSD is given by, cf. (\ref{hyxUpp_final}) and (\ref{h_yx_Lower_Gaussian})
\begin{align}
\Delta_{h'(\mathbf{y}|\mathbf{x})}&=h'_{U}(\mathbf{y}|\mathbf{x})-h'_{L}(\mathbf{y}|\mathbf{x})\nonumber\\
&=2f_{d}\left\{\log\left(1+\frac{\rho}{2f_{d}}\right)-\int_{0}^{\infty}\log\left(1+\frac{\rho}{2f_{d}} z\right)e^{-z}dz\right\}.\label{Diff_hyx_basic}
\end{align}
For asymptotically small Doppler frequencies $\Delta_{h'(\mathbf{y}|\mathbf{x})}$ approaches zero independently of the SNR. Furthermore, observing the structural similarity between (\ref{Diff_hyx_basic}) and (\ref{Diff_UppLow_hy_simple}), it can be shown that
\begin{align}
\lim_{\rho\rightarrow 0}\Delta_{h'(\mathbf{y}|\mathbf{x})}&=0
\end{align}
independently of $f_{d}$. For asymptotically high SNR and a fixed $f_{d}$ the difference is bounded by
\begin{align}
\lim_{\rho\rightarrow \infty}\Delta_{h'(\mathbf{y}|\mathbf{x})}&=2f_{d} \gamma \approx 2f_{d}\cdot 0.57721
\quad[\textrm{nat/cu}]
\end{align}
where the same limit as in (\ref{limit_tightness}) is used. Corresponding to $\Delta_{h'(\mathbf{y})}$, $\Delta_{h'(\mathbf{y}|\mathbf{x})}$ is monotonically increasing with the SNR and thus, it can be bounded by
\begin{align}
0\le \Delta_{h'(\mathbf{y}|\mathbf{x})} \le \gamma 2f_{d} \quad[\textrm{nat/cu}].
\end{align}

With $\Delta_{h'(\mathbf{y})}$ and $\Delta_{h'(\mathbf{y}|\mathbf{x})}$, the difference between the upper bound $\mathcal{I}'_{U}(\mathbf{y};\mathbf{x})$ and the lower bound
$\mathcal{I}'_{L}(\mathbf{y};\mathbf{x})$ for i.i.d.\ zero-mean proper Gaussian input symbols and a rectangular PSD is given by
\begin{align}
\Delta_{\mathcal{I}'(\mathbf{y};\mathbf{x})}&=\mathcal{I}'_{U}(\mathbf{y};\mathbf{x})-\mathcal{I}'_{L}(\mathbf{y};\mathbf{x})\big|_{\textrm{Rect}}\nonumber\\
&=\Delta_{h'(\mathbf{y})}+\Delta_{h'(\mathbf{y}|\mathbf{x})}.\label{Diff_MutInfo_SISO}
\end{align}
Hence, we get the following limits
\begin{align}
\lim_{\rho\rightarrow
0}\Delta_{\mathcal{I}'(\mathbf{y};\mathbf{x})}&=0\nonumber\\
\lim_{\rho\rightarrow \infty}\Delta_{\mathcal{I}'(\mathbf{y};\mathbf{x})}&=\gamma
(1+2f_{d})\label{Diff_MutInf_Lim}
\end{align}
and as $\Delta_{h'(\mathbf{y})}$, $\Delta_{h'(\mathbf{y})}$, and, thus, $\Delta_{\mathcal{I}'(\mathbf{y};\mathbf{x})}$ monotonically increase with the SNR, we can bound the
difference by
\begin{align}
0\le \Delta_{\mathcal{I}'(\mathbf{y};\mathbf{x})}\le \gamma(1+2f_{d})\quad
[\textrm{nat/cu}].\label{BoundednessDiffMutInfo}
\end{align}

\paragraph{Asymptotically Small Channel Dynamics}
For asymptotically small channel dynamics, i.e., $f_{d}\rightarrow 0$, the lower bound $\mathcal{I}'_{L}(\mathbf{y};\mathbf{x})$ in (\ref{MutLow_unmo}) converges to the mutual information rate in case of perfect channel knowledge in (\ref{PerfKnow_Gen})
\begin{align}
\lim_{f_{d}\rightarrow 0}
\mathcal{I}'_{L}(\mathbf{y};\mathbf{x})&=\mathcal{I}'(\mathbf{y};\mathbf{x}|\mathbf{h})
\end{align}
i.e., to the coherent capacity. This corresponds to the physical interpretation that a channel that changes arbitrarily slowly can be estimated arbitrarily well, and, therefore, the penalty term $\mathcal{I}'(\mathbf{x};\mathbf{h}|\mathbf{y})$ in (\ref{MutInf}) approaches zero. Thus, for $f_{d}\rightarrow 0$, the lower bound $\mathcal{I}'_{L}(\mathbf{y};\mathbf{x})$ is tight.

\subsection{The Asymptotic High SNR Behavior}\label{Sect_HighSNR_SISO} 
In this section, we examine the slope of the achievable rate over the SNR for asymptotically large SNRs depending on the channel dynamics. For a compactly supported PSD the lower bound on the achievable rate with i.i.d. zero-mean proper Gaussian input symbols given in (\ref{MutLow_unmo}) is characterized by the following high SNR slope\footnote{When using the term \emph{high SNR slope} we refer to the high SNR limit of the derivative of the achievable rate (bound) with respect to the logarithm of the SNR.}, which is often named \emph{pre-log}
\begin{align}
\lim_{\rho\rightarrow \infty}\frac{\partial\mathcal{I}_{L}'(\mathbf{y};\mathbf{x})}{\partial \log(\rho)}&=\lim_{\rho \rightarrow \infty}\frac{\partial}{\partial \log(\rho)}\Bigg[\int_{z=0}^{\infty}\log(\rho z
+1)e^{-z}dz-\int_{-\frac{1}{2}}^{\frac{1}{2}}\log\left(\frac{S_{h}(f)}{\sigma_{h}^{2}}\rho+1\right) df\Bigg]\nonumber\\
&=\lim_{\rho \rightarrow \infty}\left[ \int_{z=0}^{\infty}
\frac{\rho z}{\rho z+1}e^{-z}dz-\int_{-\frac{1}{2}}^{\frac{1}{2}}\frac{\frac{S_{h}(f)}{\sigma_{h}^{2}}\rho}{\frac{S_{h}(f)}{\sigma_{h}^{2}}\rho+1}df\right]\nonumber\\
&=1-2f_{d}\label{SlopeLowerBoundMutInt}
\end{align}
as $S_{h}(f)\ne 0$ for $|f|\le f_{d}$.

The upper bound $\mathcal{I}'_{U}(\mathbf{y};\mathbf{x})$ holds only for the special case of a rectangular PSD of the channel fading process. For this case the difference between the upper bound $\mathcal{I}'_{U}(\mathbf{y};\mathbf{x})$ and the lower bound $\mathcal{I}'_{L}(\mathbf{y};\mathbf{x})$ converges to a constant for high SNR, cf. (\ref{Diff_MutInf_Lim}). Thus, both bounds must have the same asymptotic high SNR slope and we conjecture that the achievable rate $\mathcal{I}'(\mathbf{y};\mathbf{x})$ is also characterized by the same asymptotic SNR slope. In \cite{La05}, it has been shown that the high SNR slope (pre-log) of the peak power constrained capacity also corresponds to $1-2f_{d}$. For a more detailed discussion on this, we refer to Section~\ref{Section_Comp_Lapidoth}.

It is interesting to note that the high SNR slope of the achievable rate is degraded by the term $2f_{d}$. Now, recall the discussion on the limitations of the discrete-time input-output relation in Section~\ref{SISOFlatModelLimit}. There it has been shown that symbol rate sampling does not yield a signal representation with a sufficient statistic, as the normalized received signal bandwidth is given by $1+2f_{d}$ (for a roll-off factor $\beta_{\textrm{ro}}=0$). The excess bandwidth leading to aliasing is given by $2f_{d}$, which exactly corresponds to the degradation of the high SNR slope of the achievable rate. Up to now, we do not know, if there is an implicit relation between these observations.

\subsection{Comparison to Asymptotes in \cite{La05}}\label{Section_Comp_Lapidoth} 
In \cite{La05}, Lapidoth gives bounds for the capacity of noncoherent Rayleigh fading channels. These bounds are mainly derived to evaluate the asymptotic high SNR behavior. He distinguishes between two cases, \emph{nonregular} and \emph{regular} fading introduced by Doob \cite{Doob90}. The case of \emph{nonregular} fading is characterized by the property that the prediction error variance of a one-step channel predictor --- having infinitely many observations in the past --- asymptotically approaches zero, when the SNR approaches infinity. As we consider the case that the PSD of the channel fading process is bandlimited with $f_{d}<0.5$, our scenario corresponds to the \emph{nonregular} case in \cite{La05}, which is also named \emph{pre-log} case. In contrast to our bounds on the achievable rate, where we assume i.i.d.\ zero-mean proper Gaussian input symbols with an average power $\sigma_{x}^{2}$, \cite{La05} does not constrain the input distribution except of a peak power constraint.

The capacity bounds in \cite{La05} are given by \cite[eq. (33) and (47)]{La05}
\begin{align}
C&\le \log\log
\tilde{\rho}-\gamma-1+\log\left(\frac{1}{\epsilon^{2}_{pred}(1/\tilde{\rho})}\right)+o(1)\label{LapUpp}\\
C &\ge \log\left(\frac{1}{\epsilon^{2}_{pred}(4/\tilde{\rho})+\frac{8}{5
\tilde{\rho}}}\right)-\gamma-\log\left(\frac{1}{1-\epsilon^{2}_{pred}(4/\tilde{\rho})}\right)-\log\left(\frac{5e}{6}\right)\label{LapLow},
\end{align}
where $\gamma\approx 0.577$ is the Euler constant and $\tilde{\rho}$ is defined as
\begin{align}
\tilde{\rho}&=\frac{P_{\textrm{peak}}\sigma_{h}^{2}}{\sigma_{n}^{2}}
\end{align}
i.e., it is an alternative definition of an SNR based on the peak power $P_{\textrm{peak}}$ instead of the average power $\sigma_{x}^{2}$ used for the definition of the average SNR $\rho$. Furthermore, $o(1)$ depends on the SNR and converges to zero for $\tilde{\rho}\rightarrow \infty$, i.e., $f(n)\in o(g(n))$ if
\begin{align}
\lim_{n\rightarrow \infty}\frac{f(n)}{g(n)}&=0.
\end{align}
In addition, the prediction error variance $\epsilon^{2}_{pred}(\delta^{2})$ is given by
\begin{align}
\epsilon^{2}_{pred}(\delta^{2})&=\exp\left(\int_{-0.5}^{0.5}\log\left(\frac{S_{h}(f)}{\sigma_{h}^{2}}+\delta^{2}\right)df\right)-\delta^{2}.
\end{align}

Although for the bounds on the peak power constrained capacity in \cite{La05} not an explicit average power constraint has been used, but only a peak power constraint, by this peak power constraint implicitly also a constraint on the average power is given. This should be obvious, as for the average power $\sigma_{x}^{2}$ the inequality $\sigma_{x}^{2}\le P_{\textrm{peak}}$ must hold. Furthermore, it has to be considered that in case of using a peak power constraint, it is in general not optimal to use the maximum average power $\sigma_{x}^{2}$. For a discussion on this see below in Section~\ref{Section_Peak_UpperBound}. In case the maximum average power $\sigma_{x}^{2}$ is not used, i.e., $\mathrm{E}\left[|x_{k}|^{2}\right]<\sigma_{x}^{2}$, the SNR $\rho$ as defined in (\ref{Def_SNR}) is not the actual average SNR. Therefore, in the case of using a peak power constraint, $\rho$ is named \emph{nominal} average SNR. However, in the case of i.i.d.\ zero-mean proper Gaussian input symbols, the achievable rate is maximized when using the maximum average power $\sigma_{x}^{2}$, i.e., in this case the nominal average SNR $\rho$ is also the actual average SNR.

As the peak power constraint that has been used for the bounds on the peak power constrained capacity in \cite{La05}, i.e., for (\ref{LapUpp}) and (\ref{LapLow}), implicitly constrains the average power to $\sigma_{x}^{2}$, for the comparison of the bounds on the achievable rate with i.i.d.\ zero-mean proper Gaussian input symbols and the bounds on the peak power constrained capacity in \cite{La05}, we choose $\tilde{\rho}$ in (\ref{LapUpp}) and (\ref{LapLow}) to be equal to the nominal average SNR $\rho$ used for the bounds on the achievable rate with i.i.d.\ Gaussian input symbols, i.e., set $\sigma_{x}^{2}=P_{\textrm{peak}}$.

\begin{figure}[t!]
\centering
 \psfrag{SNR}[cc][cc][1.2]{SNR $\rho$, $\tilde{\rho}$ [dB]}
 \psfrag{bits}[cc][cc][1.2]{[bit/cu]}
 \psfrag{new upper bound}[lc][lc][1.0]{upp. bound (\ref{MutUpp_unmo})/(\ref{MutUpp})}
 \psfrag{new lower bound}[lc][lc][1.0]{\lower4mm\hbox{low. bound (\ref{MutLow_unmo_rect})/(\ref{MutLow})}}
 \psfrag{asymp. upper bound x}[lc][lc][1.0]{asym. upp. bound (\ref{LapUpp})}
 \psfrag{asymp. lower bound}[lc][lc][1.0]{asym. low. bound (\ref{LapLow})}
 \psfrag{perfect CSI}[cc][cc][1.0]{\shortstack[l]{perfect channel\\ state information}}
 \psfrag{fd=01}[cc][cc][1.1]{\hspace{-0.2cm}$f_{d}=0.1$}
 \psfrag{fd=03}[cc][cc][1.1]{\hspace{-0.2cm}$f_{d}=0.3$}
    \includegraphics[width=0.8\columnwidth]{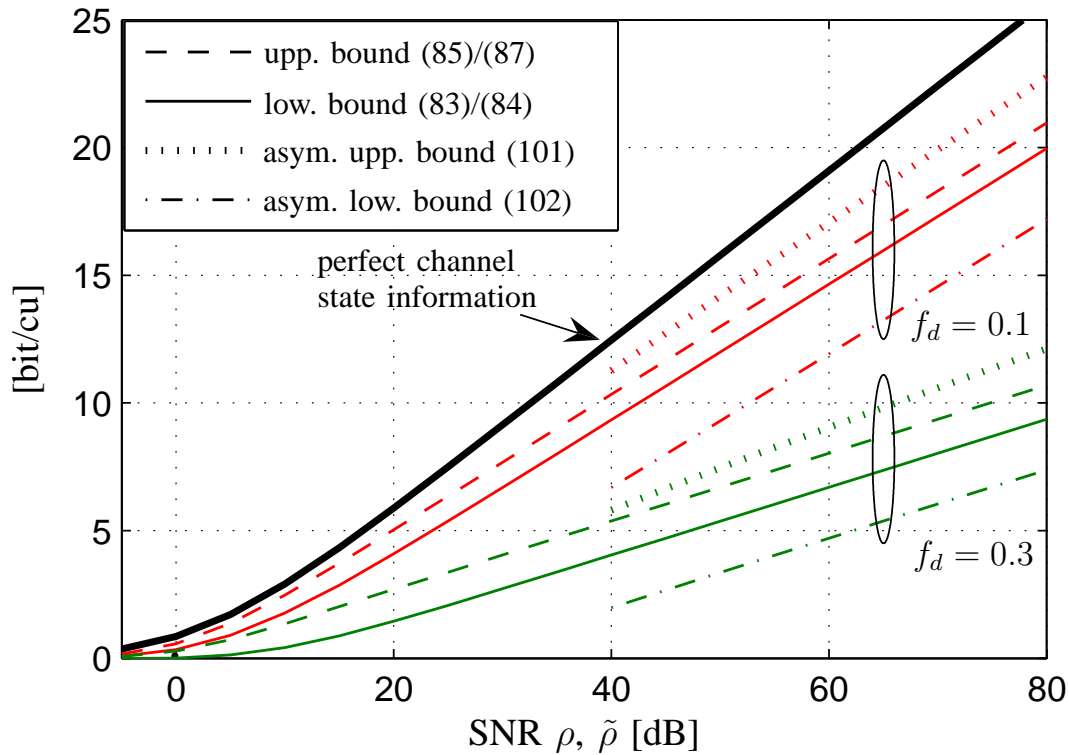}
    \caption{Comparison of the bounds on the achievable rate with i.i.d.\ zero-mean proper Gaussian inputs in (\ref{MutUpp_unmo})/(\ref{MutUpp}) and (\ref{MutLow_unmo_rect})/(\ref{MutLow}) (SNR $\rho$) with asymptotic bounds on the peak power constrained capacity in (\ref{LapUpp}) and (\ref{LapLow}) (SNR $\tilde{\rho}$), \cite[eq. (33) and (47)]{La05} (The asymptotic upper bound (\ref{LapUpp}) only holds for $\tilde{\rho}\rightarrow \infty$ as we neglect the term $o(1)$ in (\ref{LapUpp}), which approaches zero for $\tilde{\rho}\rightarrow \infty$.); rectangular PSD of the channel fading process}
    \label{CompLap}
\end{figure}

Fig.~\ref{CompLap} shows a comparison of the lower bound on the capacity in (\ref{MutLow_unmo_rect})/(\ref{MutLow}) and the upper bound on the achievable rate with i.i.d.\ zero-mean proper Gaussian inputs in (\ref{MutUpp_unmo})/(\ref{MutUpp}) with the high SNR asymptotes for the capacity in the corresponding pre-log case given in \cite{La05}, i.e., (\ref{LapUpp}) and (\ref{LapLow}). The bounds on the achievable rate with i.i.d.\ zero-mean proper Gaussian input symbols, i.e., the lower bound in (\ref{MutLow_unmo_rect})/(\ref{MutLow}) and our upper bound in (\ref{MutUpp_unmo})/(\ref{MutUpp}), are in between the asymptotes for the upper bound and the lower bound on capacity given in \cite{La05}. However, the bounds in \cite{La05} consider a peak power constrained input distribution. Therefore, this comparison is not absolutely fair. In addition, and this is the main observation from this comparison, our bounds have the same slope in the high SNR regime as the high SNR asymptotes for the peak power constrained capacity in \cite{La05}. 

\subsection{Comparison to Capacity Bounds for Peak Power Constrained Inputs in \cite{SetHaj05} and \cite{SetWanHajLap07Arxiv}}\label{SectCompPeakBounds}
In the following, we will draw the connection of the bounds on the achievable rate with i.i.d.\ zero-mean proper Gaussian input symbols given in (\ref{MutLow_unmo})/(\ref{MutLow}) and (\ref{MutUpp_unmo})/(\ref{MutUpp}) with the bounds on the peak power constrained capacity given in \cite{SetHaj05} and \cite{SetWanHajLap07Arxiv}. The peak power constrained capacity is defined by
\begin{align}
C_{\textrm{peak}}&=\lim_{N\rightarrow \infty}\sup_{\mathcal{P}^{\textrm{peak}}}\frac{1}{N}\mathcal{I}(\mathbf{y};\mathbf{x})
\end{align}
with $\mathcal{P}^{\textrm{peak}}$ being the set of peak power constrained probability density functions of the input distribution given by
\begin{align}
\mathcal{P}^{\textrm{peak}}&=\left\{p(\mathbf{x})\left|\mathbf{x}\in \mathbb{C}^{N}, \frac{1}{N}\mathrm{E}\left[\mathbf{x}^{H}\mathbf{x}\right]\le \sigma_{x}^{2}, |x_{k}|^{2}\le P_{\textrm{peak}} \hspace{0.1cm} \forall k\right.\right\}.\label{SetDef_Peak}
\end{align}
Following along the lines of the derivation of the bounds on the achievable rate with i.i.d.\ zero-mean proper Gaussian input symbols, we discuss the differences when deriving bounds on the capacity with peak power constrained input symbols. 

\subsubsection{Upper Bound}\label{Section_Peak_UpperBound}
We start with the derivation of the upper bound on the peak power constrained capacity. With (\ref{Mut_first_limes}), an upper bound on the peak power constrained capacity can be given based on an upper bound on the output entropy $h'(\mathbf{y})$ and a lower bound on the conditional output entropy $h'(\mathbf{y}|\mathbf{x})$ resulting in\footnote{Note that in (\ref{UpperAchievRate_Peak_alphaSplit1_-1}) we make a slight misuse of notation. The set $\mathcal{P}^{\textrm{peak}}$ is defined for input vectors $\mathbf{x}$ of length $N$. Therefore, the exchange of the limit and the supremum as it is used in (\ref{UpperAchievRate_Peak_alphaSplit1_-1}) while using the mutual information rate is formally not correct. However, to avoid a further complication of notation, we use the set $\mathcal{P}^{\textrm{peak}}$ also in the context of information rates. The same holds also in the following for other sets of input distributions.} 
\begin{align}
\sup_{\mathcal{P}^{\textrm{peak}}}\mathcal{I}'(\mathbf{y};\mathbf{x})&\le\sup_{\mathcal{P}^{\textrm{peak}}}\left\{h'_{U}(\mathbf{y})-h'_{L}(\mathbf{y}|\mathbf{x})\right\}.\label{UpperAchievRate_Peak_alphaSplit1_-1}
\end{align}
Note that at the moment $h'_{U}(\mathbf{y})$ and $h'_{L}(\mathbf{y}|\mathbf{x})$ are only place holders for upper and lower bounds, which are not yet further specified. In the following, we will relate the derivation of the corresponding bounds given above for i.i.d.\ zero-mean proper Gaussian input symbols to the case of peak power constrained input symbols considered here.

As the derivation of the lower bound on $h'(\mathbf{y}|\mathbf{x})$ in Section~\ref{Sect_Lowhyx} relies on the assumption on identically distributed (i.d.) input symbols, in a first step we restrict to this kind of input distributions and, therefore, define the following set of probability density functions:
\begin{align}
\mathcal{P}_{\textrm{i.d.}}^{\textrm{peak}}&=\Bigg\{p(\mathbf{x})\bigg|\mathbf{x}\in \mathbb{C}^{N}, \hspace{0.2cm} p(x_{i})=p(x_{j})\quad\forall i,j,\hspace{0.2cm}
 \left\{ \mathrm{E}[|x_{k}|^{2}]\le\sigma_{x}^{2}, \hspace{0.1cm}|x_{k}|^{2}\le P_{\textrm{peak}}\right\} \forall k\Bigg\}\label{InputSet_P_peak}
\end{align}
which corresponds to the set $\mathcal{P}^{\textrm{peak}}$ in (\ref{SetDef_Peak}) with the further restriction that the input symbols are identically distributed. I.e., we derive an upper bound on 
\begin{align}
\sup_{\mathcal{P}_{\textrm{i.d.}}^{\textrm{peak}}}\mathcal{I}'(\mathbf{y};\mathbf{x})&\le\sup_{\mathcal{P}_{\textrm{i.d.}}^{\textrm{peak}}}\left\{h'_{U}(\mathbf{y})-h'_{L}(\mathbf{y}|\mathbf{x})\right\}\nonumber\\
&=\sup_{\alpha\in[0,1]}\sup_{\mathcal{P}_{\textrm{i.d.}}^{\textrm{peak}}\big|\alpha}\left\{h'_{U}(\mathbf{y})-h'_{L}(\mathbf{y}|\mathbf{x})\right\}.\label{UpperAchievRate_Peak_alphaSplit1}
\end{align}
The calculation of the supremum in (\ref{UpperAchievRate_Peak_alphaSplit1}) is done in two steps. The inner supremum is taken over the constrained set $\mathcal{P}_{\textrm{i.d.}}^{\textrm{peak}}\big|\alpha$ being characterized by an average power $\alpha\sigma_{x}^{2}$ which holds with equality. Because of the fact that in (\ref{InputSet_P_peak}) we only use a constraint on the maximum average input power given by $\sigma_{x}^{2}$ the outer supremum is taken over $\alpha \in [0,1]$. The set $\mathcal{P}_{\textrm{i.d.}}^{\textrm{peak}}\big|\alpha$ is given by
\begin{align}
\mathcal{P}_{\textrm{i.d.}}^{\textrm{peak}}\big|\alpha&=\Bigg\{p(\mathbf{x})\bigg|\mathbf{x}\in \mathbb{C}^{N}, \hspace{0.2cm} p(x_{i})=p(x_{j})\quad\forall i,j, \hspace{0.2cm} \left\{ \mathrm{E}[|x_{k}|^{2}]=\alpha\sigma_{x}^{2}, \hspace{0.2cm} |x_{k}|^{2}\le P_{\textrm{peak}}\right\} \forall k\Bigg\}\label{InputSet_P_peak_alpha}
\end{align}
which corresponds to the set $\mathcal{P}_{\textrm{i.d.}}^{\textrm{peak}}$ except that the average power is now fixed to $\alpha\sigma_{x}^{2}$ with equality. Such a separation has also been used in \cite{SetWanHajLap07Arxiv} and in \cite{DurSchuBoelSha08IT_pub}.

For the evaluation of (\ref{UpperAchievRate_Peak_alphaSplit1}) we require an upper bound on $h'(\mathbf{y})$ and a lower bound on $h'(\mathbf{y}|\mathbf{x})$ which hold for all input distributions contained in the set $\mathcal{P}_{\textrm{i.d.}}^{\textrm{peak}}\big|\alpha$. 

All steps of the derivation of the upper bound on $h'(\mathbf{y})$ in Section~\ref{SubSect_UppBound_hy} for the case i.i.d.\ zero-mean proper Gaussian input symbols hold also for all input distributions contained in the set $\mathcal{P}_{\textrm{i.d.}}^{\textrm{peak}}\big|\alpha$, except that (a) in (\ref{UpBoundEntropProp}) is now an inequality (Hadamard's inequality), as the matrix $\mathbf{R}_{y}$ is not necessarily diagonal as we have dropped the restriction to i.i.d.\ input symbols. Furthermore, the average transmit power of the input symbols is now fixed to $\alpha\sigma_{x}^{2}$ and, thus, we get the following upper bound:
\begin{align}
h'_{U}(\mathbf{y})\big|_{p(\mathbf{x})\in\mathcal{P}_{\textrm{i.d.}}^{\textrm{peak}}|\alpha}&=\log\left(\pi e (\alpha \sigma_{x}^{2}\sigma_{h}^{2}+\sigma_{n}^{2})\right).\label{h_U_Peak_alpha}
\end{align}
Concerning the lower bound on $h'(\mathbf{y}|\mathbf{x})$ up to (\ref{hyxLow_final_}) only the assumption on i.d. input symbols has been used, which also holds for all input probability density functions contained in $\mathcal{P}_{\textrm{i.d.}}^{\textrm{peak}}\big|\alpha$. Thus, substituting (\ref{h_U_Peak_alpha}) and (\ref{hyxLow_final_}) into (\ref{UpperAchievRate_Peak_alphaSplit1}) we get
\begin{align}
\sup_{\mathcal{P}_{\textrm{i.d.}}^{\textrm{peak}}}\mathcal{I}'(\mathbf{y};\mathbf{x})&\le\sup_{\alpha\in[0,1]}\sup_{\mathcal{P}_{\textrm{i.d.}}^{\textrm{peak}}\big|\alpha}\left\{h'_{U}(\mathbf{y})-h'_{L}(\mathbf{y}|\mathbf{x})\right\}\nonumber\\
&=\sup_{\alpha\in[0,1]}\sup_{\mathcal{P}_{\textrm{i.d.}}^{\textrm{peak}}\big|\alpha}\left\{\log\left(\alpha\rho+1\right)-2f_{d}\mathrm{E}_{x}\log\left(\frac{\sigma_{h}^{2}}{2f_{d}\sigma_{n}^{2}}|x|^{2}+1\right)\right\}\nonumber\\
&=\sup_{\alpha\in[0,1]}\left\{\log\left(\alpha\rho+1\right)-2f_{d}\inf_{\mathcal{P}_{\textrm{i.d.}}^{\textrm{peak}}\big|\alpha}\mathrm{E}_{x}\log\left(\frac{\sigma_{h}^{2}}{2f_{d}\sigma_{n}^{2}}|x|^{2}+1\right)\right\}\label{UpperAchievRate_Peak_alphaSplit2}
\end{align}
with the nominal average SNR $\rho$ given in (\ref{Def_SNR}).

The term containing the infimum on the RHS of (\ref{UpperAchievRate_Peak_alphaSplit2}) can be lower-bounded in the following way:
\begin{align}
\inf_{\mathcal{P}_{\textrm{i.d.}}^{\textrm{peak}}\big|\alpha}\mathrm{E}_{x}\log\left(\frac{\sigma_{h}^{2}}{2f_{d}\sigma_{n}^{2}}|x|^{2}+1\right)&=\inf_{\mathcal{P}_{\textrm{i.d.}}^{\textrm{peak}}\big|\alpha}\int_{|x|=0}^{\sqrt{P_{\textrm{peak}}}}\frac{\log\left(\frac{\sigma_{h}^{2}}{2f_{d}\sigma_{n}^{2}}|x|^{2}+1\right)}{|x|^{2}}|x|^{2}p(|x|)d|x|\nonumber\\
&\stackrel{(a)}{\ge}\frac{\log\left(\frac{\sigma_{h}^{2}}{2f_{d}\sigma_{n}^{2}}P_{\textrm{peak}}+1\right)}{P_{\textrm{peak}}}
\inf_{\mathcal{P}_{\textrm{i.d.}}^{\textrm{peak}}\big|\alpha}\int_{|x|=0}^{\sqrt{P_{\textrm{peak}}}}|x|^{2}p(|x|)d|x|\nonumber\\
&=\frac{\log\left(\frac{\sigma_{h}^{2}}{2f_{d}\sigma_{n}^{2}}P_{\textrm{peak}}+1\right)}{P_{\textrm{peak}}}
\alpha \sigma_{x}^{2}\label{UpperAchievRate_Peak_alphaSplit3}
\end{align}
where for (a) we have used that all factors of the integrand are positive and that the term 
\begin{align}
\frac{\log\left(\frac{\sigma_{h}^{2}}{2f_{d}\sigma_{n}^{2}}|x|^{2}+1\right)}{|x|^{2}}&=\frac{1}{z}\log\left(cz+1\right)\label{FinalIntermedPeakUppTerm1}
\end{align}
with $c=\frac{\sigma_{h}^{2}}{2f_{d}\sigma_{n}^{2}}$ and $z=|x|^{2}$ is monotonically decreasing in $z$ as 
\begin{align}
\frac{\partial}{\partial z}\left\{\frac{1}{z}\log\left(cz+1\right)\right\}=\frac{c}{(c z+1)z}-\frac{\log(cz+1)}{z^{2}}&<0\nonumber\\
\Leftrightarrow \frac{cz}{cz+1}&<\log(cz+1)
\end{align}
which holds for $cz>-1$. Thus, the term in (\ref{FinalIntermedPeakUppTerm1}) is minimized for $z=|x|^{2}=P_{\textrm{peak}}$. A similar approach to calculate the infimum in (\ref{UpperAchievRate_Peak_alphaSplit3}) has been used in \cite{Verdu90} for an analogous problem. 
Notice that the result given in (\ref{UpperAchievRate_Peak_alphaSplit3}) means that the infimum on $h_{L}'(\mathbf{y}|\mathbf{x})$ for a fixed average transmit power is achieved with on-off keying.

With (\ref{UpperAchievRate_Peak_alphaSplit3}), we get the following upper bound on the RHS of (\ref{UpperAchievRate_Peak_alphaSplit2}):
\begin{align}
\sup_{\mathcal{P}_{\textrm{i.d.}}^{\textrm{peak}}}\mathcal{I}'(\mathbf{y};\mathbf{x})&\le\sup_{\alpha\in[0,1]}\left\{\log\left(\alpha\rho+1\right)-2f_{d}\frac{\alpha \sigma_{x}^{2}}{P_{\textrm{peak}}}\log\left(\frac{\sigma_{h}^{2}}{2f_{d}\sigma_{n}^{2}}P_{\textrm{peak}}+1\right) \right\}\nonumber\\
&=\sup_{\alpha\in[0,1]}\left\{\log\left(\alpha\rho+1\right)-2f_{d}\frac{\alpha}{\beta}\log\left(\frac{\rho\beta}{2f_{d}}+1\right) \right\}\label{UpperAchievRate_Peak_alphaSplit4}
\end{align}
with the \emph{nominal peak-to-average power ratio}\footnote{Instead of the common term \emph{peak-to-average power ratio} we choose the term \emph{nominal peak-to-average power ratio}, as in case of a peak power constraint it is not necessarily optimal to use the maximum average power $\sigma_{x}^{2}$. In case the actual average power is equal to the maximum average power $\sigma_{x}^{2}$, $\beta$ corresponds to the actual peak-to-average power ratio.}
\begin{align}
\beta&=\frac{P_{\textrm{peak}}}{\sigma_{x}^{2}}.\label{Def_PeakAveragePowerRatio}
\end{align}
As the argument of the supremum on the RHS of (\ref{UpperAchievRate_Peak_alphaSplit4}) is concave in $\alpha$ and, thus, there exists a unique maximum, it can easily be shown that the supremum of (\ref{UpperAchievRate_Peak_alphaSplit4}) with respect to $\alpha \in[0,1]$ is given by
\begin{align}
\alpha_{\textrm{opt}}&=\min\left\{1,\left(\frac{2f_{d}}{\beta}\log\left(\frac{\rho\beta}{2f_{d}}+1\right)\right)^{-1}-\frac{1}{\rho}\right\}\label{Eq_AlphaOpt}
\end{align}
and, thus,
\begin{align}
\sup_{\mathcal{P}_{\textrm{i.d.}}^{\textrm{peak}}}\mathcal{I}'(\mathbf{y};\mathbf{x})&\le\log\left(\alpha_{\textrm{opt}}\rho+1\right)-2f_{d}\frac{\alpha_{\textrm{opt}}}{\beta}\log\left(\frac{\rho\beta}{2f_{d}}+1\right)\nonumber\\
&=\mathcal{I}'_{U}(\mathbf{y};\mathbf{x})\big|_{P_{\textrm{peak}}}.\label{AchievRate_new_2+1}
\end{align}
With (\ref{AchievRate_new_2+1}), we have found an upper bound on the achievable rate with i.d.\ input symbols and a peak power constraint for the special case of a rectangular PSD of the channel fading process. Note that the writing $\mathcal{I}'_{U}(\mathbf{y};\mathbf{x})\big|_{P_{\textrm{peak}}}$ denotes an upper bound on the peak power constrained achievable rate. 

Note that $\alpha_{\textrm{opt}}<1$ corresponds to the case that it is not optimal to use the maximum average transmit power allowed by the set $\mathcal{P}_{\textrm{i.d.}}^{\textrm{peak}}$. This behavior is a result of the peak power constraint. Therefore, consider the extreme case $\beta=1$ and $f_{d}=0.5$, i.e., an uncorrelated channel. $\alpha=1$ then would correspond to constant modulus signaling, i.e., the transmitter puts all information into the phase of the transmitted signal. As the channel is uncorrelated from symbol to symbol and unknown to the receiver, the mutual information rate $\mathcal{I}'(\mathbf{y};\mathbf{x})$ is zero. Therefore, it is better, if the receiver does not use all its transmit power, i.e., uses an $\alpha<1$, enabling modulation of the magnitude, which leads to a positive $\mathcal{I}'(\mathbf{y};\mathbf{x})$.

\begin{figure}
\centering
 \psfrag{fd}[cc][tc][1.2]{$f_{d}$}
 \psfrag{bits}[cc][cc][1.2]{[bit/cu]}
 \psfrag{gauss xxxxxxxxxxxxxxx}[cl][cl][1.0]{\lower-2.5mm\hbox{$\mathcal{I}'_{U_{mod}}|_{\textrm{PG}}$ (\ref{MutUpp_unmo})/(\ref{MutUpp})}}
 \psfrag{peak xxxxxxxxxxxxxxx}[cl][cl][1.0]{$\mathcal{I}'_{U_{mod}}|_{P_{peak}}$ (\ref{AchievRate_new_2+1})/(\ref{MutUpp})}
 \psfrag{LowPeak xxxxxxxxxxx}[cl][cl][1.0]{\lower4mm\hbox{$\mathcal{I}'_{L}|_{\textrm{CM},\sigma_{x}^{2}}$ (\ref{MutInfLowCM_Num})/(\ref{MutLow})}} 
 \psfrag{0dB}[cc][cc][1.1]{$0$\,dB}
 \psfrag{12dB}[cc][cc][1.1]{$12$\,dB}
 \psfrag{beta=1}[cc][cc][1.1]{$\beta=1$}
 \psfrag{beta=2}[cc][cc][1.1]{$\beta=2$}
 \psfrag{beta=4}[cc][cc][1.1]{$\beta=4$}
    \includegraphics[width=0.8\columnwidth]{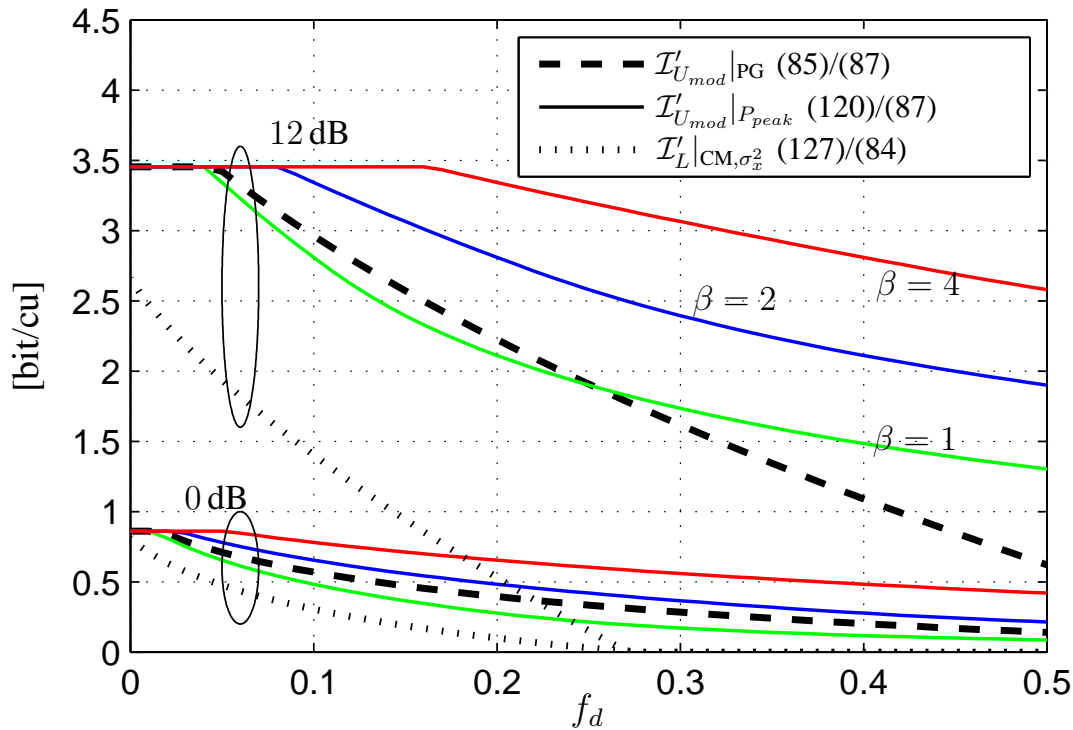}
    \caption{Comparison of the upper bounds on the achievable rate for i.d.\ input symbols with a peak power constraint in (\ref{AchievRate_new_2+1})/(\ref{MutUpp}) and with i.i.d.\ zero-mean proper Gaussian inputs (PG) in (\ref{MutUpp_unmo})/(\ref{MutUpp}); (Note, (\ref{MutUpp_unmo}) also holds for i.d.\ zero-mean proper Gaussian input symbols); in addition, the capacity lower bound (\ref{MutInfLowCM_Num})/(\ref{MutLow}) is shown, which is achievable with i.i.d.\ input symbols and a peak power constraint}
    \label{Fig_Comp_Upp_PeakGauss}
\end{figure}

The choice $\alpha_{\textrm{opt}}=1$, corresponding to the case that it is optimal to use the maximum possible average transmit power, can be shown to be optimal, on the one hand, if
\begin{align}
1&\le \rho \le \frac{2f_{d}}{\beta}\left[\exp\left(\frac{1}{2}\frac{\beta}{2f_{d}}\right)-1\right] \label{alphaOptCond1}
\end{align}
or, on the other hand, if
\begin{align}
2f_{d}&\le \frac{\beta}{\rho+2} \textrm{ for } \rho\le 1.\label{alphaOptCond2}
\end{align}
For a proof of these conditions see Appendix~\ref{AppendixAlpha1Conditions}. As in realistic scenarios $f_{d}$ is close to zero, the conditions (\ref{alphaOptCond1}) and (\ref{alphaOptCond2}) are typically fulfilled. However, for the parameter range displayed in Fig.~\ref{Fig_Comp_Upp_PeakGauss} the conditions in (\ref{alphaOptCond1}) and (\ref{alphaOptCond2}) are not always fulfilled.\footnote{\begin{samepage}Note that in case of i.i.d.\ zero-mean proper Gaussian input symbols as discussed before, it is not necessary to use the factor $\alpha$, i.e., consider cases where the actual average transmit power is smaller than the maximum average transmit power, as it can be shown that the upper bound in (\ref{MutUpp_unmo}) is always maximized while using the maximum available average transmit power. The proof is based on the fact that (\ref{MutUpp_unmo}) monotonically increases with $\rho$ as
\begin{align}
\frac{\partial}{\partial \rho}\left\{\log\left(\rho+1\right)-2f_{d}\int_{0}^{\infty}\log\left(\frac{\rho}{2f_{d}}z+1\right)e^{-z}dz\right\}&=\frac{1}{\rho+1}-2f_{d}\int_{0}^{\infty}\frac{\frac{1}{2f_{d}}z}{\frac{\rho}{2f_{d}}z+1}e^{-z}dz\nonumber\\
&\stackrel{(b)}{\ge}\frac{1}{\rho+1}-2f_{d}\frac{\frac{1}{2f_{d}}}{\frac{\rho}{2f_{d}}+1}\ge 0\label{AlternativeProofEquationDelta2}
\end{align}
where for (b) we use that $\frac{\frac{1}{2f_{d}}z}{\frac{\rho}{2f_{d}}z+1}$ is concave in $z$ and, thus, we can apply Jensen's inequality. This indicates that in case of the lack of a peak power constraint it is optimal to use the maximum average transmit power $\sigma_{x}^{2}$.\end{samepage}}

In terms of the analytical expression, the upper bound on the achievable rate with i.d.\ input symbols in (\ref{AchievRate_new_2+1}) is equal to the upper bound on the peak power constrained capacity given in \cite[Prop. 2.2]{SetWanHajLap07Arxiv}. However, the upper bound in \cite[Prop. 2.2]{SetWanHajLap07Arxiv} is, on the one hand, an upper bound on capacity as, except of the peak and average power constraints, no further assumptions on the input distributions have been made. On the other hand, the upper bound in \cite[Prop. 2.2]{SetWanHajLap07Arxiv} holds for arbitrary PSDs of the channel fading process, while the derivation of the upper bound in (\ref{AchievRate_new_2+1}) is based on the assumption of a rectangular PSD of the channel fading process. However, the approach of the derivation of the upper bound on the capacity given in \cite[Prop. 2.2]{SetWanHajLap07Arxiv} is completely different to our approach and is inherently based on the peak power constraint, while we use this peak power constraint only in the last step of the derivation. Therefore, our lower bound on $h'(\mathbf{y}|\mathbf{x})$ in (\ref{hyxLow_final_}) also enables to give an upper bound on the achievable rate for non-peak power constrained input symbols like proper Gaussian input symbols.

As stated, we made the assumption on identically distributed (i.d.) input symbols in the derivation of our upper bound. We do not know if this assumption poses a real restriction in the sense of excluding the capacity-achieving input distribution. Therefore, it would be necessary to know if the capacity-achieving input distribution is characterized by identically distributed input symbols. We have no answer to this question. However, as in case of a peak power constraint our upper bound on the achievable rate given in (\ref{AchievRate_new_2+1}) corresponds to the upper bound on the peak power constrained capacity given in \cite[Prop. 2.2]{SetWanHajLap07Arxiv}, the restriction to identically distributed inputs seems not to be a severe restriction in the sense that it leads to an upper bound being lower than the capacity.

However, in \cite{SetWanHajLap07Arxiv} it is shown that i.i.d.\ inputs, i.e., with an additional constraint on independent input symbols, are not capacity achieving in general. Based on the parameter
\begin{align}
\lambda&=\int_{-\frac{1}{2}}^{\frac{1}{2}}|S_{h}(f)|^{2}df
\end{align}
it has been shown in \cite{SetWanHajLap07Arxiv} that under the assumption of an absolutely summable autocorrelation function $r_{h}(l)$, see (\ref{CorrAbsoluteSummableNew}), in the asymptotic low SNR limit i.i.d.\ inputs are only capacity-achieving in the following two cases
\begin{itemize}
\item if $\lambda=\sigma_{h}^{4}$, corresponding to a memoryless channel,
\item or with a nominal peak-to-average power
ratio of $\beta=1$ and $\lambda\ge 2 \sigma_{h}^{4}$, i.e., when the fading process is \emph{nonephemeral}.
\end{itemize}
Notice that the proof in \cite{SetWanHajLap07Arxiv} is explicitly based on the asymptotic low SNR limit. On the other hand, for the high SNR case i.i.d.\ zero-mean proper Gaussian inputs achieve the same asymptotic high SNR behavior, in terms of the slope (pre-log), as the peak power constrained channel capacity, as it has been discussed in Section~\ref{Section_Comp_Lapidoth}.

In Fig.~\ref{Fig_Comp_Upp_PeakGauss}, the upper bound on the achievable rate with a peak power constraint in (\ref{AchievRate_new_2+1}) is shown for different nominal peak-to-average power ratios $\beta$ in comparison to the upper bound on the achievable rate for zero-mean proper Gaussian input symbols in (\ref{MutUpp_unmo}) (both combined with (\ref{MutUpp})). This comparison shows that except for $\beta$ close to $1$ and a small to average SNR or sufficiently small channel dynamics the upper bound on the achievable rate for proper Gaussian inputs is lower than the upper bound for peak power constrained input symbols in (\ref{AchievRate_new_2+1}).

\paragraph{High Nominal Peak-to-Average Power Ratios}
Considering higher order modulation, the nominal peak-to-average power ratio $\beta$ may become relatively large. For proper Gaussian inputs it is in fact infinite. Obviously, for large peak powers $P_{\textrm{peak}}$, the second term in the upper bound on the RHS of (\ref{AchievRate_new_2+1}) approaches zero and, thus 
\begin{align}
\lim_{\beta\rightarrow \infty}\mathcal{I}_{U}'(\mathbf{y};\mathbf{x})\big|_{P_{\textrm{peak}}}&=\log\left(\rho+1\right)
\end{align}
which obviously is loose as this is the capacity of an AWGN channel being already larger than the coherent capacity of the fading channel. This underlines the value of the upper bound on the achievable rate with i.i.d.\ zero-mean proper Gaussian input symbols, which are not peak power limited and serve well to upper-bound the achievable rate with practical modulation and coding schemes.

It can be shown that the upper bound in (\ref{AchievRate_new_2+1}) becomes loose for $\beta>1$ and high SNR. Therefore, we calculate the asymptotic high SNR slope of the peak power constrained upper bound given in (\ref{AchievRate_new_2+1}), where for the moment we restrict to the case of using the maximum average power, i.e., $\alpha=1$, although this is in general not an upper bound on the achievable rate. The motivation for this will become obvious afterwards. For the peak power constrained upper bound given in (\ref{UpperAchievRate_Peak_alphaSplit4}) and for the special case $\alpha=1$ the derivative with respect to $\log(\rho)$ in the high SNR limit is given by
\begin{align}
\lim_{\rho\rightarrow
\infty}\frac{\partial\mathcal{I}_{U}'(\mathbf{y};\mathbf{x})\big|_{P_{\textrm{peak}},\alpha=1}}{\partial \log(\rho)}&= \lim_{\rho\rightarrow
\infty}\frac{\partial}{\partial\log(\rho)}\left[\log(\rho+1)-\frac{2f_{d}}{\beta}\cdot\log\left(\frac{\rho \beta}{2f_{d}}+1\right)\right]\nonumber\\
&=\lim_{\rho\rightarrow
\infty}\left[\frac{\rho}{\rho+1}-\frac{2f_{d}}{\beta}\frac{\frac{\beta}{2f_{d}}\rho}{\frac{\beta}{2f_{d}}\rho+1}\right]\nonumber\\
&=1-\frac{2f_{d}}{\beta}.\label{EqAsym_UppPeakSlope}
\end{align}
Obviously, if the nominal peak-to-average power ratio $\beta$ is not equal to one, the slope of the peak power constrained upper bound with the constraint $\alpha=1$ is higher than the slope of the high SNR asymptote on the peak power constraint capacity given in \cite{La05}, see (\ref{LapUpp}), which is given by $1-2f_{d}$. The asymptotic bound in (\ref{LapUpp}) holds for an arbitrary nominal peak-to-average power ratio $\beta$. As an optimized $\alpha$ will lead to a larger upper bound, this unveils that the peak power constrained upper bound on the achievable rate in (\ref{AchievRate_new_2+1}) is loose for $\beta>1$ and high SNR.

\subsubsection{Lower Bound}\label{Sect_LowerBoundPeak}
As done before for the case of the upper bound, now we discuss the relation between the lower bound on the achievable rate with i.i.d.\ zero-mean proper Gaussian input symbols in (\ref{MutLow_unmo}) and the lower bound on the capacity for peak power constrained input symbols given in \cite[(34)]{SetHaj05}. The lower bound on the achievable rate given in (\ref{MutLow_unmo}) obviously does not hold in case of a peak power constrained input, as in this case the coherent mutual information rate $\mathcal{I}'(\mathbf{y};\mathbf{x}|\mathbf{h})$, being used in (\ref{BoundCondEntrop}) to calculate a lower bound on $h'(\mathbf{y})$, is smaller than (\ref{PerfKnow_Gen}), which holds for the case of i.i.d.\ zero-mean proper Gaussian inputs, the capacity-achieving input distribution in the coherent case.

As the mutual information for an arbitrary input distribution in the set $\mathcal{P}^{\textrm{peak}}$ defined in (\ref{SetDef_Peak}) is a lower bound on the capacity with a peak power constrained input distribution, in a first step we assume a constant modulus (CM) input distribution. I.e., all input symbols have power $\sigma_{x}^{2}$ and a uniformly distributed phase. 

Based on (\ref{Mut_first_limes}) and (\ref{BoundCondEntrop}), a lower bound on the mutual information rate with constant modulus input symbols is given by\looseness-1
\begin{align}
\sup_{\mathcal{P}^{\textrm{peak}}}\mathcal{I}'(\mathbf{y};\mathbf{x})
&\ge\sup_{\mathcal{P}^{\textrm{peak}}}\left\{\mathcal{I}(y;x|h)+h'(\mathbf{y}|\mathbf{x},\mathbf{h})-h'(\mathbf{y}|\mathbf{x})\right\}\nonumber\\
&\ge \left\{\mathcal{I}(y;x|h)+h'(\mathbf{y}|\mathbf{x},\mathbf{h})-h'(\mathbf{y}|\mathbf{x})\right\}\big|_{\textrm{CM},\sigma_{x}^{2}}\nonumber\\
&\stackrel{(a)}{=}\mathcal{I}(y;x|h)\big|_{\textrm{CM}, \sigma_{x}^{2}}-\int_{f=-\frac{1}{2}}^{\frac{1}{2}}\log\left(\frac{\sigma_{x}^{2}}{\sigma_{n}^{2}}S_{h}(f)+1\right)df\label{MutInfLowCM_Num}
\end{align}
where for (a) we have used (\ref{NoiseEntropy}), and the fact that for constant modulus input symbols $h'(\mathbf{y}|\mathbf{x})$ is equal to (\ref{hyxUpp_final}), see (\ref{hyx_CM_U_equal}). Furthermore, $\mathcal{I}(y;x|h)\big|_{\textrm{CM}, \sigma_{x}^{2}}$ corresponds to the coherent mutual information using circularly symmetric constant modulus input symbols with power $\sigma_{x}^{2}$. 

Hence, we have found a lower bound on the capacity that is achievable with i.i.d.\ constant modulus input symbols with a uniformly distributed phase. However, as far as we know there is no closed form solution for the first term in (\ref{MutInfLowCM_Num}), i.e., $\mathcal{I}(y;x|h)\big|_{\textrm{CM}, \sigma_{x}^{2}}$, so it has to be calculated numerically. In addition, for nominal peak-to-average power ratios $\beta>1$ this bound is in general not tight. The lower bound (\ref{MutInfLowCM_Num}) in combination with (\ref{MutLow}) is shown in Fig.~\ref{Fig_Comp_Upp_PeakGauss}. As it is based on constant modulus signaling, this bound becomes loose with an increasing SNR. The lower bound in (\ref{MutInfLowCM_Num}) corresponds to the lower bound on the peak power constrained capacity given in \cite[(34)]{SetHaj05}.

Using the well known time-sharing argument, the lower bound on capacity given in (\ref{MutInfLowCM_Num}) can be enhanced. The time-sharing argumentation is based on the fact that, while keeping the average transmit power constant, using the channel only during a fraction of the time might lead to a higher achievable rate. Using this time-sharing argument, a lower bound on the peak power constrained capacity for input distributions with an average power $\sigma_{x}^{2}$ and a nominal peak-to-average power ratio $\beta$ is consequently given by the following expression:
\begin{align}
\sup_{\mathcal{P}^{\textrm{peak}}}\mathcal{I}'(\mathbf{y};\mathbf{x})
&\ge\max_{\gamma \in [1,\beta]}\left\{
\frac{1}{\gamma}\mathcal{I}(y;x|h)\big|_{\textrm{CM}, \gamma\sigma_{x}^{2}}-\frac{1}{\gamma}\int_{-\frac{1}{2}}^{\frac{1}{2}}\log\left(\frac{\gamma\sigma_{x}^{2}}{\sigma_{n}^{2}}S_{h}(f)+1\right)df\right\}.\label{LowMutInf_Peak}
\end{align}
This lower bound exactly corresponds to the lower bound on the peak power constrained capacity given in \cite[(34)/(29)]{SetHaj05}.\footnote{Note that it would also be possible to enhance the lower bound on the capacity for zero-mean proper Gaussian inputs in (\ref{MutLow_unmo}) based on the time-sharing argument, i.e., by discarding the restriction to identically distributed input symbols. However, as for the derivation of the upper bound on the achievable rate in (\ref{MutUpp_unmo}) we need the restriction to i.d.\ input symbols, such a lower bound without the assumption on i.d.\ input symbols would not match this upper bound. Therefore, we do not consider this further.} As the lower bound in (\ref{LowMutInf_Peak}) does not hold for i.i.d.\ input symbols due to the application of the time-sharing argument, it would be unfair to use it for comparison in Fig.~\ref{Fig_Comp_Upp_PeakGauss}. Thus, in Fig.~\ref{Fig_Comp_Upp_PeakGauss} (\ref{MutInfLowCM_Num}) is shown.

Note that in contrast to the lower bound on the achievable rate for i.i.d.\ zero-mean proper Gaussian input distributions in (\ref{MutLow_unmo}), the lower bound in (\ref{LowMutInf_Peak}) does not converge to the coherent capacity for asymptotically small channel dynamics, i.e., $f_{d}\rightarrow 0$, as the coherent mutual information rate, which equals $\mathcal{I}(y;x|h)$, with any peak power limited input distribution is smaller than the coherent capacity, which is achieved for i.i.d.\ zero-mean proper Gaussian input symbols, cf. (\ref{PerfKnow_Gen}). This is also one advantage of our study of bounds on the achievable rate with i.i.d.\ zero-mean proper Gaussian input symbols. As the coherent capacity is achieved by this input distribution, this approach allows to give a lower bound on the achievable rate, which becomes tight for asymptotically small channel dynamics.

\section{Alternative Upper Bound on the Achievable Rate with I.I.D.\ Input Symbols Based on the One-Step Channel Prediction Error Variance}\label{Section_AlternativeIIDBound}
In Section~\ref{SectionBoundsMain}, we have derived bounds on the achievable rate with i.i.d.\ zero-mean proper Gaussian input symbols and also have discussed their link to capacity bounds for peak power constrained input symbols given in \cite{SetHaj05} and \cite{SetWanHajLap07Arxiv}. These bounds are based on a purely mathematical derivation and do not give any link to a physical interpretation like the channel prediction error variance as it has been used in \cite{La05}. In the present section, we give a new upper bound on the achievable rate which is based on the channel prediction error variance and is also not restricted to peak power constrained input symbols. In contrast, for the derivation of the channel prediction based capacity bounds in \cite{La05}, the peak power constraint has been required for technical reasons. However, for this derivation we have to restrict to i.i.d.\ input symbols, which has not been required for the derivation of the upper bound on the achievable rate in Section~\ref{SectionBoundsMain}\footnote{Note, in Section~\ref{SectionBoundsMain} the upper bound has only been evaluated for i.i.d.\ zero-mean proper Gaussian input symbols in the final step, a restriction to independent input symbols is not required for the derivation itself.}. As no peak power constraint is required for the derivation of the upper bound in the present section, we are able to evaluate the new upper bound also for i.i.d.\ zero-mean proper Gaussian input symbols. Additionally, we will also evaluate the upper bound for peak power constrained input symbols. However, due to the required restriction to i.i.d.\ input symbols, the resulting upper bound with peak power constrained input symbols is not an upper bound on the peak power constrained capacity, but only on the achievable rate with i.i.d.\ input symbols and a peak power constraint. In contrast to the upper bound on the achievable rate with i.i.d.\ zero-mean proper Gaussian input symbols given in Section~\ref{Sect_UpperAchieRateMath}, which holds only for a rectangular PSD of the fading process, the upper bound given now holds for an arbitrary PSD with compact support. 

In the first part of the following derivation, we only restrict to i.i.d.\ input symbols with a maximum average power $\sigma_{x}^{2}$. Any other restriction on the input symbols, either to zero-mean proper Gaussian symbols or a peak power constraint will be applied later. Thus, we define the set of all i.i.d.\ input distributions with a maximum average power $\sigma_{x}^{2}$ as
\begin{align}
\mathcal{P}_{\textrm{i.i.d.}}&=\bigg\{p(\mathbf{x})\bigg| \mathbf{x}\in \mathbb{C}^{N},\hspace{0.1cm} p(\mathbf{x})=\prod_{i=1}^{N}p(x_{i}),\hspace{0.1cm} p(x_{i})=p(x_{j}) \forall i,j,\hspace{0.1cm} \mathrm{E}[|x_{k}|^{2}]\le \sigma_{x}^{2}\quad\forall k\bigg\}.\label{SetDef_Piid2}
\end{align}

\subsection{Achievable Rate based on Channel Prediction}
Corresponding to Section~\ref{SectionBoundsMain}, we express the mutual information rate $\mathcal{I}'(\mathbf{y};\mathbf{x})$ based on the separation in (\ref{Mut_first_limes}). As previously stated, we construct an upper bound on the achievable rate based on channel prediction. As the channel fading process is stationary and ergodic, and as we assume i.i.d.\ input symbols, we can rewrite $h'(\mathbf{y}|\mathbf{x})$ as follows:
\begin{align}
h'(\mathbf{y}|\mathbf{x})&=\lim_{N\rightarrow
\infty}\frac{1}{N}h(\mathbf{y}|\mathbf{x})\nonumber\\
&\stackrel{(a)}{=}\lim_{N\rightarrow \infty}\frac{1}{N}\sum_{k=1}^{N}h(y_{k}|\mathbf{x},\mathbf{y}_{1}^{k-1})\nonumber\\
&\stackrel{(b)}{=}\lim_{N\rightarrow \infty}\frac{1}{N}\sum_{k=1}^{N}h(y_{k}|\mathbf{x}_{1}^{k},\mathbf{y}_{1}^{k-1})\nonumber\\
&\stackrel{(c)}{=}\lim_{N\rightarrow
\infty}h(y_{N}|\mathbf{x}_{1}^{N},\mathbf{y}_{1}^{N-1})\label{SepPredInter1}
\end{align}
where, e.g., the vector $\mathbf{y}_{1}^{N-1}$ contains all channel output symbols from the time instant $1$ to the time instant $N-1$. Here, for (a) we have used the chain rule for differential entropy, (b) is based on the fact that $y_{k}$ conditioned on $\mathbf{y}_{1}^{k-1}$ and $\mathbf{x}_{1}^{k}$ is independent of the symbols $\mathbf{x}_{k+1}^{N}$ due to the independency of the transmit symbols. Equality (c) follows from the ergodicity and stationarity of the channel fading process and the assumption on independent transmit symbols, see \cite[Chapter 4.2]{CoverBook2}. Correspondingly, $h'(\mathbf{y})$ can be rewritten as follows: 
\begin{align}
h'(\mathbf{y})&=\lim_{N\rightarrow \infty}h(y_{N}|\mathbf{y}_{1}^{N-1}).\label{PredSepoutputentropy}
\end{align}
Thus, based on (\ref{SepPredInter1}) and (\ref{PredSepoutputentropy}), the achievable rate is given by\looseness-1
\begin{align}
\mathcal{I}'(\mathbf{y};\mathbf{x})&=\lim_{N\rightarrow \infty}\left\{h(y_{N}|\mathbf{y}_{1}^{N-1})-h(y_{N}|\mathbf{x}_{1}^{N},\mathbf{y}_{1}^{N-1})\right\}\label{MutInfoPred}
\end{align}
which we name \emph{prediction separation} of the mutual information rate.

\subsection{An Upper Bound based on the Channel Prediction Error Variance}
Now, we will upper-bound the achievable rate based on the expression in (\ref{MutInfoPred}). 

\subsubsection{Upper Bound on $h'(\mathbf{y})$}\label{Sect_Pred_simple_Upper} 
As conditioning reduces entropy, we can upper bound $h(y_{N}|\mathbf{y}_{1}^{N-1})$ in (\ref{PredSepoutputentropy}) by
\begin{align}
h(y_{N}|\mathbf{y}_{1}^{N-1})&\le h(y_{N}).\label{OneStepPred_hy_402}
\end{align}
Using (\ref{PredSepoutputentropy}), (\ref{OneStepPred_hy_402}), ergodicity, and stationarity, we get
\begin{align}
h'(\mathbf{y})&\le h(y_{N}) \stackrel{(a)}{\le} \log\left(\pi e
\left(\alpha \sigma_{x}^{2}\sigma_{h}^{2}+\sigma_{n}^{2}\right)\right)=h'_{U}(\mathbf{y})\label{OneStepPred_hy_403}
\end{align}
where for (a) we used the fact that proper Gaussian distributions maximize entropy and that the average transmit power is given by $\alpha \sigma_{x}^{2}$ with $\alpha \in [0,1]$. Using an average transmit power of $\alpha\sigma_{x}^{2}$ still enables to choose  average transmit powers smaller than the maximum average transmit power $\sigma_{x}^{2}$.

Obviously, the upper bound on $h'(\mathbf{y})$ in (\ref{OneStepPred_hy_403}) is equal to the upper bound in (\ref{h_U2}), except of the factor $\alpha$, which we introduced here to account for average transmit powers smaller than the maximum average transmit power $\sigma_{x}^{2}$. This is relevant in case of peak power constrained input symbols, see Section~\ref{Section_Peak_UpperBound}.

\subsubsection{The Entropy Rate $h'(\mathbf{y}|\mathbf{x})$}\label{Sect_OneStepPred_hyx}
In the following, we lower-bound $h'(\mathbf{y}|\mathbf{x})$ based on the channel prediction representation in (\ref{SepPredInter1}). This lower-bounding approach of $h'(\mathbf{y}|\mathbf{x})$ is completely different to the one used in Section~\ref{Sect_Lowhyx}. Therefore, within the present section we express $h(y_{N}|\mathbf{x}_{1}^{N},\mathbf{y}_{1}^{N-1})$ at the RHS of (\ref{SepPredInter1}) based on the one-step channel prediction  error variance. As the following argumentation will show, the channel output $y_{N}$ conditioned on $\mathbf{x}_{1}^{N},\mathbf{y}_{1}^{N-1}$ is proper Gaussian and, thus, fully characterized by its conditional mean and conditional variance. The conditional mean is given by
\begin{align}
\textrm{E}\left[y_{N}|\mathbf{x}_{1}^{N},\mathbf{y}_{1}^{N-1}\right]&=\textrm{E}\left[x_{N}h_{N}+n_{N}|\mathbf{x}_{1}^{N},\mathbf{y}_{1}^{N-1}\right]\nonumber\\
&=x_{N}\textrm{E}\left[h_{N}|\mathbf{x}_{1}^{N-1},\mathbf{y}_{1}^{N-1}\right]\nonumber\\
&=x_{N}\hat{h}_{N}\label{Mean_yN_Pred}
\end{align}
where $\hat{h}_{N}$ is the MMSE estimate of $h_{N}$ based on the channel output observations at all previous time instances and the channel input symbols at these time instances. Based on $\hat{h}_{N}$, the channel output $y_{N}$ can be written as
\begin{align}
y_{N}&=x_{N}h_{N}+n_{N}=x_{N}\left(\hat{h}_{N}+e_{N}\right)+n_{N}\label{Outputbasedhath_Pred}
\end{align}
with the prediction error given by
\begin{align}
e_{N}&=h_{N}-\hat{h}_{N}.\label{Sep_h_hat_h_e_sm_pred}
\end{align}
As both, the noise as well as the fading process, are jointly proper Gaussian, the MMSE estimate is equivalent to the linear minimum mean squared error (LMMSE). Thus, $\hat{h}_{N}$ and $h_{N}$ are jointly proper Gaussian and it follows that the estimation error $e_{N}$ is zero-mean proper Gaussian. Note that here $\hat{h}_{N}$ also has zero mean.

As $e_{N}$ is proper Gaussian, it can be easily seen by (\ref{Outputbasedhath_Pred}) that $y_{N}$ conditioned
on $\mathbf{x}_{1}^{N},\mathbf{y}_{1}^{N-1}$ is also proper Gaussian. Thus, for the evaluation of $h(y_{N}|\mathbf{x}_{1}^{N},\mathbf{y}_{1}^{N-1})$, we calculate the conditional variance of the channel output $y_{N}$ which is given by\looseness-1
\begin{align}
\textrm{var}\left[y_{N}|\mathbf{x}_{1}^{N},\mathbf{y}_{1}^{N-1}\right]&=\mathrm{E}\left[\left|y_{N}-\mathrm{E}\left[y_{N}|\mathbf{x}_{1}^{N},\mathbf{y}_{1}^{N-1}\right]\right|^{2}\Big|\mathbf{x}_{1}^{N},\mathbf{y}_{1}^{N-1}\right]\nonumber\\
&=\mathrm{E}\left[\left|x_{N}(h_{N}-\hat{h}_{N})+n_{N}\right|^{2}\Big|\mathbf{x}_{1}^{N},\mathbf{y}_{1}^{N-1}\right]\nonumber\\
&=|x_{N}|^{2}\mathrm{E}\left[\left|e_{N}\right|^{2}\Big|\mathbf{x}_{1}^{N-1},\mathbf{y}_{1}^{N-1}\right]+\sigma_{n}^{2}\nonumber\\
&=|x_{N}|^{2}\sigma_{e_{\textrm{pred}}}^{2}(\mathbf{x}_{1}^{N-1})+\sigma_{n}^{2}
\end{align}
where
\begin{align}
\sigma_{e_{\textrm{pred}}}^{2}(\mathbf{x}_{1}^{N-1})&=\mathrm{E}\left[\left|h_{N}-\hat{h}_{N}\right|^{2}\Big|\mathbf{x}_{1}^{N-1},\mathbf{y}_{1}^{N-1}\right]\nonumber\\
&=\mathrm{E}\left[\left|e_{N}\right|^{2}\Big|\mathbf{x}_{1}^{N-1},\mathbf{y}_{1}^{N-1}\right]\nonumber\\
&\stackrel{(a)}{=}\mathrm{E}\left[\left|e_{N}\right|^{2}\Big|\mathbf{x}_{1}^{N-1}\right]\label{PredErr_Var}
\end{align}
is the prediction error variance of the MMSE estimator for $\hat{h}_{N}$. For (a) we have used the fact that the zero-mean estimation error $e_{N}$ is orthogonal to and, thus, independent of the observations $\mathbf{y}_{1}^{N-1}$. However, the prediction error variance depends on the input symbols $\mathbf{x}_{1}^{N-1}$, which is indicated by writing $\sigma_{e_{\textrm{pred}}}^{2}(\mathbf{x}_{1}^{N-1})$.

Based on the channel prediction error variance, we can rewrite the entropy $h(y_{N}|\mathbf{x}_{1}^{N},\mathbf{y}_{1}^{N-1})$ as
\begin{align}
h(y_{N}|\mathbf{x}_{1}^{N},\mathbf{y}_{1}^{N-1})&=\mathrm{E}_{\mathbf{x}}\left[\log\left(\pi e
\textrm{ var}\left[y_{N}|\mathbf{x}_{1}^{N},\mathbf{y}_{1}^{N-1}\right]\right)\right]\nonumber\\
&=\mathrm{E}_{\mathbf{x}}\left[\log\left(\pi e
\left(\sigma_{n}^{2}+\sigma_{e_{\textrm{pred}}}^{2}(\mathbf{x}_{1}^{N-1})|x_{N}|^{2}\right)\right)\right].\label{SepPredInter3_}
\end{align}

With (\ref{SepPredInter1}) and (\ref{SepPredInter3_}), we get for i.i.d.\ input
symbols 
\begin{align}
h'(\mathbf{y}|\mathbf{x})&=\mathrm{E}_{x_{k}}\left[\mathrm{E}_{\mathbf{x}_{-\infty}^{k-1}}\left[\log\left(\pi
e
\left(\sigma_{n}^{2}+\sigma_{e_{\textrm{pred},\infty}}^{2}(\mathbf{x}_{-\infty}^{k-1})|x_{k}|^{2}\right)\right)\right]\right]\label{SepPredInter3}
\end{align}
where $\sigma_{e_{\textrm{pred},\infty}}^{2}(\mathbf{x}_{-\infty}^{k-1})$ is the prediction error variance in (\ref{PredErr_Var}) for an infinite number of channel observations in the past, i.e.,
\begin{align}
\sigma_{e_{\textrm{pred},\infty}}^{2}(\mathbf{x}_{-\infty}^{k-1})&=\lim_{N\rightarrow
\infty}\sigma_{e_{\textrm{pred}}}^{2}(\mathbf{x}_{1}^{N-1})
\label{enhanc_upper_hy_fin_403}
\end{align}
which is indicated by writing $\sigma_{e_{\textrm{pred},\infty}}^{2}(\mathbf{x}_{-\infty}^{k-1})$. Note that we have switched the notation and now predict at the time instant $k$ instead of predicting at the time instant $N$. This is possible, as the channel fading process is stationary, the input symbols are assumed to be i.i.d., and as we consider an infinitely long past.\looseness-1

\subsubsection{Upper Bound on the Achievable Rate}
With (\ref{MutInfoPred}), (\ref{OneStepPred_hy_402}), (\ref{OneStepPred_hy_403}), and (\ref{SepPredInter3_})/(\ref{SepPredInter3}), we can give the following upper bound on the achievable rate with i.i.d.\ input symbols: 
\begin{align} 
\mathcal{I}'(\mathbf{y};\mathbf{x})&\le\log\left(\alpha\rho+1\right)-\mathrm{E}_{x_{k}}\left[\mathrm{E}_{\mathbf{x}_{-\infty}^{k-1}}\left[\log
\left(1+\frac{\sigma_{e_{\textrm{pred},\infty}}^{2}(\mathbf{x}_{-\infty}^{k-1})}{\sigma_{n}^{2}}|x_{k}|^{2}\right)\right]\right]\label{enhanc_upper_hy_fin_402_simp}
\end{align}
where $\rho$ is the nominal average SNR, see (\ref{Def_SNR}). Obviously, the upper bound in (\ref{enhanc_upper_hy_fin_402_simp}) still depends on the channel prediction error variance $\sigma_{e_{\textrm{pred},\infty}}^{2}(\mathbf{x}_{-\infty}^{k-1})$ given in (\ref{enhanc_upper_hy_fin_403}), which itself depends on the distribution of the input symbols in the past. Effectively $\sigma_{e_{\textrm{pred},\infty}}^{2}(\mathbf{x}_{-\infty}^{k-1})$ is itself a random quantity. For infinite transmission lengths, i.e., $N\rightarrow \infty$, its distribution is independent of the time instant $k$, as the channel fading process is stationary and as the transmit symbols are i.i.d..

\subsubsection{The Prediction Error Variance
$\sigma_{e_{\textrm{pred},\infty}}^{2}(\mathbf{x}_{-\infty}^{k-1})$}\label{Sect_ArgChannelPredlowerOneStep}
The prediction error variance $\sigma_{e_{\textrm{pred},\infty}}^{2}(\mathbf{x}_{-\infty}^{k-1})$ in (\ref{enhanc_upper_hy_fin_403}) depends on the distribution of the input symbols $\mathbf{x}_{-\infty}^{k-1}$. To construct an upper bound on the RHS of (\ref{enhanc_upper_hy_fin_402_simp}), we need to find a distribution of the transmit symbols in the past, i.e., $\mathbf{x}_{-\infty}^{k-1}$, which leads to a distribution of $\sigma_{e_{\textrm{pred},\infty}}^{2}(\mathbf{x}_{-\infty}^{k-1})$ that maximizes the RHS of
(\ref{enhanc_upper_hy_fin_402_simp}). Therefore, we have to express the channel prediction error variance $\sigma_{e_{\textrm{pred},\infty}}^{2}(\mathbf{x}_{-\infty}^{k-1})$ as a function of the transmit symbols in the past, i.e., $\mathbf{x}_{-\infty}^{k-1}$. In a first step, we give such an expression for the case of a finite past time horizon, i.e., for $\sigma_{e_{\textrm{pred}}}^{2}(\mathbf{x}_{1}^{N-1})$ in (\ref{PredErr_Var}) which can be expressed by
\begin{align}
\sigma_{e_{\textrm{pred}}}^{2}(\mathbf{x}_{1}^{N-1})&=
\sigma_{h}^{2}-\mathbf{r}_{\mathbf{y}_{1}^{N-1}h_{N}|\mathbf{x}_{1}^{N-1}}^{H}\mathbf{R}_{\mathbf{y}_{1}^{N-1}|\mathbf{x}_{1}^{N-1}}^{-1}\mathbf{r}_{\mathbf{y}_{1}^{N-1}h_{N}|\mathbf{x}_{1}^{N-1}}\label{PredErrQuadartGen}
\end{align}
where $\mathbf{R}_{\mathbf{y}_{1}^{N-1}|\mathbf{x}_{1}^{N-1}}$ is the correlation matrix of the observations $\mathbf{y}_{1}^{N-1}$ while the past transmit symbols $\mathbf{x}_{1}^{N-1}$ are known, i.e.,
\begin{align}
\mathbf{R}_{\mathbf{y}_{1}^{N-1}|\mathbf{x}_{1}^{N-1}}&=\mathrm{E}\left[\mathbf{y}_{1}^{N-1}(\mathbf{y}_{1}^{N-1})^{H}\big|\mathbf{x}_{1}^{N-1}\right]\nonumber\\
&=\mathbf{X}_{N-1}\mathbf{R}_{h}\mathbf{X}_{N-1}^{H}+\sigma_{n}^{2}\mathbf{I}_{N-1}\label{PredQuadObsCorr}
\end{align}
with $\mathbf{X}_{N-1}$ being a diagonal matrix containing the past transmit symbols such that $\mathbf{X}_{N-1}=\mathrm{diag}\left(\mathbf{x}_{1}^{N-1}\right)$.
In addition, $\mathbf{R}_{h}$ is the autocorrelation matrix of the channel fading process 
\begin{align}
\mathbf{R}_{h}&=\mathrm{E}\left[\mathbf{h}_{1}^{N-1}(\mathbf{h}_{1}^{N-1})^{H}\right]
\end{align}
where $\mathbf{h}_{1}^{N-1}$ contains the fading weights from time instant $1$ to $N-1$. The cross correlation vector $\mathbf{r}_{\mathbf{y}_{1}^{N-1}h_{N}|\mathbf{x}_{1}^{N-1}}$ between the observation vector $\mathbf{y}_{1}^{N-1}$ and the fading weight $h_{N}$ while knowing the past transmit symbols $\mathbf{x}_{1}^{N-1}$ is given by
\begin{align}
\mathbf{r}_{\mathbf{y}_{1}^{N-1}h_{N}|\mathbf{x}_{1}^{N-1}}&=\mathrm{E}\left[\mathbf{y}_{1}^{N-1}h_{N}^{*}\big|\mathbf{x}_{1}^{N-1}\right]=\mathbf{X}_{N-1}\mathbf{r}_{h,\textrm{pred}}\label{PredQuadCrossCorr}
\end{align}
with $\mathbf{r}_{h,\textrm{pred}}=\left[r_{h}(-(N-1)) \hdots r_{h}(-1)\right]^{T}$
where $r_{h}(l)$ is the autocorrelation function as defined in (\ref{Corr_h}).

Substituting (\ref{PredQuadObsCorr}) and (\ref{PredQuadCrossCorr}) into (\ref{PredErrQuadartGen}) yields
\begin{align}
\sigma_{e_{\textrm{pred}}}^{2}(\mathbf{x}_{1}^{N-1})&=\sigma_{h}^{2}-\mathbf{r}_{h,\textrm{pred}}^{H}\mathbf{X}_{N-1}^{H}\left(\mathbf{X}_{N-1}\mathbf{R}_{h}\mathbf{X}_{N-1}^{H}+\sigma_{n}^{2}\mathbf{I}_{N-1}\right)^{-1}\mathbf{X}_{N-1}\mathbf{r}_{h,\textrm{pred}}\nonumber\\
&=\sigma_{h}^{2}-\mathbf{r}_{h,\textrm{pred}}^{H}\left(\mathbf{R}_{h}+\sigma_{n}^{2}\left(\mathbf{X}_{N-1}^{H}\mathbf{X}_{N-1}\right)^{-1}\right)^{-1}\mathbf{r}_{h,\textrm{pred}}\nonumber\\
&\stackrel{(a)}{=}\sigma_{h}^{2}-\mathbf{r}_{h,\textrm{pred}}^{H}\left(\mathbf{R}_{h}+\sigma_{n}^{2}\mathbf{Z}^{-1}\right)^{-1}\mathbf{r}_{h,\textrm{pred}}\label{PredEstErrVarDef}
\end{align}
where for (a) we have used $\mathbf{Z}=\mathbf{X}_{N-1}^{H}\mathbf{X}_{N-1}$, i.e., $\mathbf{Z}$ is a diagonal matrix containing the powers of the individual transmit symbols in the past from time instant $1$ to $N-1$. For ease of notation, we omit the index $N-1$.\footnote{Note that the inverse of $\mathbf{Z}$ in (\ref{PredEstErrVarDef}) does not exist, if a diagonal element $z_{i}$ of the diagonal matrix $\mathbf{Z}$ is zero, i.e., one transmit symbol has zero power. However, as the prediction error variance is continuous in $z_{i}=0$ for all $i$ this does not lead to problems in the following derivation.}

Remember that we want to derive an upper bound on the achievable rate with i.i.d.\ input symbols by maximizing the RHS of (\ref{enhanc_upper_hy_fin_402_simp}) over all i.i.d.\ distributions of the transmit symbols in the past with an average power $\alpha\sigma_{x}^{2}$. Obviously, the distribution of the phases of the past transmit symbols $\mathbf{x}_{1}^{N-1}$ has no influence on the channel prediction error variance $\sigma_{e_{\textrm{pred}}}^{2}(\mathbf{x}_{1}^{N-1})$. Thus, it rests to evaluate, for which distribution of the power of the past transmit symbols the RHS of (\ref{enhanc_upper_hy_fin_402_simp}) is maximized. In the following, we will show that the RHS of (\ref{enhanc_upper_hy_fin_402_simp}) is maximized in case the past transmit symbols have a constant power $\alpha\sigma_{x}^{2}$. I.e., calculation of the prediction error variance under the assumption that the past transmit symbols are constant modulus symbols with transmit power $|x_{k}|^{2}=\alpha\sigma_{x}^{2}$ maximizes the RHS of (\ref{enhanc_upper_hy_fin_402_simp}) over all i.i.d.\ input distributions for the given average power constraint of $\alpha\sigma_{x}^{2}$.

To prove this statement, we use the fact that the expression in the expectation operation at the RHS of (\ref{enhanc_upper_hy_fin_402_simp}) (but here for the case of a finite past time horizon) with (\ref{PredEstErrVarDef}), i.e.,
\begin{align}
\log\left(1+\frac{|x_{N}|^{2}}{\sigma_{n}^{2}}\left(\sigma_{h}^{2}-\mathbf{r}_{h,\textrm{pred}}^{H}\left(\mathbf{R}_{h}+\sigma_{n}^{2}\mathbf{Z}^{-1}\right)^{-1}\mathbf{r}_{h,\textrm{pred}}\right)\right)\label{EquatForConvexityProof}
\end{align}
is convex with respect to each individual element of the diagonal of $\mathbf{Z}$, which we name $\mathbf{z}$. The proof of convexity of (\ref{EquatForConvexityProof}) is given in Appendix~\ref{App_ConvexityProof}. Based on this convexity and Jensen's inequality, we get
\begin{align}
&\mathrm{E}_{\mathbf{z}}\left[\log
\left(1+\frac{|x_{N}|^{2}}{\sigma_{n}^{2}}\left(\sigma_{h}^{2}-\mathbf{r}_{h,\textrm{pred}}^{H}\left(\mathbf{R}_{h}+\sigma_{n}^{2}\mathbf{Z}^{-1}\right)^{-1}\mathbf{r}_{h,\textrm{pred}}\right)\right)\right]\nonumber\\
&\quad\ge \log
\left(1+\frac{|x_{N}|^{2}}{\sigma_{n}^{2}}\left(\sigma_{h}^{2}-\mathbf{r}_{h,\textrm{pred}}^{H}\left(\mathbf{R}_{h}+\sigma_{n}^{2}\left(\mathrm{E}_{\mathbf{z}}\left[\mathbf{Z}\right]\right)^{-1}\right)^{-1}\mathbf{r}_{h,\textrm{pred}}\right)\right)\nonumber\\
&\quad=\log\left(1+\frac{|x_{N}|^{2}}{\sigma_{n}^{2}}\left(\sigma_{h}^{2}-\mathbf{r}_{h,\textrm{pred}}^{H}\left(\mathbf{R}_{h}+\frac{\sigma_{n}^{2}}{\alpha\sigma_{x}^{2}}\mathbf{I}_{N-1}\right)^{-1}\mathbf{r}_{h,\textrm{pred}}\right)\right)\nonumber\\
&\quad=\log\left(1+\frac{|x_{N}|^{2}}{\sigma_{n}^{2}}\sigma_{e_{\textrm{pred},\textrm{CM}}}^{2}\right)
\end{align}
where $\sigma_{e_{\textrm{pred},\textrm{CM}}}^{2}$ is the channel prediction error variance in case all past transmit symbols are constant modulus symbols with power $\alpha\sigma_{x}^{2}$. Here, the index CM denotes constant modulus.

As this lower-bounding of (\ref{EquatForConvexityProof}) can be performed for an arbitrary $N$, i.e., for an arbitrarily long past, we can also conclude that the RHS of (\ref{enhanc_upper_hy_fin_402_simp}) is upper-bounded by
\begin{align}
\mathcal{I}'(\mathbf{y};\mathbf{x})&\le\log\left(\alpha\rho+1\right)-\mathrm{E}_{x_{k}}\left[\log
\left(1+\frac{\sigma_{e_{\textrm{pred},\textrm{CM},\infty}}^{2}}{\sigma_{n}^{2}}|x_{k}|^{2}\right)\right]\label{IYX_final_upp_pred}
\end{align}
where $\sigma_{e_{\textrm{pred},\textrm{CM},\infty}}^{2}$ is the channel prediction error variance in case all past transmit symbols are constant modulus symbols with power $\alpha\sigma_{x}^{2}$ and an infinitely long past observation horizon. In this case, the prediction error variance is no longer a random quantity but is constant for all time instances $k$. 

Constant modulus symbols are in general not the capacity maximizing input distribution. However, we only use them to find a distribution of
$\sigma_{e_{\textrm{pred},\infty}}^{2}(\mathbf{x}_{-\infty}^{k-1})$ that maximizes (\ref{enhanc_upper_hy_fin_402_simp}).

For constant modulus input symbols with power $\alpha\sigma_{x}^{2}$ and an infinitely long past, the prediction error variance is given by, cf. \cite{La05}
\begin{align}
\sigma_{e_{\textrm{pred},\textrm{CM},\infty}}^{2}&=\frac{\sigma_{n}^{2}}{\alpha\sigma_{x}^{2}}\left\{\exp\left(\int_{-\frac{1}{2}}^{\frac{1}{2}}\log\left(1+\frac{\alpha\sigma_{x}^{2}}{\sigma_{n}^{2}}S_{h}(f)\right)df\right)-1\right\}\label{InfEstErrVar}
\end{align}

As far as we know, the upper bound on the achievable rate in (\ref{IYX_final_upp_pred}) is new. The innovation in the derivation of this bound lies in the fact that we separate the input symbols into the one at the time instant $x_{k}$ and the previous input symbols contained in $\mathbf{x}_{-\infty}^{k-1}$. The latter ones are only relevant to calculate the prediction error variance, which itself is a random variable depending on the distribution of the past transmit symbols. To derive an upper bound on the achievable rate with i.i.d.\ input distributions, we have shown that the achievable rate is upper-bounded if the prediction error variance is calculated under the assumption that all past transmit symbols are constant modulus input symbols. As the assumption on constant modulus symbols is only used in the context of the prediction error variance, the upper bound on the achievable rate still holds for any i.i.d.\ input distribution with the given average power constraint. This allows us to evaluate this bound also for the case of i.i.d.\ zero-mean proper Gaussian input symbols.

\subsubsection{Proper Gaussian Input Symbols}
Evaluating (\ref{IYX_final_upp_pred}) for i.i.d.\ zero-mean proper Gaussian (PG) input symbols yields
\begin{align}
\mathcal{I}'(\mathbf{y};\mathbf{x})\big|_{\textrm{PG}}&\le \sup_{\alpha\in [0,1]}\left\{\log\left(\alpha \rho+1\right)-\int_{0}^{\infty}\log\left(1+\frac{\sigma_{e_{\textrm{pred},\textrm{CM},\infty}}^{2}}{\sigma_{h}^{2}}\alpha\rho z\right)e^{-z}dz\right\}\nonumber\\
&\stackrel{(a)}{\le}\sup_{\alpha\in [0,1]}\left\{\log\left(\alpha \rho+1\right)-\int_{0}^{\infty}\log\left(1+\frac{\sigma_{e_{\textrm{pred},\textrm{CM},\infty}}^{2}\big|_{\alpha=1}}{\sigma_{h}^{2}}\alpha\rho z\right)e^{-z}dz\right\}\nonumber\\
&\stackrel{(b)}{=}\log\left(\rho+1\right)-\int_{0}^{\infty}\log\left(1+\frac{\sigma_{e_{\textrm{pred},\textrm{CM},\infty}}^{2}\big|_{\alpha=1}}{\sigma_{h}^{2}}\rho z\right)e^{-z}dz\nonumber\\
&=\mathcal{I}_{U}'(\mathbf{y};\mathbf{x})\big|_{\textrm{pred},\textrm{PG}}\label{IYX_final_upp_pred_PG}
\end{align}
where (a) is based on the fact that $\sigma_{e_{\textrm{pred},\textrm{CM},\infty}}^{2}$ monotonically decreases with an increasing $\alpha$, and, thus, that the term in the second line of (\ref{IYX_final_upp_pred_PG}) is maximized if the prediction error variance is calculated for $\alpha=1$, which is denoted by writing $\sigma_{e_{\textrm{pred},\textrm{CM},\infty}}^{2}\big|_{\alpha=1}$. Furthermore, (b) follows from the monotonicity of the argument of the supremum in the third line of (\ref{IYX_final_upp_pred_PG}) in $\alpha$, which can be shown analogously to the monotonicity of (\ref{Diff_UppLow_hy_simple}) based on (\ref{AlternativeProofEquationDelta}). In conclusion, this means that the upper bound for zero-mean proper Gaussian input symbols is maximized for the maximum average transmit power $\sigma_{x}^{2}$. 

As the coherent mutual information rate $\mathcal{I}'(\mathbf{y};\mathbf{x}|\mathbf{h})$ upper-bounds $\mathcal{I}'(\mathbf{y};\mathbf{x})$, we can enhance the upper bound in (\ref{IYX_final_upp_pred_PG}) analogously to (\ref{MutUpp}) with $\mathcal{I}'(\mathbf{x};\mathbf{y}|\mathbf{h})$ given in (\ref{PerfKnow_Gen}).

\begin{figure}
\centering
 \psfrag{fd}[cc][cc][1.2]{$f_{d}$}
 \psfrag{bits}[cc][cc][1.2]{[bit/cu]}
 \psfrag{gauss xxxxxxxxxxxxxxxxxxxxxx}[cl][cl][1.1]{\lower-2.5mm\hbox{$\mathcal{I}'_{U}(\mathbf{y};\mathbf{x})|_{\textrm{PG}}$ (\ref{MutUpp_unmo})/(\ref{MutUpp})}}
 \psfrag{gaussLowxxxxxxxxxxxxxx}[cl][cl][1.1]{\lower-1.5mm\hbox{$\mathcal{I}'_{L}(\mathbf{y};\mathbf{x})|_{\textrm{PG}}$ (\ref{MutLow_unmo_rect})/(\ref{MutLow})}}
 \psfrag{gauss predxxxxxxxxxxxx}[cl][cl][1.1]{$\mathcal{I}'_{U}(\mathbf{y};\mathbf{x})|_{\textrm{pred},\textrm{PG}}$ (\ref{IYX_final_upp_pred_PG})/(\ref{MutUpp})}
 \psfrag{0dB}[cc][cc][1.1]{$0$\,dB}
 \psfrag{12dB}[cc][cc][1.1]{$12$\,dB}
\includegraphics[width=0.8\columnwidth]{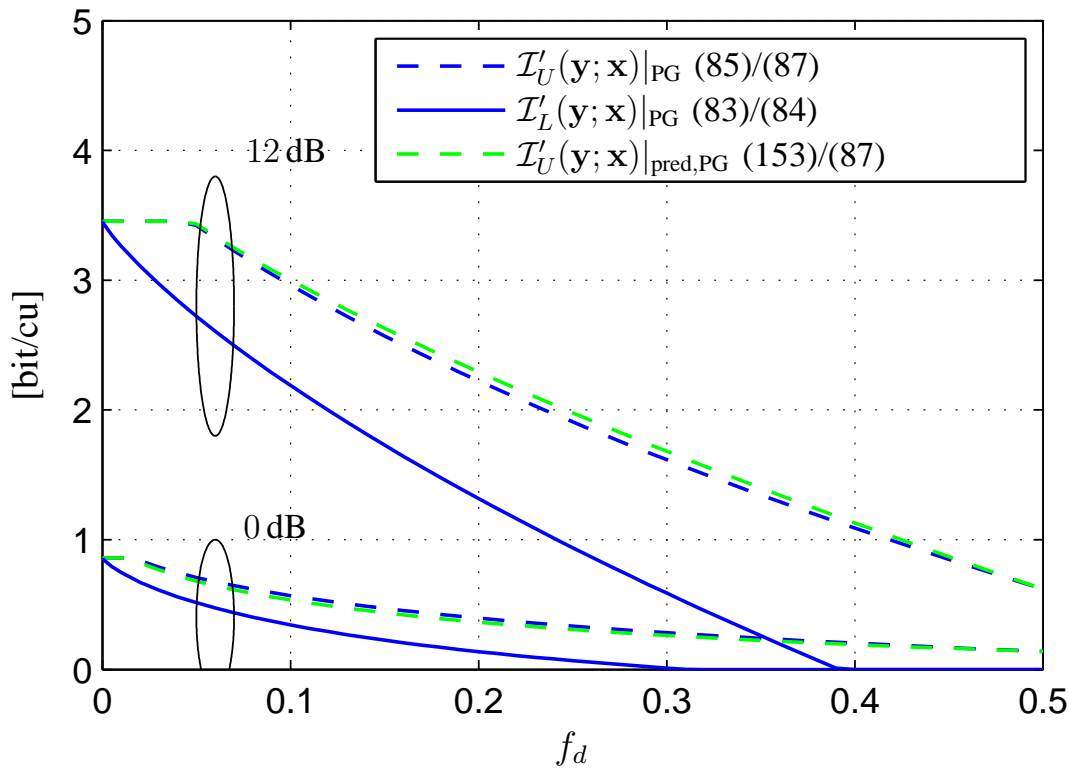}
\caption{Comparison of the upper bound on the achievable rate with i.i.d.\ zero-mean proper Gaussian inputs based on channel prediction (\ref{IYX_final_upp_pred_PG})/(\ref{MutUpp}) with the upper bound given in (\ref{MutUpp_unmo})/(\ref{MutUpp}); in addition the lower bound on the achievable rate with i.i.d.\ zero-mean proper Gaussian inputs (\ref{MutLow_unmo_rect})/(\ref{MutLow}) is shown; rectangular PSD $S_{h}(f)$}
    \label{Upper_Gauss_pred}
\end{figure}

Fig.~\ref{Upper_Gauss_pred} shows the prediction based upper bound on the achievable rate with i.i.d.\ zero-mean proper Gaussian input symbols given in (\ref{IYX_final_upp_pred_PG}) in comparison to the upper and lower bound on the achievable rate with i.i.d.\ zero-mean proper Gaussian inputs given Section~\ref{Sect_AchievRate_SISO} for a rectangular PSD of the channel fading process. Both upper bounds are shown in combination with the coherent upper bound, i.e., (\ref{MutUpp}) and (\ref{PerfKnow_Gen}).
A comparison of the prediction based upper bound (\ref{IYX_final_upp_pred_PG})/(\ref{MutUpp}) and the bound given in (\ref{MutUpp_unmo})/(\ref{MutUpp}) shows, that it depends on the channel parameters which one is tighter. It can easily be shown that for $f_{d}\rightarrow 0$ and for $f_{d}=0.5$ both bounds, i.e., (\ref{MutUpp_unmo}) and (\ref{IYX_final_upp_pred_PG}), are equal. For other $f_{d}$ it depends on the SNR $\rho$ which bound is tighter. An analytical comparison turns out to be difficult as in both cases we use a different way of lower-bounding $h'(\mathbf{y}|\mathbf{x})$.

\subsubsection{Peak Power Constrained Input Distributions}\label{Section_FinalBounds_Pred}
Now, we consider the case of a peak power constrained to $P_{\textrm{peak}}$ in addition to the average power constraint. With the nominal peak-to-average power ratio $\beta=P_{\textrm{peak}}/\sigma_{x}^{2}$, with (\ref{IYX_final_upp_pred}) we get the following upper bound on the achievable rate with i.i.d.\ input symbols:
\begin{align}
\sup_{\mathcal{P}_{\textrm{i.i.d.}}^{\textrm{peak}}}\mathcal{I}'(\mathbf{y};\mathbf{x})&\le\sup_{\alpha\in[0,1]}\sup_{\mathcal{P}_{\textrm{i.i.d.}}^{\textrm{peak}}\big|\alpha}\left\{\log\left(\alpha\rho+1\right)-\mathrm{E}_{x_{k}}\left[\log
\left(1+\frac{\sigma_{e_{\textrm{pred},\textrm{CM},\infty}}^{2}}{\sigma_{n}^{2}}|x_{k}|^{2}\right)\right]\right\}\nonumber\\
&\stackrel{(a)}{\le}\sup_{\alpha\in[0,1]}\left\{\log\left(\alpha\rho+1\right)-
\frac{\alpha}{\beta}\log\left(1+\frac{\sigma_{e_{\textrm{pred},\textrm{CM},\infty}}^{2}}{\sigma_{h}^{2}}\rho \beta\right)\right\}\label{IYX_final_upp_pred_Peak-1}
\end{align}
where $\mathcal{P}_{\textrm{i.i.d.}}^{\textrm{peak}}$ corresponds to $\mathcal{P}_{\textrm{i.i.d.}}$ but with the additional peak power constraint $|x_{k}|^{2}\le\beta \sigma_{x}^{2}$. $\mathcal{P}_{\textrm{i.i.d.}}^{\textrm{peak}}|\alpha$ corresponds to 
$\mathcal{P}_{\textrm{i.i.d.}}^{\textrm{peak}}$ in (\ref{SetDef_Piid2}) but with the average transmit power fixed to $\alpha\sigma_{x}^{2}$. Inequality (a) can be shown following an analogous argumentation as in Section~\ref{Section_Peak_UpperBound} from (\ref{UpperAchievRate_Peak_alphaSplit2}) to (\ref{UpperAchievRate_Peak_alphaSplit4}). Note that the prediction error variance $\sigma_{e_{\textrm{pred},\textrm{CM},\infty}}^{2}$ depends on $\alpha$. Now, we would have to calculate the supremum of the RHS of (\ref{IYX_final_upp_pred_Peak-1}) with respect to $\alpha$ which turns out to be difficult due to the dependency of $\sigma_{e_{\textrm{pred},\textrm{CM},\infty}}^{2}$ on $\alpha$. However, as $\sigma_{e_{\textrm{pred},\textrm{CM},\infty}}^{2}$ monotonically decreases with an increasing $\alpha$, and as the RHS of (\ref{IYX_final_upp_pred_Peak-1}) monotonically increases with a decreasing $\sigma_{e_{\textrm{pred},\textrm{CM},\infty}}^{2}$, we can upper-bound (\ref{IYX_final_upp_pred_Peak-1}) by setting $\alpha=1$ within $\sigma_{e_{\textrm{pred},\textrm{CM},\infty}}^{2}$ in (\ref{InfEstErrVar}), i.e.,\looseness-1
\begin{align}
\sup_{\mathcal{P}_{\textrm{i.i.d.}}^{\textrm{peak}}}\mathcal{I}'(\mathbf{y};\mathbf{x})&\le \sup_{\alpha\in[0,1]}\left\{\log\left(\alpha\rho+1\right)-\frac{\alpha}{\beta}\log\left(1+\frac{\sigma_{e_{\textrm{pred},\textrm{CM},\infty}}^{2}\big|_{\alpha=1}}{\sigma_{h}^{2}}\rho \beta\right)\right\}\nonumber\\
&=\log\left(\alpha_{\textrm{opt}}\rho+1\right)-
\frac{\alpha_{\textrm{opt}}}{\beta}\log\left(1+\frac{\sigma_{e_{\textrm{pred},\textrm{CM},\infty}}^{2}\big|_{\alpha=1}}{\sigma_{h}^{2}}\rho \beta\right)\nonumber\\
&=\mathcal{I}_{U}'(\mathbf{y};\mathbf{x})\big|_{\textrm{pred},P_{\textrm{peak}}}\label{IYX_final_upp_pred_Peak}
\end{align}
with
\begin{align}
\alpha_{\textrm{opt}}&=\min\left\{1,\left(\frac{1}{\beta}\log\left(1+\frac{\sigma_{e_{\textrm{pred},\textrm{CM},\infty}}^{2}\big|_{\alpha=1}}{\sigma_{h}^{2}}\rho\beta\right)\right)^{-1}-\frac{1}{\rho}\right\}.\label{alpha_opt_pred_}
\end{align}
For (\ref{alpha_opt_pred_}), we have used the fact that the argument of the supremum in the first line of (\ref{IYX_final_upp_pred_Peak}) is concave in $\alpha$, and, thus, there exists a unique maximum. 

\paragraph{Comparison to Capacity Bounds in \cite{SetWanHajLap07Arxiv} and \cite{SetHaj05}}
In the following, we will compare the upper bound on the achievable rate with peak power constrained i.i.d. input symbols in (\ref{IYX_final_upp_pred_Peak}) with the upper bound on the peak power constrained capacity given in \cite[Prop. 2.2]{SetWanHajLap07Arxiv}, which is, modified to our notation, given by
\begin{align}
C&\le\log\left(\alpha_{\textrm{opt}}\rho+1\right)-\frac{\alpha_{\textrm{opt}}}{\beta}\int_{-\frac{1}{2}}^{\frac{1}{2}}\log\left(\rho\beta\frac{S_{h}(f)}{\sigma_{h}^{2}}+1\right)df\label{SethurUpp}
\end{align}
with
\begin{align}
\alpha_{\textrm{opt}}&=\min\left\{1,\left(\frac{1}{\beta}\int_{-\frac{1}{2}}^{\frac{1}{2}}\log\left(\rho\beta\frac{S_{h}(f)}{\sigma_{h}^{2}}+1\right)df\right)^{-1}-\frac{1}{\rho}\right\}.\label{SethurUpp_AlphaOpt}
\end{align}
Note that, in terms of the analytical expression, (\ref{SethurUpp}) corresponds to (\ref{AchievRate_new_2+1}) for the special case of a rectangular PSD $S_{h}(f)$, see the discussion in Section~\ref{Section_Peak_UpperBound}.

On the other hand, we compare the upper bound on the achievable rate with peak power constrained i.i.d. input symbols in (\ref{IYX_final_upp_pred_Peak}) with the lower bound on the peak power constrained capacity given in \cite[(35)]{SetHaj05}
\begin{align}
C_{l1}(\rho)&=h(y_{k}|\hat{h}_{k})-\int_{-\frac{1}{2}}^{\frac{1}{2}}\log\left(\pi e\sigma_{n}^{2}\left(1+\rho \frac{S_{h}(f)}{\sigma_{h}^{2}}\right)\right)df\label{SethurLow}
\end{align}
where $k$ is an arbitrary chosen time instant with an infinitely long past and $h(y_{k}|\hat{h}_{k})$ is the differential output entropy while conditioning on the channel estimate $\hat{h}_{k}$, being given by the MMSE estimate $\mathrm{E}\left[h_{k}\big|\mathbf{x}_{-\infty}^{k-1},\mathbf{y}_{-\infty}^{k-1}\right]$, which is linear due to the fact that this estimation problem is jointly proper Gaussian. Based on the time-sharing argumentation, see Section~\ref{Sect_LowerBoundPeak}, an enhanced lower bound on the peak power constrained capacity is given by \cite[(29)/(35)]{SetHaj05}
\begin{align}
C&\ge \max_{\gamma \in[1,\beta]}\frac{1}{\gamma}C_{l1}(\rho\gamma).\label{SethurLow_TimeS}
\end{align}

\paragraph{Numerical Evaluation}
\begin{figure}
\centering
 \psfrag{SNR}[cc][cc][1.1]{SNR [dB]}
 \psfrag{bits}[cc][cc][1.1]{[bit/cu]}
 \psfrag{peak xxxxxxxxxxxxxxxxxxxxx}[cl][cl][1.06]{UB capacity, (\ref{SethurUpp})/(\ref{MutUpp})}
 \psfrag{peak predxxxxxxxxxxxxxx}[cl][cl][1.06]{UB i.i.d., (\ref{IYX_final_upp_pred_Peak})/(\ref{MutUpp})}
 \psfrag{PredLowxxxxxxxxxxxxxxxx}[cl][cl][1.06]{LB no time-share, (\ref{SethurLow})}
 \psfrag{PredLowTSxxxxxxxxxxxxxxxxx}[cl][cl][1.06]{\lower4mm\hbox{LB time-share $\gamma_{\textrm{opt}}$, (\ref{SethurLow_TimeS})/(\ref{SethurLow})}}
 \psfrag{CSI}[cc][rc][1.0]{\lower-4mm\hbox{\hspace{0.6cm}\shortstack[l]{perfect channel\\ state information}}}
 \psfrag{fd=0.1}[cc][cc][1.1]{\hspace{0.4cm}$f_{d}=0.1$}
 \psfrag{fd=0.01}[cc][cc][1.1]{\hspace{0.4cm}$f_{d}=0.01$}
 \psfrag{fd=0.001}[cc][cc][1.1]{\hspace{0.4cm}$f_{d}=0.001$}
\includegraphics[width=0.8\columnwidth]{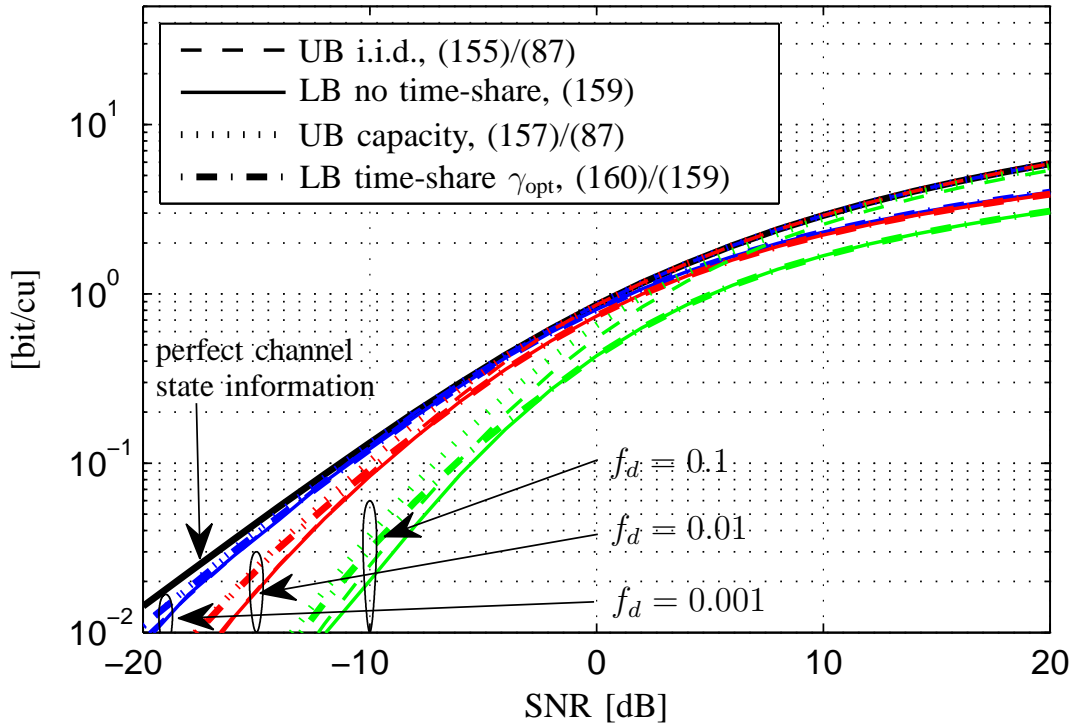}
    \caption{Comparison of the upper bound on the achievable rate with i.i.d.\ symbols and a peak power constraint given in (\ref{IYX_final_upp_pred_Peak})/(\ref{MutUpp}) based on channel prediction to the upper bound on the peak power constrained capacity given in \cite[Proposition 2.2]{SetWanHajLap07Arxiv}, i.e., (\ref{SethurUpp})/(\ref{MutUpp}), for $\beta=2$; in addition, the lower bound on the peak power constrained capacity \cite[(35)]{SetHaj05}, i.e., (\ref{SethurLow}), is shown for a constant modulus (CM) input distribution with $100$ signaling points, without and with time-sharing (\ref{SethurLow_TimeS}) ($\gamma_{\textrm{opt}}$); rectangular PSD $S_{h}(f)$} \label{Upper_Peak_pred}
\end{figure}

Fig.~\ref{Upper_Peak_pred} shows the upper bound on the achievable rate with i.i.d.\ input symbols and a peak power constraint based on the channel prediction error variance in (\ref{IYX_final_upp_pred_Peak})/(\ref{MutUpp}) in comparison to the upper bound on the peak power constrained capacity given in \cite[Prop. 2.2]{SetWanHajLap07Arxiv}, i.e., (\ref{SethurUpp}), combined with (\ref{MutUpp}), with $\beta=2$ for both. For comparison we use the lower bound on the peak power constrained capacity given in \cite[(35)]{SetHaj05}, i.e, (\ref{SethurLow}) based on a constant modulus input distribution with $100$ discrete signaling points with a uniform angular spacing. This approximates the case of a uniformly distributed phase. This lower bound is shown without time-sharing and with time-sharing ($\gamma_{\textrm{opt}}$), see (\ref{SethurLow_TimeS}) \footnote{Concerning the relation of the lower bound on the peak power constrained capacity in \cite[(35)/(29)]{SetHaj05}, i.e., (\ref{SethurLow_TimeS})/(\ref{SethurLow}), and the one used for comparison in Section~\ref{Sect_LowerBoundPeak}, i.e., (\ref{MutInfLowCM_Num}) respectively (\ref{LowMutInf_Peak}), see \cite{SetHaj05}.}. Note that the lower bound in (\ref{SethurLow}) is achievable with constant modulus input symbols with a uniformly distributed phase. Recall that time-sharing means, that the transmitter uses the channel only a $1/\gamma$ part of the time. Obviously, time-sharing is not in accordance with the assumption on i.i.d.\ input symbols. Therefore, the lower bound without time-sharing matches the new upper bound on the achievable rate with i.i.d.\ input symbols in (\ref{IYX_final_upp_pred_Peak})/(\ref{MutUpp}), while the lower bound with time-sharing ($\gamma_{\textrm{opt}}$) only matches the capacity upper bound in \cite[Prop. 2.2]{SetWanHajLap07Arxiv}, i.e., (\ref{SethurUpp})/(\ref{MutUpp}). From Fig.~\ref{Upper_Peak_pred} it can be seen that the upper bound on the achievable rate with i.i.d.\ input symbols in (\ref{IYX_final_upp_pred_Peak})/(\ref{MutUpp}) is lower or equal than the capacity upper bound in \cite[Prop. 2.2]{SetWanHajLap07Arxiv}, i.e., (\ref{SethurUpp})/(\ref{MutUpp}). However, (\ref{IYX_final_upp_pred_Peak})/(\ref{MutUpp}) is only an upper bound on the achievable rate with i.i.d.\ input symbols and not an upper bound on the capacity, as i.i.d.\ input symbols are in general not capacity achieving, see \cite{SetWanHajLap07Arxiv} and Section~\ref{Section_Peak_UpperBound}. This can also be seen, as the lower bound on the achievable rate with time-sharing is larger than the upper bound on the achievable rate with i.i.d.\ input symbols (\ref{IYX_final_upp_pred_Peak})/(\ref{MutUpp}) for very low SNRs. Furthermore, it is worth mentioning that for the case of a nominal peak-to-average power ratio $\beta=1$, the upper bound in (\ref{IYX_final_upp_pred_Peak}) and the one given in \cite[Prop. 2.2]{SetWanHajLap07Arxiv}, i.e., (\ref{SethurUpp}), coincide. In addition, the prediction based upper bound on the achievable rate in (\ref{IYX_final_upp_pred_Peak}) as well as the capacity upper bound in \cite[Prop. 2.2]{SetWanHajLap07Arxiv}, i.e., (\ref{SethurUpp}), both become loose for $\beta>1$ and high SNR or $\beta$ very large.

\section{Comparison to Synchronized Detection with a Pilot Based Channel Estimation}\label{Sect_SyncDet}
In typical mobile communication systems periodical pilot symbols are introduced into the transmit data sequence. The pilot symbol spacing $L$ is chosen such that the channel fading process is sampled at least with Nyquist frequency, i.e., $L<\lfloor 1/(2f_{d})\rfloor$. Based on these pilot symbols the channel is estimated, allowing for a coherent detection (synchronized detection). In conventional receivers, the channel estimation and the detection/decoding are two separate steps, such that the channel is estimated solely based on the pilot symbols. The resulting channel estimation error process is temporally correlated. However, performing coherent detection, the information contained in this temporal correlation is discarded. For a detailed discussion on this, we refer to \cite{Doer_IZS2010_acc}. The  channel estimation error leads to an SNR degradation. Bounds on the achievable rate for this separate processing have been given in \cite{BaFoMe01a}. For i.i.d.\ zero-mean proper Gaussian data symbols these bounds become
\begin{align}
\mathcal{R}_{\textrm{sep}}\ge\mathcal{R}_{L,\textrm{sep}}&=\frac{L-1}{L}\int_{z=0}^{\infty}\log\left(1+\rho\frac{1-\frac{\sigma_{e_{\textrm{pil}}}^{2}}{\sigma_{h}^{2}}}{1+\rho\frac{\sigma_{e_{\textrm{pil}}}^{2}}{\sigma_{h}^{2}}} z\right)e^{-z}dz\label{LowerBoundSD}\\
\mathcal{R}_{\textrm{sep}}\le\mathcal{R}_{U,\textrm{sep}}&=\mathcal{R}_{L,\textrm{sep}}+\frac{L-1}{L}\Bigg(\log\left(1+\rho\frac{\sigma_{e_{\textrm{pil}}}^{2}}{\sigma_{h}^{2}}\right)-\int_{z=0}^{\infty}\log\left(1+\rho\frac{\sigma_{e_{\textrm{pil}}}^{2}}{\sigma_{h}^{2}} z\right)e^{-z}dz\Bigg).\label{UpperBoundSD}
\end{align}
where $\sigma_{e_{\textrm{pil}}}^{2}$ is the channel estimation error variance when estimating the channel solely based on pilot symbols which is given by
\begin{align}
\sigma_{e_{\textrm{pil}}}^{2}&=\int_{f=-\frac{1}{2}}^{\frac{1}{2}}\frac{S_{h}(f)}{\frac{\rho}{L}\frac{S_{h}(f)}{\sigma_{h}^{2}}+1}df.
\end{align}
Based on the lower bound in (\ref{LowerBoundSD}) it can easily be seen that the achievable rate is decreased in comparison to perfect channel knowledge by two factors. First, symbol time instances that are used for pilot symbols are lost for data symbols leading to the pre-log factor $\frac{L-1}{L}$, and secondly, the average SNR is decreased by the factor $\left(1-\frac{\sigma_{e_{\textrm{pil}}}^{2}}{\sigma_{h}^{2}}\right)/\left(1+\rho\frac{\sigma_{e_{\textrm{pil}}}^{2}}{\sigma_{h}^{2}}\right)$ due to the channel estimation error variance. 

\begin{figure}
\centering
 \psfrag{fd}[cc][cc][1.2]{$f_{d}$}
 \psfrag{bits}[cc][cc][1.2]{[bit/cu]}
 \psfrag{12dB}[cc][cc][1.1]{$12\textrm{dB}$}
 \psfrag{6dB}[cc][cc][1.1]{$6\textrm{dB}$}
 \psfrag{0dB}[cc][cc][1.1]{$0\textrm{dB}$}
 \psfrag{UBIID}[lc][lc][1.1]{\lower3.5mm\hbox{UB i.i.d.\ PG (\ref{MutUpp_unmo})/(\ref{MutUpp})}}
 \psfrag{LBIID}[lc][lc][1.1]{\lower3.5mm\hbox{LB i.i.d.\ PG (\ref{MutLow_unmo_rect})/(\ref{MutLow})}}
 \psfrag{LBSep xxxxxxxxxxxxxxxxxx}[lc][lc][1.1]{SD $\mathcal{R}_{L,\textrm{sep}}$ (\ref{LowerBoundSD})}
 \psfrag{UBSep xxxxxxxxxxxxxxxxxx}[lc][lc][1.1]{SD $\mathcal{R}_{U,\textrm{sep}}$ (\ref{UpperBoundSD})}
\includegraphics[width=0.8\columnwidth]{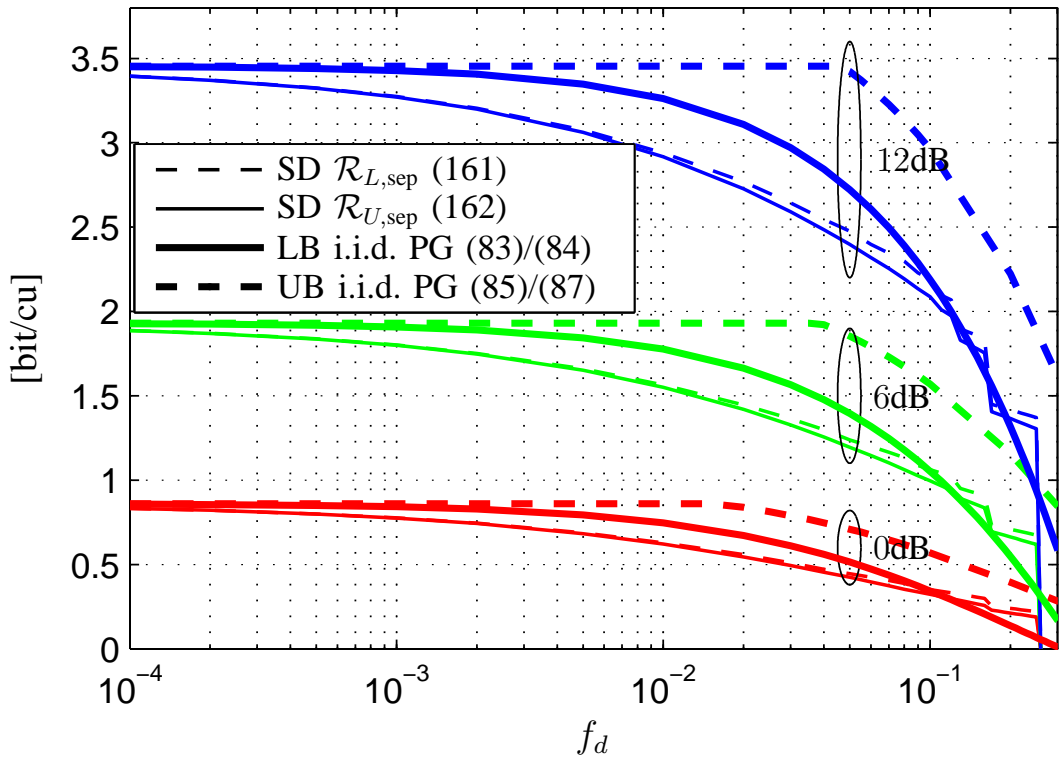}
    \caption{Comparison of the bounds on the achievable rate with synchronized detection and a solely pilot based channel estimate (SD) and the bounds on the achievable rate with i.i.d.\ zero-mean proper Gaussian (PG) input symbols given in (\ref{MutUpp_unmo})/(\ref{MutUpp}) and (\ref{MutLow_unmo_rect})/(\ref{MutLow}); the pilot spacing $L$ for synchronized detection is chosen such that $\mathcal{R}_{L,\textrm{sep}}$ is maximized; rectangular PSD $S_{h}(f)$} \label{FigCompIID_SDsep}
\end{figure}

Fig.~\ref{FigCompIID_SDsep} shows a comparison of the bounds on the achievable rate with synchronized detection based on a solely pilot based channel estimate in (\ref{LowerBoundSD}) and (\ref{UpperBoundSD}) with the bounds on the achievable rate with i.i.d.\ zero-mean proper Gaussian input symbols given in Section~\ref{Sect_AchievRate_SISO}. For synchronized detection with a solely pilot based channel estimation the pilot spacing has been chosen such that the lower bound on the achievable rate in (\ref{LowerBoundSD}) is maximized. As this lower bound is relatively tight, the chosen pilot spacing should be close to the one that maximizes the achievable rate with synchronized detection using a solely pilot based channel estimation. Obviously, for the practical important range of small channel dynamics, i.e., $f_{d}\ll0.1$, the achievable rate with synchronized detection using a solely pilot based channel estimation stays below the achievable rate with i.i.d.\ zero-mean proper Gaussian input symbols, indicating the possible gain when using enhanced receiver structures. Even in case of using pilot symbols, the receiver performance can be enhanced by using a joint processing of pilot and data symbols instead of a separate processing. For a more detailed discussion on the difference between separate and joint processing we refer to \cite{Doer_IZS2010_acc}. In this work also a lower bound on the achievable rate with joint processing of pilot and data symbols is given. One possibility of such a joint processing is to use an iterative code-aided channel estimation, where the channel estimation is enhanced based on reliability information on the data symbols delivered by the decoder. Based on this enhanced channel estimation detection and decoding is performed again, see e.g., \cite{Valenti2001} and \cite{SchmittMeyrZhang09TCOMAC}.\looseness-1

\section{Conclusion}\label{Sect_Conclusion}
The main focus of the present paper is the study of the achievable rate with i.i.d.\ zero-mean proper Gaussian input symbols on stationary Rayleigh flat-fading channels, where it is assumed that the receiver is aware of the law of the channel, but does not know its realization. We are interested in the achievable rate with i.i.d.\ zero-mean proper Gaussian input symbols, as this input distribution serves well to upper-bound the achievable rate with practical modulation and coding schemes. 

In the first part of this paper, i.e., in Section~\ref{SectionBoundsMain}, we have given a new upper bound on the achievable rate for i.i.d.\ zero-mean proper Gaussian input symbols, which holds in case of a rectangular PSD of the channel fading process. Furthermore, we also give a lower bound on the capacity which is achievable with i.i.d.\ zero-mean proper Gaussian input symbols. This lower bound is already known from \cite{DengHaim04}. With the upper and lower bound on the achievable rate for i.i.d.\ zero-mean proper Gaussian inputs, we have found a set of bounds, which is tight in the sense that their difference is bounded. We are able to bound this gap analytically by $(1+2f_{d})\gamma$ with the Euler constant $\gamma \approx 0.577 [\textrm{nat/cu}]$. Thus, for the specific case of proper Gaussian inputs we give bounds, which are tight (in the sense given above) over the whole SNR range. In contrast, available bounds on capacity often focus only on a specific SNR range, e.g., \cite{SetWanHajLap07Arxiv} discusses the low SNR regime whereas \cite{La05} considers the high SNR regime. 

The main novelty in this part of the paper lies in the new upper bound. It is based on a new lower bound on the conditional channel output entropy rate $h'(\mathbf{y}|\mathbf{x})$ for the special case of a rectangular PSD of the channel fading process. This bound is not based on a peak power constraint, and, therefore, allows to give an upper bound on the achievable rate with i.i.d.\ zero-mean proper Gaussian inputs. To the best of our knowledge, this is the only known upper bound on the achievable rate without a peak power constraint, which is tight in the sense that its slope (pre-log) corresponds to the slope of the lower bound on the capacity. However, for the derivation of our upper bound on the achievable rate we need the restriction to a rectangular PSD of the channel fading process.

Furthermore, the comparison of the bounds on the achievable rate with i.i.d.\ zero-mean proper Gaussian input symbols with the asymptotic bounds on the peak power constrained capacity given in \cite{La05} shows the interesting fact that the achievable rate with i.i.d.\ zero-mean proper Gaussian inputs is characterized by the same asymptotic high SNR slope as the peak power constrained capacity. This shows that this kind of input distribution is not highly suboptimal with respect to its high SNR performance.

Moreover, we have discussed the relation of the bounds on the achievable rate with i.i.d.\ zero-mean proper Gaussian input symbols to known bounds on the capacity with peak power constrained input symbols given in \cite{SetHaj05} and \cite{SetWanHajLap07Arxiv}. With respect to this, based on the given lower bound on $h'(\mathbf{y}|\mathbf{x})$, we have also derived an upper bound on the achievable rate with identically distributed (i.d.) peak power constrained input symbols, which is identified to be similar to an upper bound on capacity given in \cite{SetWanHajLap07Arxiv}. The assumption on i.d.\ input symbols is required in the derivation of our lower bound on $h'(\mathbf{y}|\mathbf{x})$. However, due to this restriction, with our derivation we are not able to show that the given upper bound on the achievable rate is an upper bound on the peak power constrained capacity. Furthermore, our derivation is restricted to a rectangular PSD of the channel fading process, whereas the upper bound on capacity given in \cite{SetWanHajLap07Arxiv} holds for an arbitrary PSD of the channel fading process. Concerning the lower bounds, the difference of the lower bound on the achievable rate with i.i.d.\ zero-mean proper Gaussian input symbols and the lower bound on the peak power constrained capacity given in \cite{SetHaj05} results mainly from the coherent mutual information, which is part of the lower bound in both cases.

In the second part of the present paper, i.e., in Section~\ref{Section_AlternativeIIDBound}, we have derived an alternative upper bound on the achievable rate with i.i.d.\ input symbols based on a prediction separation of the mutual information rate. Based on this separation, the conditional channel output entropy rate $h'(\mathbf{y}|\mathbf{x})$ can be expressed by the one-step channel prediction error variance, which is a well known result, see, e.g., \cite{La05}. We show for i.i.d. input symbols that the calculation of the prediction error variance $\sigma_{e_{\textrm{pred},\infty}}^{2}(\mathbf{x}_{-\infty}^{k-1})$ under the assumption of constant modulus symbols yields an upper bound on the achievable rate. As the constant modulus assumption is only used in the context of $\sigma_{e_{\textrm{pred},\infty}}^{2}(\mathbf{x}_{-\infty}^{k-1})$, we can still give upper bounds on the achievable rate for general i.i.d.\ input symbol distributions, even for the case without a peak power constraint. On the one hand, we evaluate this upper bound for i.i.d.\ zero-mean proper Gaussian input symbols. It depends on the channel parameters if this upper bound based on the channel prediction error variance given in (\ref{IYX_final_upp_pred_PG}) or the upper bound derived in Section~\ref{Sect_UpperAchieRateMath} is tighter. However, the prediction based bound is more general as it holds for arbitrary PSDs of the fading process with compact support and is not limited to rectangular PSDs as the one given in Section~\ref{Sect_UpperAchieRateMath}. On the other hand, we have evaluated the upper bound on the achievable rate based on the prediction error variance for peak power constrained input symbols. In this regard, we have observed that for nominal peak-to-average power ratios of $\beta=2$ and $\beta=1$ this upper bound on the achievable rate with i.i.d.\ input symbols is lower than or equal to the capacity upper bound in \cite[Prop. 2.2]{SetWanHajLap07Arxiv}. But, it is not an upper bound on the capacity due to the restriction to i.i.d.\ input symbols. We do not know if this ordering holds in general. 

Finally, in Section~\ref{Sect_SyncDet}, we have compared the bounds on the achievable rate with i.i.d.\ zero-mean proper Gaussian input symbols to bounds on the achievable rate with synchronized detection and a solely pilot based channel estimation. This comparison gives an indication of the possible gain when using enhanced receivers, e.g., receivers based on iterative code-aided channel estimation.

\appendices
\section{Approximation of a Rectangular PSD by an Absolutely Summable Autocorrelation Function}\label{Appendix_Approx_RectRC}
In this appendix, we show that the rectangular PSD in (\ref{Rect_PSD}), whose autocorrelation function is not absolutely summable but only square summable, see (\ref{CorrSquareSummableNew}), can be arbitrarily closely approximated by a PSD with an absolutely summable autocorrelation function, see (\ref{CorrAbsoluteSummableNew}).

The discrete-time autocorrelation function $r_{h}(l)$ corresponding to the rectangular PSD $S_{h}(f)\big|_{\textrm{Rect}}$ in (\ref{Rect_PSD}), which is given by
\begin{align}
r_{h}(l)&=\sigma_{h}^{2}\mathrm{sinc}(2f_{d}l)
\end{align}
is not absolutely summable. However, the rectangular PSD can be arbitrarily closely approximated by a PSD
with a shape corresponding to the transfer function of a raised cosine filter, i.e.,\looseness-1
\begin{align}
S_{h}(f)\big|_{\textrm{RC}}&=\left\{\begin{array}{ll}
\frac{\sigma_{h}^{2}}{2f_{d}} & |f|\le (1-\beta_{\textrm{ro}})f_{d}\\
\frac{\sigma_{h}^{2}}{4f_{d}}\left[1-\sin\left(\frac{\pi(f-f_{d})}{2\beta_{\textrm{ro}}{f_{d}}}\right)\right] &  (1-\beta_{\textrm{ro}})f_{d}< f\le (1+\beta_{\textrm{ro}})f_{d}\\
0 & f_{d}(1+\beta_{\textrm{ro}})< |f| \le 0.5
\end{array}\right.
\end{align}
which for $\beta_{\textrm{ro}}\rightarrow 0$ approaches the rectangular PSD $S_{h}(f)\big|_{\textrm{Rect}}$. Furthermore, the discrete-time autocorrelation function corresponding to $S_{h}(f)\big|_{\textrm{RC}}$ is given by
\begin{align}
r_{h}(l)&=\sigma_{h}^{2}\mathrm{sinc}(2f_{d}l)\frac{\cos\left(\beta_{\textrm{ro}} \pi 2f_{d}
l\right)}{1-4\beta_{\textrm{ro}}^{2}4f_{d}^{2}l^{2}}
\end{align}
which for $\beta_{\textrm{ro}}>0$ is absolutely summable. Thus, the rectangular PSD in (\ref{Rect_PSD}) can be arbitrarily closely approximated by a PSD with an absolutely summable autocorrelation function.

\section{Modified Upper Bound on $h'(\mathbf{y})$ for Gaussian Inputs}\label{hy_upp_num_uncorr}
In this appendix, we derive an alternative upper bound on the channel output entropy rate $h'(\mathbf{y})$ for the case of i.i.d.\ zero-mean proper Gaussian input symbols, which is tighter than the one given in (\ref{h_U2}). This derivation is based on work given in \cite{Perera}, \cite{PereraDiss06}. As its evaluation requires more complex numerical methods, we do not further use this bound, but give it for completeness of presentation.

Obviously, an upper bound on the entropy rate $h'(\mathbf{y})$ is given by assuming an uncorrelated channel fading process, i.e., its correlation matrix is assumed to be diagonal. This can be easily shown based on the chain rule for differential entropy
\begin{align}
h'(\mathbf{y})=\lim_{N\rightarrow \infty}\frac{1}{N}h(\mathbf{y})&=\lim_{N\rightarrow \infty}\frac{1}{N}\sum_{k=1}^{\infty}h(y_{k}|\mathbf{y}_{1}^{k-1})\nonumber\\
&\stackrel{(a)}{\le} \lim_{N\rightarrow \infty}\frac{1}{N}\sum_{k=1}^{\infty}h(y_{k})\nonumber\\
&\stackrel{(b)}{=}h(y_{k})\label{AppUpp_hy_gen}
\end{align}
where for (a) we have used the fact that conditioning reduces entropy and (b) is due to the ergodicity and the stationarity of the channel fading process and the assumption on i.i.d.\ input symbols. The major difference between this upper bound and the upper bound given in (\ref{h_U2}) is that the latter one implicitly corresponds to the case that the channel observations $y_{k}$ are proper Gaussian, while the RHS of (\ref{AppUpp_hy_gen}) still corresponds to the actual channel output entropy of the individual time instances. The upper bounding in (\ref{AppUpp_hy_gen}) only discards the temporal dependencies between the different observations.

In the following, we calculate the entropy $h(y_{k})$ for the case of zero-mean proper Gaussian input symbols with an average power $\sigma_{x}^{2}$
\begin{align}
h(y_{k})&=-\mathrm{E}_{y_{k}}\left[\log(p(y_{k}))\right]\nonumber\\
&=-\int_{\mathbb{C}}\int_{\mathbb{C}}p(y_{k}|x_{k})p(x_{k})dx_{k}\log\left(\int_{\mathbb{C}}p(y_{k}|x_{k})p(x_{k})dx_{k}\right)dy_{k}\nonumber\\
&=-\int_{0}^{\infty}\left[\int_{0}^{\infty}\frac{2|y|}{\sigma_{h}^{2}|x|^2+\sigma_{n}^{2}}e^{-\frac{|y|^2}{\sigma_{h}^{2}|x|^2+\sigma_{n}^{2}}}\frac{2|x|}{\sigma_{x}^{2}}e^{-\frac{|x|^2}{\sigma_{x}^{2}}}d|x|\right]\nonumber\\
&\qquad\times\log\left(\int_{0}^{\infty}\frac{2|y|}{\sigma_{h}^{2}|x|^2+\sigma_{n}^{2}}e^{-\frac{|y|^2}{\sigma_{h}^{2}|x|^2
+\sigma_{n}^{2}}}\frac{2|x|}{\sigma_{x}^{2}}e^{-\frac{|x|^2}{\sigma_{x}^{2}}}d|x|\right)d|y|\nonumber\\
&\quad+\log(2\pi)-\frac{\gamma}{2}+\frac{1}{2}\int_{z=0}^{\infty}\log\left(\sigma_{h}^{2}\sigma_{x}^{2}z+\sigma_{n}^{2}\right)e^{-z}dz\label{App_h_U}\\
&=h_{U_{2}}'(\mathbf{y})
\end{align}
where $\gamma\approx 0.57721$ is the Euler constant. To the best of our knowledge, the first integral in (\ref{App_h_U}) cannot be calculated analytically. However, it can be evaluated numerically using Hermite polynomials and Simpson's rule, see \cite{Perera}, \cite{PereraDiss06}, \cite{Steen}, or by Monte Carlo integration.

\begin{figure}
\centering
 \psfrag{SNRdB}[cc][cc][1.2]{$\rho$ [dB]}
 \psfrag{Diff}[cc][cc][1.2]{[nat/cu]}
 \psfrag{Delta1xxxx}[lc][lc][1.2]{$\Delta_{h'(\mathbf{y})}$}
 \psfrag{Delta2xxxx}[lc][lc][1.2]{$\Delta_{h'(\mathbf{y}),2}$}
 \psfrag{gam}[cc][cc][1.2]{$\gamma$}
    \includegraphics[width=0.8\columnwidth]{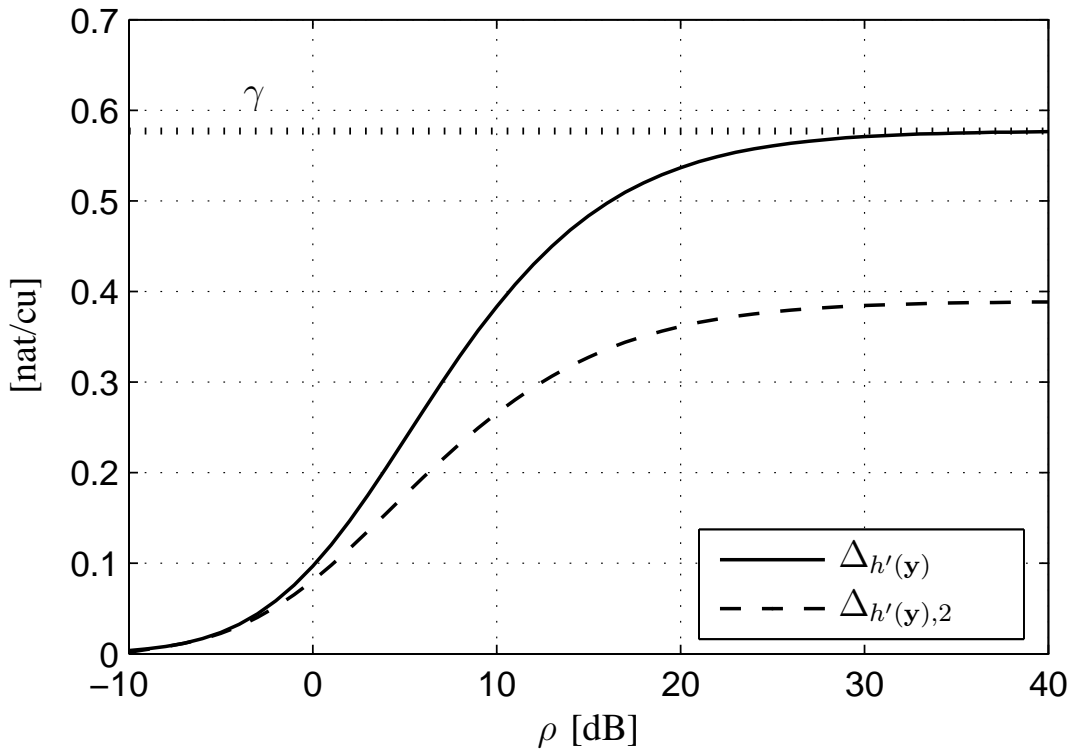}
    \caption{Comparison of $\Delta_{h'(\mathbf{y}),2}$ with $\Delta_{h'(\mathbf{y})}$}
    \label{Fig_Dif_hyU_hyL_new_ana_mod_bound_py}
\end{figure}

For the evaluation of the tightness of $h'_{U_{2}}(\mathbf{y})$, in Fig.~\ref{Fig_Dif_hyU_hyL_new_ana_mod_bound_py} the difference
\begin{align}
\Delta_{h'(\mathbf{y}),2}&=h'_{U_{2}}(\mathbf{y})-h'_{L}(\mathbf{y})
\end{align}
is shown in comparison to the difference $\Delta_{h'(\mathbf{y})}$ given in (\ref{Diff_UppLow_hy_simple}). Obviously, the upper bound $h_{U_{2}}'(\mathbf{y})$ is tighter than the upper bound $h_{U}'(\mathbf{y})$ given in (\ref{h_U2}).

\section{Sufficient Conditions for $\alpha_{\textrm{opt}}=1$ in (\ref{Eq_AlphaOpt})}\label{AppendixAlpha1Conditions}
In this appendix, we give conditions on the parameters $f_{d}$, $\rho$, and $\beta$ such that $\alpha_{\textrm{opt}}=1$ in (\ref{Eq_AlphaOpt}), i.e., the upper bound in (\ref{UpperAchievRate_Peak_alphaSplit4}) is maximized by choosing the maximum average power $\sigma_{x}^{2}$. Therefore, we have to evaluate for which parameter choice the following inequality holds:
\begin{align}
\left(\frac{2f_{d}}{\beta}\log\left(\frac{\rho\beta}{2f_{d}}+1\right)\right)^{-1}-\frac{1}{\rho}&\ge 1\nonumber\\
\Leftrightarrow \frac{2f_{d}}{\beta}\log\left(\frac{\rho\beta}{2f_{d}}+1\right)&\le \frac{\rho}{1+\rho}.\label{EqAppAlpha1Cond1}
\end{align}
The following calculations are closely related to a corresponding problem in \cite[Appendix C]{DurSchuBoelSha08IT_pub}. We divide the evaluation into the two cases $\rho>1$ and $\rho\le 1$.

For $\rho>1$, the RHS of (\ref{EqAppAlpha1Cond1}) can be lower-bounded by
\begin{align}
\frac{\rho}{1+\rho}&\ge \frac{1}{2}
\end{align}
yielding the following sufficient condition for (\ref{EqAppAlpha1Cond1}) to hold:
\begin{align}
\frac{2f_{d}}{\beta}\log\left(\frac{\rho\beta}{2f_{d}}+1\right)&\le \frac{1}{2}\nonumber\\
\Leftrightarrow \rho&\le \frac{2f_{d}}{\beta}\left[\exp\left(\frac{1}{2}\frac{\beta}{2f_{d}}\right)-1\right].
\end{align}
Thus, $\alpha_{\textrm{opt}}=1$ holds if
\begin{align}
1&<\rho\le \frac{2f_{d}}{\beta}\left[\exp\left(\frac{1}{2}\frac{\beta}{2f_{d}}\right)-1\right].\label{Con1_Appalpha}
\end{align}

Now, we discuss the case $\rho\le 1$. Using the inequality $\frac{1}{x}\log(x+1)\le \frac{1}{\sqrt{x+1}}$ for $x\ge 0$, for $\rho\le 1$ the LHS of (\ref{EqAppAlpha1Cond1}) can be upper-bounded by
\begin{align}
\frac{2f_{d}}{\beta}\log\left(\frac{\rho\beta}{2f_{d}}+1\right)&\le \frac{\rho}{\sqrt{\frac{\rho\beta}{2f_{d}}+1}}.\label{EqAppAlpha1Cond2}
\end{align}
Based on (\ref{EqAppAlpha1Cond2}), inequality (\ref{EqAppAlpha1Cond1}) holds if the following sufficient condition is fulfilled:
\begin{align}
\frac{\rho}{\sqrt{\frac{\rho\beta}{2f_{d}}+1}}&\le  \frac{\rho}{1+\rho}\nonumber\\
\Leftrightarrow 2f_{d}&\le \frac{\beta}{\rho+2}
\end{align}
so that we get the second condition
\begin{align}
2f_{d}&\le \frac{\beta}{\rho+2} \textrm{ for } \rho\le 1.\label{Con2_Appalpha}
\end{align}

Thus, if (\ref{Con1_Appalpha}) or (\ref{Con2_Appalpha}) is fulfilled, (\ref{Eq_AlphaOpt}) yields $\alpha_{\textrm{opt}}=1$.

\section{Convexity of (\ref{EquatForConvexityProof})}\label{App_ConvexityProof}
To prove that (\ref{EquatForConvexityProof}) is convex with respect to the individual diagonal elements of $\mathbf{Z}$, we rewrite the prediction error variance $\sigma_{e_{\textrm{pred}}}^{2}(\mathbf{x}_{1}^{N-1})=\sigma_{e_{\textrm{pred}}}^{2}(\mathbf{z})$ as follows:
\begin{align}
\sigma_{e_{\textrm{pred}}}^{2}(\mathbf{z})&=\sigma_{h}^{2}-\mathbf{r}_{h,\textrm{pred}}^{H}\left(\mathbf{R}_{h}+\sigma_{n}^{2}\mathbf{Z}^{-1}\right)^{-1}\mathbf{r}_{h,\textrm{pred}}\nonumber\\
&\stackrel{(a)}{=}\sigma_{h}^{2}-\mathbf{r}_{h,\textrm{pred}}^{H}\left(\mathbf{R}_{h}^{-1}-\mathbf{R}_{h}^{-1}\left(\frac{\mathbf{Z}}{\sigma_{n}^{2}}+\mathbf{R}_{h}^{-1}\right)^{-1}\mathbf{R}_{h}^{-1}\right)\mathbf{r}_{h,\textrm{pred}}\nonumber\\
&\stackrel{(b)}{=}\sigma_{h}^{2}-\mathbf{r}_{h,\textrm{pred}}^{H}\left(\mathbf{R}_{h}^{-1}-\mathbf{R}_{h}^{-1}\left(\frac{z_{i}\mathbf{V}_{i}+\mathbf{Z}_{\backslash i}}{\sigma_{n}^{2}}+\mathbf{R}_{h}^{-1}\right)^{-1}\mathbf{R}_{h}^{-1}\right)\mathbf{r}_{h,\textrm{pred}}\nonumber\\
&=\sigma_{h}^{2}-\mathbf{r}_{h,\textrm{pred}}^{H}\left(\mathbf{R}_{h}^{-1}\!-\mathbf{R}_{h}^{-1}\Bigg[\!\Bigg(\frac{1}{\sigma_{n}^{2}}\mathbf{Z}_{\backslash i}+\mathbf{R}_{h}^{-1}\right)\left\{\!\left(\frac{1}{\sigma_{n}^{2}}\mathbf{Z}_{\backslash i}+\mathbf{R}_{h}^{-1}\right)^{-1}\!\frac{z_{i}}{\sigma_{n}^{2}}\mathbf{V}_{i}+\mathbf{I}\right\}\!\Bigg]^{-1}\mathbf{R}_{h}^{-1}\Bigg)\mathbf{r}_{h,\textrm{pred}}\nonumber\\
&=\sigma_{h}^{2}-\mathbf{r}_{h,\textrm{pred}}^{H}\Bigg(\mathbf{R}_{h}^{-1}-\mathbf{R}_{h}^{-1}\left\{\!\left(\frac{1}{\sigma_{n}^{2}}\mathbf{Z}_{\backslash i}+\mathbf{R}_{h}^{-1}\!\right)^{\!-1}\frac{z_{i}}{\sigma_{n}^{2}}\mathbf{V}_{i}+\mathbf{I}\right\}^{\!-1}\left(\frac{1}{\sigma_{n}^{2}}\mathbf{Z}_{\backslash i}+\mathbf{R}_{h}^{-1}\right)^{-1}\mathbf{R}_{h}^{-1}\Bigg)\mathbf{r}_{h,\textrm{pred}}\nonumber\\
&\stackrel{(c)}{=}\sigma_{h}^{2}-\mathbf{r}_{h,\textrm{pred}}^{H}\Bigg(\mathbf{R}_{h}^{-1}\!-\mathbf{R}_{h}^{-1}
\Bigg\{\mathbf{I}-\frac{z_{i}}{1+z_{i}\lambda_{\textrm{max}}}\!\left(\frac{\mathbf{Z}_{\backslash
i}}{\sigma_{n}^{2}}+\mathbf{R}_{h}^{-1}\!\right)^{-1}\frac{\mathbf{V}_{i}}{\sigma_{n}^{2}}\Bigg\}\!\left(\frac{\mathbf{Z}_{\backslash i}}{\sigma_{n}^{2}}+\mathbf{R}_{h}^{-1}\right)^{-1}\!\mathbf{R}_{h}^{-1}\!\Bigg)\mathbf{r}_{h,\textrm{pred}}\nonumber\\
&=\sigma_{h}^{2}-\mathbf{r}_{h,\textrm{pred}}^{H}\left(\mathbf{R}_{h}^{-1}-\mathbf{R}_{h}^{-1}
\left(\frac{\mathbf{Z}_{\backslash i}}{\sigma_{n}^{2}}+\mathbf{R}_{h}^{-1}\right)^{-1}\mathbf{R}_{h}^{-1}\right)\mathbf{r}_{h,\textrm{pred}}\nonumber\\
&\quad
-\frac{z_{i}\mathbf{r}_{h,\textrm{pred}}^{H}\mathbf{R}_{h}^{-1}\left(\frac{\mathbf{Z}_{\backslash
i}}{\sigma_{n}^{2}}+\mathbf{R}_{h}^{-1}\right)^{-1}\frac{\mathbf{V}_{i}}{\sigma_{n}^{2}}\left(\frac{\mathbf{Z}_{\backslash i}}{\sigma_{n}^{2}}+\mathbf{R}_{h}^{-1}\right)^{-1}\mathbf{R}_{h}^{-1}\mathbf{r}_{h,\textrm{pred}}}{1+z_{i}\lambda_{\textrm{max}}}\nonumber\\
&\stackrel{(d)}{=}\sigma_{e_{\textrm{pred}}}^{2}(\mathbf{z}_{\backslash
i})-\frac{z_{i}\cdot a}{1+z_{i}\lambda_{\textrm{max}}}\label{PredErrVarIsolatedzi}
\end{align}
where for (a) we have used the matrix inversion lemma, and for (b) we have separated the diagonal matrix $\mathbf{Z}$ as follows:
\begin{align}
\mathbf{Z}&=\mathbf{Z}_{\backslash i}+z_{i}\mathbf{V}_{i}
\end{align}
where $\mathbf{Z}_{\backslash i}$ corresponds to $\mathbf{Z}$ except that the $i$-th diagonal element is set
to $0$, $\mathbf{V}_{i}$ is a matrix with all elements zero except of the $i$-th diagonal element being equal
to $1$, and $z_{i}$ is the $i$-th diagonal element of the matrix $\mathbf{Z}$. In addition, for (c) we have used the Sherman-Morrison formula and $\lambda_{\textrm{max}}$ is the non-zero eigenvalue of the rank one matrix
\begin{align}
\mathbf{B}&=\left(\frac{1}{\sigma_{n}^{2}}\mathbf{Z}_{\backslash
i}+\mathbf{R}_{h}^{-1}\right)^{-1}\frac{1}{\sigma_{n}^{2}}\mathbf{V}_{i}.
\end{align}
Furthermore, for (d) we substituted $\sigma_{e_{\textrm{pred}}}^{2}(\mathbf{z}_{\backslash
i})$ for
\begin{align}
\sigma_{e_{\textrm{pred}}}^{2}(\mathbf{z}_{\backslash
i})&=\sigma_{h}^{2}-\mathbf{r}_{h,\textrm{pred}}^{H}\left(\mathbf{R}_{h}^{-1}-\mathbf{R}_{h}^{-1}\left(\frac{\mathbf{Z}_{\backslash i}}{\sigma_{n}^{2}}+\mathbf{R}_{h}^{-1}\right)^{-1}\mathbf{R}_{h}^{-1}\right)\mathbf{r}_{h,\textrm{pred}}\nonumber
\end{align}
which is the prediction error variance if the observation at the $i$-th time instant is not used for channel prediction. Additionally, for (d) we have also used the substitution
\begin{align}
a&=\mathbf{r}_{h,\textrm{pred}}^{H}\mathbf{R}_{h}^{-1}\left(\frac{\mathbf{Z}_{\backslash
i}}{\sigma_{n}^{2}}+\mathbf{R}_{h}^{-1}\right)^{-1}\frac{\mathbf{V}_{i}}{\sigma_{n}^{2}}\left(\frac{\mathbf{Z}_{\backslash i}}{\sigma_{n}^{2}}+\mathbf{R}_{h}^{-1}\right)^{-1}\mathbf{R}_{h}^{-1}\mathbf{r}_{h,\textrm{pred}}\nonumber\\
&\ge 0\label{aIsPositiveProofPred}
\end{align}
where the nonnegativity follows as $\mathbf{V}_{i}$ is positive semidefinite.

Thus, with (\ref{PredErrVarIsolatedzi}) we have found a separation of the channel prediction error variance $\sigma_{e_{\textrm{pred}}}^{2}(\mathbf{z})$ into the term $\sigma_{e_{\textrm{pred}}}^{2}(\mathbf{z}_{\backslash
i})$ being independent of $z_{i}$, and an additional term, which depends on $z_{i}$. Note that $a$ and $\lambda_{\textrm{max}}$ in the second term on the RHS of (\ref{PredErrVarIsolatedzi}) are independent of $z_{i}$ and that the element $i$ is an arbitrarily chosen element. I.e., we can use this separation for each diagonal element of the matrix $\mathbf{Z}$.

By substituting the RHS of (\ref{PredErrVarIsolatedzi}) into (\ref{EquatForConvexityProof}) we get
\begin{align}
\log\left(1+\frac{|x_{N}|^{2}}{\sigma_{n}^{2}}\left(\sigma_{e_{\textrm{pred}}}^{2}(\mathbf{z}_{\backslash
i})-\frac{z_{i}\cdot a}{1+z_{i}\lambda_{\textrm{max}}}\right)\right)&=K.\label{ProofPredConvexFirstEquat_mod}
\end{align}

Recall that we want to show the convexity of (\ref{ProofPredConvexFirstEquat_mod}) with respect to the element $z_{i}$. Therefore, we calculate its second derivative with respect to $z_{i}$ which is given by
\begin{align}
\frac{\partial^{2}K}{(\partial z_{i})^{2}}&=
\frac{\frac{|x_{N}|^{2}}{\sigma_{n}^{2}}\frac{a 2\lambda_{\textrm{max}}(1+z_{i}\lambda_{\textrm{max}})}{(1+z_{i}\lambda_{\textrm{max}})^{4}}\left\{1+\frac{|x_{N}|^{2}}{\sigma_{n}^{2}} \left(\sigma_{e_{\textrm{pred}}}^{2}(\mathbf{z}_{\backslash
i})-\frac{a\left(z_{i}+\frac{1}{2\lambda_{\textrm{max}}}\right)}{1+z_{i}\lambda_{\textrm{max}}}\right)\right\}} {\left(1+\frac{|x_{N}|^{2}}{\sigma_{n}^{2}}\left(\sigma_{e_{\textrm{pred}}}^{2}(\mathbf{z}_{\backslash
i})-\frac{a z_{i}}{1+z_{i}\lambda_{\textrm{max}}}\right)\right)^{2}}\nonumber
\end{align}
and we will show that it is nonnegative, i.e.,
\begin{align}
\frac{\partial^{2}K}{(\partial z_{i})^{2}}&\ge 0.\label{PositivitySecondDeiv_ConvexityProofPred}
\end{align}
Therefore, first we show that $\lambda_{\textrm{max}}$ is nonnegative. This can be done based on the definition of the eigenvalues of the matrix $\mathbf{B}$\looseness-1
\begin{align}
\mathbf{B}\mathbf{u}&=\left(\frac{1}{\sigma_{n}^{2}}\mathbf{Z}_{\backslash
i}+\mathbf{R}_{h}^{-1}\right)^{-1}\frac{1}{\sigma_{n}^{2}}\mathbf{V}_{i}\mathbf{u}=\lambda_{\textrm{max}}\mathbf{u}\nonumber\\
\Rightarrow \frac{1}{\sigma_{n}^{2}}\mathbf{u}^{H}\mathbf{V}_{i}\mathbf{u}&= \lambda_{\textrm{max}}\mathbf{u}^{H}\left(\frac{1}{\sigma_{n}^{2}}\mathbf{Z}_{\backslash
i}+\mathbf{R}_{h}^{-1}\right)\mathbf{u}\nonumber\\
\stackrel{(a)}{\Rightarrow} \lambda_{\textrm{max}}&\ge 0\nonumber
\end{align}
where (a) follows from the fact that the eigenvalues of $\left(\frac{1}{\sigma_{n}^{2}}\mathbf{Z}_{\backslash
i}+\mathbf{R}_{h}^{-1}\right)$ are nonnegative, as $\mathbf{R}_{h}$ is positive definite and the diagonal entries of the diagonal matrix $\mathbf{Z}_{\backslash
i}$ are also nonnegative. In addition, $\mathbf{V}_{i}$ is also positive semidefinite.

With $\lambda_{\textrm{max}}$, $z_{i}$, and $a$ being nonnegative, for the proof of (\ref{PositivitySecondDeiv_ConvexityProofPred}), it rests to show that
\begin{align}
\sigma_{e_{\textrm{pred}}}^{2}(\mathbf{z}_{\backslash
i})-\frac{a}{1+z_{i}\lambda_{\textrm{max}}}\left(z_{i}+\frac{1}{2\lambda_{\textrm{max}}}\right)&\ge 0.\label{InequalityKeyProofConvexityPred}
\end{align}
To prove this inequality, we calculate the derivative of the LHS of (\ref{InequalityKeyProofConvexityPred}) with respect to $z_{i}$, which is given by
\begin{align}
\frac{\partial }{\partial z_{i}}\left\{\sigma_{e_{\textrm{pred}}}^{2}(\mathbf{z}_{\backslash
i})-\frac{a\left(z_{i}+\frac{1}{2\lambda_{\textrm{max}}}\right)}{1+z_{i}\lambda_{\textrm{max}}}\right\}
&=-\frac{a}{2(1+z_{i}\lambda_{\textrm{max}})^{2}}\le 0
\end{align}
where for the last inequality we have used (\ref{aIsPositiveProofPred}). I.e., the LHS of (\ref{InequalityKeyProofConvexityPred}) monotonically decreases in $z_{i}$. Furthermore, for $z_{i}\rightarrow \infty$ the LHS of (\ref{InequalityKeyProofConvexityPred}) becomes
\begin{align}
\lim_{z_{i}\rightarrow \infty} \left\{\sigma_{e_{\textrm{pred}}}^{2}(\mathbf{z}_{\backslash
i})-\frac{a\left(z_{i}+\frac{1}{2\lambda_{\textrm{max}}}\right)}{1+z_{i}\lambda_{\textrm{max}}} \right\} 
&\stackrel{(a)}{=}\lim_{z_{i}\rightarrow \infty} \sigma_{e_{\textrm{pred}}}^{2}(\mathbf{z})\stackrel{(b)}{\ge} 0
\end{align}
where (a) follows due to (\ref{PredErrVarIsolatedzi}), and where (b) holds as the prediction error variance must be nonnegative. As the LHS of (\ref{InequalityKeyProofConvexityPred}) is monotonically decreasing in $z_{i}$ and as its limit for $z_{i}\rightarrow \infty$ is nonnegative, (\ref{InequalityKeyProofConvexityPred}) must hold.

Thus, with (\ref{InequalityKeyProofConvexityPred}) inequality (\ref{PositivitySecondDeiv_ConvexityProofPred}) holds and, thus, (\ref{ProofPredConvexFirstEquat_mod}) is convex in $z_{i}$. As the element $i$ has been chosen arbitrarily, in conclusion, we have shown that (\ref{EquatForConvexityProof}) is convex in each $z_{i}$ for $i=1, \hdots, N-1$.

\ifCLASSOPTIONcaptionsoff
  \newpage
\fi

\bibliographystyle{IEEEtran}
\bibliography{Bib_mod_IEEE}

\end{document}